\documentclass[a4paper,11pt]{article}
\pdfoutput=1
\usepackage{graphicx}
\usepackage{amsfonts,amssymb,slashed,latexsym,setspace,cancel,subfigure}
\usepackage{jheppub}

\def\nn{\nonumber\\*}

\newcommand{\beqa}{\begin{eqnarray}}
\newcommand{\eeqa}{\end{eqnarray}}

\newcommand{\be}{\begin{equation}}
\newcommand{\ee}{\end{equation}}
\def\fig#1{Fig.~\ref{#1}}
\def\tabl#1{Table~\ref{#1}}
\def\Eqn#1{Eq.\ (\ref{#1})}
\def\rpla#1{}
\def\rplb#1{#1}

\def\OOL{-\frac{\alpha_\lambda}{8\pi}x^2 h(x)}
\def\TwOL{-\frac{\alpha_\lambda}{4\pi}x^2 h(x)}
\def\ThOL{-3 \frac{\alpha_\lambda}{8\pi}x^2 h(x)}

\def\OOLo{-\frac{\alpha_\lambda}{24\pi}x^2 h(x)}
\def\TwOLo{-\frac{\alpha_\lambda}{12\pi}x^2 h(x)}
\def\ThOLo{- \frac{\alpha_\lambda}{8\pi}x^2 h(x)}

\def\SOLo{-\frac{\alpha_\lambda}{4\pi}x^2 h(x)}

\title{\bf Higgs mass from neutrino-messenger mixing}

\author[a,b]{Pritibhajan Byakti,}
\author[a]{Charanjit K Khosa,}
\author[c]{V. S. Mummidi,}
\author[a]{Sudhir K Vempati}

\affiliation[a]{Center for High Energy Physics,\\
Indian Institute of Science, Bangalore 560012, INDIA}
\affiliation[b]{Department of Theoretical Physics,\\
              Indian Association for the Cultivation of Science,\\
              2A $\&$ 2B Raja S.C. Mullick Road,
              Kolkata 700 032, INDIA}
\affiliation[c]{Harish-Chandra Research Institute,\\ 
Chhatnag Road, Jhusi, Allahabad 211019, INDIA}

\emailAdd{tppb@iacs.res.in}
\emailAdd{khosacharanjit@chep.iisc.ernet.in}
\emailAdd{venkatasuryanarayana@hri.res.in}
\emailAdd{vempati@chep.iisc.ernet.in}

\date{}

\abstract{ The discovery of the Higgs particle at 125 GeV has put strong
constraints on minimal messenger models of gauge mediation, pushing the
stop masses into the multi-TeV regime.  Extensions of these models with  
matter-messenger mixing terms have been proposed to generate a large  trilinear 
parameter, $A_t$, relaxing these constraints. The detailed survey of these 
models \cite{Byakti:2013ti,Evans:2013kxa} so far considered messenger mixings
 with only MSSM superfields. In the present work, we extend the survey to MSSM 
with inverse-seesaw mechanism. The neutrino-sneutrino corrections to the Higgs 
mass in the inverse seesaw model are not significant in the minimal gauge 
mediation model, unless one considers messenger-matter interaction terms. We  
classify all possible models with  messenger-matter  interactions and perform 
thorough numerical analysis to find out the promising  models. We found that out 
of the 17 possible models 9 of them can lead to Higgs mass within the 
observed 
value without raising the sfermion masses significantly. The successful  models 
have  stop masses $\sim $1.5 TeV with small or negligible mixing
and yet a light CP even Higgs at 125 GeV.} 
\begin{document}
\maketitle

\section{Introduction}
Supersymmetry (SUSY) \cite{Salam:1974yz, Fayet:1976cr, Nilles:1983ge, 
Wess:1992cp, Martin:1997ns} offers one of the most elegant solutions to the 
hierarchy problem. In the Minimal Supersymmetric Standard Model (MSSM) 
\cite{Fayet:1976et, Martin:1997ns, Vempati:2012np}, the Higgs mass is protected 
from the dangerous UV sensitive radiative corrections. However, for various 
reasons, supersymmetry breaking cannot be incorporated in the MSSM in a 
straightforward way. SUSY is spontaneously broken  in a remote sector and the 
information is then passed to the MSSM sector through mediators.  Among the 
different types of mediation mechanisms, gauge mediation  (GMSB in short, 
for a review see \cite{Giudice:1998bp}) is interesting as  it  generates  soft 
parameters which  
are flavor diagonal.

The discovery of   $\sim$125 GeV CP  even neutral Higgs boson\cite{Aad:2012tfa, 
Chatrchyan:2012xdj} has however imposed strong constraints on GMSB models. To 
accommodate the light CP-even Higgs boson of that mass range,  the  spectrum of 
GMSB models had to become heavy \cite{Draper:2011aa}. Such 
a heavy spectrum is not `natural' as it leads to larger fine tuning. Secondly,  
there is a bleak chance to discover any such particle at Large Hadron Collider 
(LHC). This is true for all the GMSB models which are characterized by small 
A-terms, including the most general one of general gauge 
mediation (GGM) \cite{Intriligator:2010be,Grajek:2013ola}. 

Several solutions have been put forward to remedy this situation. They can  
mainly be divided into two classes : (a) models which  generate  large A-terms 
through some mechanism\cite{Byakti:2013ti, Evans:2013kxa, Kang:2012ra, 
Albaid:2012qk, Han:1998xy, Craig:2012xp, Evans:2011bea, Martin:2012dg, 
Casas:2016xnl, Ding:2015vla, Delgado:2015bwa, Jelinski:2015gsa, Evans:2015swa, 
Backovic:2015rwa, Basirnia:2015vga, 
Fischler:2013tva, Jelinski:2013kta, Calibbi:2013mka} 
(b)  models which augment to the Higgs mass through 
additional 
contributions while keeping the A-terms small.  The former class is dominated by 
models which contain new interactions between messenger and matter fields. These 
generate the required A-terms for the stop sector, though some of them could 
suffer from other problems like $A_t/m^2 $ problem \cite{Craig:2012xp}.  In the second 
category several strategies are proposed,  for example,  $U(1)$ gauge group 
extension\cite{Mummidi:2013hba},   NMSSM and/or vector matter \cite{Babu:2008ge, 
Endo:2011mc, Endo:2011xq, Nakayama:2012zc, Craig:2012xp, 
 Asano:2015ana, Mittag:2015qpa, Kowalska:2015wua, 
Domingo:2015qaa, Cahill-Rowley:2014ora, Ellwanger:2014hia, Athron:2014yba, 
Allanach:2013kza, Yanagida:2012ef, deLaPuente:2013uba},  SO(10) D terms \cite{Krauss:2013jva}
 to name a few. Another 
way is to have an additional source of supersymmetry breaking, preferably  
mediated by gravitational interactions, such that  it dominantly generates $A_t$ 
and other related soft terms\cite{Iyer:2014iha}. 

In the present work we will focus on the first class of models with  
messenger-matter interactions. A classification of all such models has been 
presented in Refs. \cite{ Byakti:2013ti,  Evans:2013kxa}) for MSSM.  The 
classification in \cite{Evans:2013kxa} concentrated on the messenger 
interactions with hadronic matter  fields $Q, U^c$ and the Higgs 
field $H_u$ which are relevant for  the $A_t$ and other trilinear parameters. 
In Ref.\cite{Byakti:2013ti}, messenger  and matter fields interact SU(5) 
multiplet-wise. As a consequence  other fields like $D^c, L, E^c$ and $H_d$ also 
interact with the messengers in 
the studied models. In MSSM, the messenger matter interactions involving
leptonic fields will not play any role in the generation of $A_t$ or  on the 
Higgs mass. However, the  situation changes in the presence of an  `inverse' 
seesaw mechanism \cite{Elsayed:2011de}. 

The standard seesaw mechanism with right handed neutrinos can have large Yukawa 
couplings $\sim \mathcal{O}(y_t)$, the corrections to the Higgs mass are tiny 
as the right handed neutrinos are very heavy, close to the GUT scale to give 
the correct neutrino masses (see Ref. \cite{Draper:2013ava} and references 
there in). On the other hand, inverse seesaw mechanism has 
 additional singlets by which the right handed neutrino masses need not 
be very  heavy and this enables corrections to the Higgs mass 
\cite{Elsayed:2011de, Guo:2013sna, Chun:2014tfa} which can be significant in 
some regions of the parameter space.  There is however a caveat: the 
neutrino-sneutrino radiative  corrections to the Higgs 
 mass are different compared to the top-stop corrections.  In the limit  of 
large right handed neutrino masses, $m_R \gg m_{\tilde{\nu}}$ the  neutrino and 
the sneutrino corrections  to the  Higgs mass cancel each other leading to 
negligible enhancement to the Higgs mass\cite{Guo:2013sna, Chun:2014tfa}. However, 
there are two situations when the 
corrections to the Higgs mass can be significant: (a)  if the slepton and 
sneutrino masses are  comparable to  $\sim m_R$, typically in the multi-TeV 
regime and (b) the trilinear parameter associated  with the neutrino Yukawa, 
$X_N$ is large, leading to large mixing in the sneutrino sector. \rplb{For 
case: (a) large slepton  masses are  not natural  in minimal versions of GMSB. 
One possibility is to consider GGM boundary conditions with a  separate and 
large slepton masses as boundary conditions\cite{Chun:2014tfa}.  For case:(b), 
trilinear parameters are generally small in minimal messenger models of gauge 
mediation.  To generate large trilinear sneutrino mixing parameter,  we 
consider matter-messenger  mixing in the present paper.}

As mentioned earlier, we extended the classification of the messenger-matter
interaction models to the lepton and neutrino fields.  We found that there are 
17 models which are tabulated in \tabl{t:class}. Considering GMSB 
boundary conditions along with these neutrino-messenger couplings we show 
that light stops can give Higgs mass $\sim 125$ GeV in nine of these models. In these 
mixing models, only the third generation is allowed to couple with messengers. 
Hence we are safe from flavor constraints.

The paper organizes as follows. In \S\ \ref{s:recap} we summarize  
gauge mediated SUSY breaking with and without mixing. Then we discuss inability 
of inverse seesaw models to produce correct Higgs mass in \S\ \ref{s:mes-mix} 
and motivate  the study of messenger-matter interaction involving leptons  and 
right handed neutrino fields. We classify the models based on the 
messenger-matter interactions. In \S\ \ref{s:analysis} classified models are 
analyzed in detail. Finally in \S\ \ref{s:con} we conclude.

\section{Recap of GMSB with and without matter-messenger mixing terms}\label{s:recap}
Gauge mediated SUSY breaking models consist of three sectors: (a) visible 
sector, (b) messenger sector, and (c) hidden sector. We do not know much about 
the hidden sector. However it is assumed that SUSY is spontaneously broken 
there and information of SUSY breaking is encoded in the spurion field $X$. 
Vacuum expectation value (VEV) of this spurion field is: $\langle X \rangle = M 
+ \theta^2 F$ where $M$ is the messenger scale and $F$ is SUSY breaking VEV. 
The spurion field has superpotential level interaction with the 
messenger fields $\Phi_m$ as follows:
\begin{equation}\label{e:mspot}
W_{\rm mes} = f X \sum_i \bar \Phi_{im} \Phi_{im},
\end{equation}
where superfield $\bar\Phi_m$ is conjugate representation of $\Phi_m$ under SM 
gauge group. In principle one can have complicated version of the above model; 
however, it is the simplest one and  is called minimal GMSB (mGMSB) model. In 
general, messenger fields are multiplets of SU(5) like 5, 10 and 15 dimensional 
representations. Messenger fields are not, in general, considered to be 
incomplete multiplets of SU(5) as it may destroy one of beautiful features of 
MSSM, which is the unification of gauge coupling constants. However one can use incomplete 
multiplets as messengers  without spoiling unification in special cases 
\cite{Calibbi:2009cp, Byakti:2012qk}.

Because of non-zero F-term VEV of $X$, messenger sector is not supersymmetric. 
As the messenger fields are charged under gauge groups, the SUSY breaking 
information passes to the visible sector through gauge interactions. Gaugino 
masses are generated at 1-loop level:
\begin{equation}\label{e:gaugino}
M_r = \frac{\alpha_r}{4\pi} d\,N\, \Lambda\, g(x),
\end{equation}
where $r=1,2,3$ represents U(1), SU(2) and SU(3) groups respectively,
 $\alpha_r=  g_r^2/4\pi $,  $\Lambda= F/M$, $x=F/M^2$,  d is the Dynkin index, $N$ is the 
number of messengers and the 
function 
$g(x)$ has the following form:
\begin{equation}
g(x) = \frac{1}{x^2} \left[(1+x)\ln(1+x)    \right] + \left( x\to -x \right).
\end{equation}
Scalar soft mass squares are generated at 2-loop level,
\begin{eqnarray}\label{e:sfermion}
M_{\tilde a}^2{}_{\rm usual} &=& 2 N\, d\,\Lambda^2\,\left[\sum_r\, 
C_r(a)\,\left(\frac{ 
\alpha_r}{4 \pi}\right)^2\right] f(x),
\end{eqnarray}
where $C_r(a)$ is the quadratic Casimir of the representation of the MSSM field 
labeled by `$a$' and the group corresponding to $r$ , $\tilde a$ is the 
super-partner of the field $a$, and the function $f(x)$ has the following 
form:
\begin{eqnarray}
f(x) = \frac{1+x}{x^2} \left[\ln(1+x) - 2 Li_2\left(\frac{x}{1+x}\right)  +  
\frac12 Li_2\left(\frac{2x}{1+x}\right) \right] + \left( x\to -x \right).
\end{eqnarray}
Note that gaugino masses are proportional to $N$ whereas the sfermion 
masses are proportional to $\sqrt{N}$. Thus one can have heavier gauginos with 
fixed sfermion masses for a larger $N$. Same is also true for the Dynkin 
index $d$. One gets heavier gauginos with fixed sfermion masses for a
$10\oplus\overline{10}$ messenger as compared to a copy of $5\oplus\bar5$ 
messenger field.

The spectrum changes in accordance with the messenger sector. In 
Ref.\cite{Meade:2008wd, Dumitrescu:2010ha} expressions for soft masses were 
derived  without considering any model for the messenger and the hidden 
sector. This model, as it encompasses all the GMSB models, is known as general 
gauge mediation or GGM. Expressions for the soft 
masses are as follows: 
\begin{eqnarray}
M_r &=& \alpha_r B_r, \\*
M_{\tilde a}^2{}_{\rm usual} &=& \sum_r\, \alpha_r^2 C_r(a)\, A_r.
\end{eqnarray}
Now we see that instead of one scale $\Lambda$ there are six 
dimensionful parameters, $B_r$ and $A_r$. In principle they can be arbitrary. 
GGM thus predicts non-universal gauginos without spoiling the gauge coupling 
unification.

As A-terms are not generated even in GGM, none of the  pure GMSB 
models can explain the Higgs mass with a light stop spectrum. In order to 
explain the Higgs mass one either requires stop masses $\gtrsim 4$ TeV or 
maximum mixing 
in the stop sector \cite{Draper:2011aa}. One way of generating mixing term or  
A-term at the boundary is to consider messenger-matter interactions 
\cite{Dine:1996xk, Han:1998xy}.

\subsection{Matter-Messenger Interactions}

The idea of extending GMSB models by considering messenger-matter interactions 
is not new \cite{Dine:1996xk, Han:1998xy}. In particular, to solve the severe 
$\mu$-$B_\mu$ problem in GMSB, one needs to couple the Higgs sector with the 
messenger 
sector\cite{Dvali:1996cu, Delgado:2007rz, Komargodski:2008ax}. In Ref. 
\cite{Chacko:2001km}, contribution of the messenger-matter interactions to other soft masses and the 
A-terms was calculated using wavefunction renormalization technique 
\cite{Giudice:1997ni}. After the discovery of the Higgs particle this idea 
became popular as it can save GMSB models by generating large stop mixing 
parameter. Several works were presented in this idea \cite{Kang:2012ra, 
Albaid:2012qk, Craig:2012xp, Evans:2011bea, Martin:2012dg}. In Ref. 
\cite{Byakti:2013ti, Evans:2013kxa},  messenger-matter interactions were 
classified and  thoroughly studied in MSSM. In \cite{Byakti:2013ti} messengers 
are assumed to be 1, 5 and 10 dimensional 
representations of SU(5) and these messengers are interacting with the MSSM 
SU(5) multiplet-wise. In principle messenger-matter
 interactions can introduce  flavor violation. Peccei-Quinn symmetry was used to 
suppress the flavor violation as well as to classify the models. The general  
classification 
reproduced  older models \cite{Kang:2012ra, Albaid:2012qk, Han:1998xy, 
Craig:2012xp, Evans:2011bea, Martin:2012dg} and  as 
well as found some new models. On the other hand in Ref.\cite{Evans:2013kxa} messengers 
are allowed to interact with MSSM fields by SM multiplet-wise.   More recent 
works in this direction can be found in the Refs.\cite{Mariotti:2015qcj, 
Casas:2016xnl, Ding:2015vla, Delgado:2015bwa, Jelinski:2015gsa, Evans:2015swa, 
Backovic:2015rwa, Basirnia:2015vga, Fischler:2013tva, Jelinski:2013kta, 
Calibbi:2013mka, Krauss:2013jva}. Models with explicit flavour violation can be found in \cite{Calibbi:2014yha, Calibbi:2013mka}.

Messenger-matter interactions are classified into two types depending 
on the number of matter fields in the interaction: (a) Type I where one matter 
field interacts with two messenger fields, and (b) Type II where one messenger
 field interacts with two matter fields. Superpotentials for these two types 
are given as:
\begin{eqnarray}
W_{\rm mix} = \left\{\begin{array}{l}
\frac12 \lambda_{aAB}\Phi_a \Phi_{Am} \Phi_{Bm}\,\, \mbox{ Type I}, \\    
      \frac12 \lambda_{abA}\Phi_a \Phi_b \Phi_{Am}\,\hphantom{AM} 
 \mbox{Type II},
 \end{array}
   \right.
\end{eqnarray}
where $a, b, c\cdots$ is used to indicate visible sector fields and the capital 
indices $A, B, C,\cdots$ are used to indicate messenger fields. Because of 
presence of these couplings, one gets 1-loop 
correction to the soft scalar mass-squared as follows
\begin{equation}
\delta_{\mbox{1-loop}} M_{\tilde a}^2 = - \frac{x^2 \Lambda^2 
h(x)}{96\pi^2}  \left\{ \begin{array}{c}
\sum_{BC} d_a^{BC}\, |\lambda_{aBC}|^2 \,{\rm Type\,I},\\
\vphantom{aa}\\
\sum_{bB} d_a^{bB}\, |\lambda_{abB}|^2 \,{\rm Type\,II},
\end{array}
     \right.
\end{equation}
$d_{\rm indices}$ is a group theoretical factor which 
appears in beta functions and $h(x)$ has the following form:
\begin{equation}
h(x) =  3 \frac{(x-2)\ln(1-x)}{x^4} + (x\to -x).
\end{equation}
Note that 1-loop correction is always negative and it contributes  only to the 
fields which are directly coupled to the messenger fields. Another point is 
that these contribution are suppressed for small values of $x\lesssim 0.1$ 
and  dominant for $x\sim 0.5$. The A-terms and 2-loop corrections are usually 
calculated using wavefunction  renormalization technique\cite{Giudice:1997ni, 
Chacko:2001km, Evans:2013kxa}. For Type I and Type II 
models these corrections are as follows \cite{Evans:2013kxa}:

\paragraph{Type I models}
\begin{eqnarray}\label{e:type1}
A_a &=&- \frac{1}{8\pi}\sum_{B,C} \, d_{a}^{BC} |\alpha_{a BC}| \, \Lambda,\\
\delta_{\mbox{2-loop}} M_{\tilde a}^2 &=& \frac{1}{16\pi^2}  
\Big[ \sum_{B,C,D,c} d_a^{BC}d_{B}^{cD}|\alpha_{aBC}| | \alpha_{c B D}|  + 
\frac14 \sum_{B,C,D,E} d_{a}^{BC}d_{a}^{DE} |\alpha_{aBC}| 
|\alpha_{aDE}|\nonumber\\*
& &\qquad - \frac12 \sum_{B,C,c,d} d_a^{cd} d_c^{BC}|\alpha_{acd}||\alpha_{c 
BC}| - \sum_{B,C} d_{a}^{BC} C_r^{aBC} \alpha_r  |\alpha_{aBC}|  \Big] 
\, \Lambda^2,
\end{eqnarray}
where we used $\alpha_{\rm indices}$ to denote $\lambda^2_{\rm 
indices}/4 \pi$.

\paragraph{Type II models}
\begin{eqnarray}
A_a &=& - \frac{1}{4\pi}\sum_{B,c} d_{a}^{c B}  |\alpha_{acB}|\, \Lambda,\\
\delta_{\mbox{2-loop}} M_{\tilde a}^2 &=& \frac{1}{16\pi^2}  
\Big[ 
\frac12 \sum_{B,c,d,e} d_a^{ c B }d_{B}^{de} |\alpha_{a c B}| |\alpha_{de 
B}| + \sum_{B,C,c,d} d_a^{c B}d_c^{dC}|\alpha_{acB}| |\alpha_{c dC}| 
\nonumber \\*
& +& \sum_{B,C,c,d} d_{a}^{cB}d_{a}^{dC}|\alpha_{acB}| |\alpha_{adC}|
  - \sum_{B,c,d,f} d_a^{cd}d_c^{f B} | \alpha_{acd}| |\alpha_{c f 
B}|\nonumber\\*
&  +&  \frac{1}{32\pi^2} \sum_{B,c,d,e,f} d_a^{cd}d_c^{ef}  y_{acd}^* 
y_{cef}\lambda_{a dB}\lambda_{efB}^*  + \frac{1}{32\pi^2} \sum_{B,c,d,e,f} 
d_a^{cB}d_{B}^{ef} \lambda_{ac B}^*\lambda_{efB}y_{acd} y_{d ef}^* \nonumber\\*
 &-& 2 \sum_{B,c} d_{a}^{cB} \left( C_r^{a} +  C_r^{c}+  
C_r^{B}\right)\alpha_r  |\alpha_{a cB}|  + 
  \frac12 \sum_{B,c,e,f} d_a^{c B}d_c^{ef} |\alpha_{cef}| 
|\alpha_{acB}|\Big]\,\Lambda^2.
\end{eqnarray}
Thus the total soft masses at the boundary are
\begin{equation}
M_{\tilde a}^2 = M_{\tilde a}^2{}_{\rm usual} + \delta_{\mbox{1-loop}} M_{ 
\tilde a}^2 + \delta_{\mbox{2-loop}} M_{\tilde a}^2.
\end{equation}


\section{Messenger-matter interactions involving leptons and 
neutrinos}\label{s:mes-mix}

Before studying the messenger-matter mixing terms involving leptonic fields and right handed neutrinos, 
 we are going to review the inverse seesaw model. 
 \subsection{Inverse seesaw model}
The canonical seesaw mechanism requires extension of the MSSM with a heavy 
field which could be right handed neutrino or triplet Higgs or a triplet 
fermion (for a review see \cite{Mohapatra:1998rq}). The smallness of the 
neutrino mass is associated with the heaviness of the additional particle. In
 the canonical seesaw mechanism, the corrections to the Higgs mass are
 typically very tiny as the right handed neutrino scale is
 very heavy $\geq$ $10^{14}$ GeV. Presence of matter messenger mixing terms will
 not improve the situation. Note that 
the right handed neutrino ($N^c$) mass must be less than the messenger scale, 
otherwise at the messenger scale, $N^c$ fields will be integrated out 
and messenger-matter interactions involving $N^c$ will now reduce to higher 
dimensional operator at the messenger scale. If we get any trilinear scalar 
coupling from these operators then they must be suppressed not only by the 
$N^c$ mass but also from loop factors. However, for $N^c$ mass up to 
$10^5$ GeV, the allowed value of the Yukawa coupling ($y_N$) can be utmost $~$ $10^{-5}$ to get 
neutrino mass of ${\cal O({\rm eV})}$. Because of such a small value of $y_N$, 
contribution to the Higgs mass  from the neutrino sector is negligibly small.

The situation drastically improves in inverse seesaw model \cite{Mohapatra:1986bd}. Supersymmetric 
version of this model has the following superpotential \cite{Guo:2013sna,Gogoladze:2012jp}: 
\begin{eqnarray}\label{e:superpot}
W &=&  U^c {\bf Y_u} Q H_u - D^c {\bf Y_d} Q H_d - E^c {\bf Y_e} L H_d +
\mu H_u H_d \nonumber\\*
&+& N^c y_N L H_u + m_R N^c S + \frac12 \mu_s S^2, 
\end{eqnarray}
where the  MSSM fields are in  standard notation with $\mathbf{Y_u}$ etc, 
representing the Yukawa matrices for three generations and 
the  $N^c$ and $S$ are new fields added to the MSSM field content. 
These are  singlet scalar superfields.  Note that if we set  $\mu_s$  to zero, 
the above superpotential enjoys U(1) lepton number symmetry.  Its presence 
implies that  this symmetry is softly broken.  As $\mu_s\to 0$ restores the 
symmetry, it can be (technically) naturally small.  In the above superpotential we have
 considered only one generation (third) for the inverse seesaw sector. The 
generalisation to three generations is straight forward and has minor impact on our analysis.

In the basis $ \{{\nu}_L, N^c,S \}$, the mass matrix $M_{\nu}$ of the 
neutral leptons for one generation, is given by
\begin{equation}
M_{{\nu}} = \left( \begin{array}{ccc} 
0 & m_D & 0 \\
m_D&0&  m_R\\ 
0& m_R& \mu_s \end{array} \right),
\end{equation}
where $ m_D=y_N\, \langle H_u\rangle $. The eigenvalues of the above mass 
matrix are as follows:
\begin{eqnarray}
m_{\nu_1} &\approx& {m_D^2\, \mu_s\over m_R^2},\nn
m_{\nu_2} &\approx&- \left({m_D^2\over 2\, m_R} + m_R\right),\nn
m_{\nu_3} &\approx& \left({m_D^2\over 2\, m_R}+ m_R\right).\label{nueigen}
\end{eqnarray} 
Here $ m_{\nu_1}$, the lightest neutrino eigenvalue, is proportional to 
the parameter $\mu_s$.  From electroweak precision data \cite{delAguila:2008pw},  
$m_D \lesssim  0.05 ~m_R$ and thus last two eigenvalues of the mass matrix are 
degenerate.  As $m_D$ and $m_R$ related,  
$y_N$ and $m_R$ are also related:
\begin{equation}
 y_N = \sqrt{\frac{m_{\nu_1}}{\mu_s}} \frac{\sqrt2}{v} {\rm cosec}\beta\, m_R,
\end{equation}
where $v$ is the electroweak VEV of the Higgs fields: $v=\sqrt{\langle H_u 
\rangle^2 + \langle H_d \rangle^2}$= 246 GeV. For a fixed $\mu_s$, which 
we fix it to be electron mass, we see that $y_N$ scales as $m_R$. 

The scalar potential for this model is  given below which contains SUSY preserving as well as 
SUSY breaking soft terms:
\begin{equation}\label{e:Spot}
V_S = V_F + V_D + V_{\rm{soft}},
\end{equation}
where 
\begin{eqnarray}\label{e:Vexplicit}
V_F &=& | Y_e \tilde E^c H_d + y_N H_u \tilde N|^2 + | Y_u \tilde Q \tilde U^c 
+ \mu H_d + y_N \tilde L \tilde N^c|^2 + 
|y_N \tilde L H_u + m_R \tilde S|^2 \nonumber\\*
&+& | m_R \tilde N^c + \mu_s\, \tilde S|^2 + \ldots, \\* 
V_D &=&\frac{1}{8}(g^2+g'^2)\,(|H_u|^2-|H_d|^2), \\*
V_{\rm{soft}} &=& A_N y_N \tilde L H_u \tilde N^c +B_R \tilde N^c\tilde S + 
B_{S}\tilde S^\dagger\tilde S + h. c. + M_N 
\tilde N^{c\dagger} \tilde N^c + \cdots  \,.
\end{eqnarray}
To calculate the  neutrino-sneutrino correction to Higgs mass, one needs to 
calculate the sneutrino mass matrix  which has the form: 
\begin{equation}
\mathcal{M}_{\tilde{\nu}}^2 = \left( \begin{array}{ccc} 
M_{\tilde L}^2 + D_L + m_D^2  & m_D\, ( A_N - \mu \cot \beta)  & m_R\, m_D \\
* &  m_D^2 + M_{N}^2 + m_R^2 & B_R + m_R\, \mu_s \\ 
* & * & m_R^2 + \mu_s^2 + m_{\tilde{S}}^2 \end{array} \right),
\end{equation}
where the basis is $ \{\tilde{\nu}_L, \tilde{N^c}, \tilde{S} \} $, $M_{\tilde 
L}^2$ is the slepton mass, and $m_{\tilde{S}} $ is the soft mass of $S$. As the 
mass matrix is symmetric, terms omitted can be easily understood. As the field 
$N^c$ and $S$ are gauge singlets, soft masses $B_R, B_S$ and $M_N$ are zero at 
the boundary, the messenger scale. Assuming these are small and $m_D/m_R < 
1$, one obtains the following eigenvalues \cite{Chun:2014tfa}:
\begin{eqnarray}
m_{\tilde \nu_1}^2&\approx&M_{\tilde L}^2 + m_D^2 \left(1+{m_R^2\over d_2 }+
{X_N^2\over d_1}\right),\nonumber\\*
m_{\tilde \nu_2}^2&\approx& M_N^2 + m_R^2+m_D^2 \left(1-
{X_N^2\over d_1}\right),\nonumber\\*
m_{\tilde \nu_3}^2&\approx& m_{ \tilde S}^2 + m_R^2-
{m_R^2\, m_D^2\over d_2},\label{snueigen}
\end{eqnarray}
where
\begin{eqnarray}
d_1=M_{\tilde L}^2-M_N^2- m_R^2,\label{d1e}\\*
d_2=M_{\tilde L}^2- m_R^2-m_{\tilde S}^2,\label{d2e} \\*
X_N=A_N - \mu \cot\beta \,.
\end{eqnarray}
To compute the corrections to the Higgs mass, we use
 the effective potential method \cite{Coleman:1973jx}. The one-loop effective
 potential  for neutrino-sneutrino sector is \cite{Chun:2014tfa, Guo:2013sna}:    
\begin{equation}
V_{1-loop}^{\nu / \tilde\nu} (Q^2) = {2 \over 64 \pi^2} \left[ \sum_{i=1}^3 
m_{\tilde \nu_i}^4\left( \log { m^2_{\tilde \nu_i} \over Q^2} - {3 \over 2} 
\right) - \sum_{i=1}^3 m_{ \nu_i}^4  \left( \log { m^2_{ \nu_i} \over Q^2} 
- 
{3 \over 2} \right) \right],
\end{equation} 
where first and second term represent the contribution of sneutrino and  
neutrino mass eigenstates respectively. An overall factor 2 takes care of the 
degrees of freedom for the complex scalar and Weyl fermion. The complete 
calculation of the correction to the  Higgs mass is given in the Appendix A. It
 should be noted that the calculation presented in Appendix A is a slight
 generalisation of the one presented in Ref. \cite{Chun:2014tfa} as we
 relaxed the assumption that $X_N$ is a small parameter.

Without going into details of sneutrino-neutrino sector
 corrections to the Higgs mass, we can make the following observations:
\begin{enumerate}
\item If $m_R$ $>>$ $M_{SUSY}$, the sneutrino (\Eqn{snueigen}) and
 neutrino (\Eqn{nueigen}) eigenvalues are degenerate and are of order
 of $m_R$. There will be a complete cancellation between the scalar
 and fermion sector contributions  and consequently no significant correction to the Higgs mass.
\item To have a significant corrections to the Higgs mass, one
 should have heavy sleptons\cite{Chun:2014tfa}. In this case  the degeneracy between the scalar and fermion 
eigenvalues breaks and therefore  cancellation will not be exact. \rplb{ 
Since these corrections are inversely proportional to $d_{1, 2}$, for 
heavy sleptons (comparable to $m_R$), $d_{1,2}$ will be small and consequently 
one will get significant enhancement to the Higgs mass. Such large sleptons 
masses can be generated in a framework like general gauge mediation. An 
alternative way of enhancing the Higgs mass is through matter-messenger 
corrections which can generate a large $X_N$ parameter and/or significant corrections to $M_{\tilde L}^2$. 
In this case, the sneutrino derivatives 
are now proportional to $X_N$ (see Appendix A for explicit expressions), 
the sneutrino contribution will be more as compared to neutrino contribution.}
\end{enumerate}

Both the above conditions (large $X_N$ and very heavy sleptons) 
are not met in the minimal gauge mediation model.  Thus the Higgs mass 
corrections remain small. Clearly  both the scenarios with enhanced corrections 
are not applicable in  minimal GMSB.  One could however argue to increase the
messenger scale, but this would only increase the mass of the stops which is
contrary of our philosophy of keeping stops light.

\rplb{The possibility of increasing the sneutrino/neutrino contributions by 
increasing the
 slepton  mass in a  general gauge mediation model was discussed in 
 Ref. \cite{Chun:2014tfa}. In the present work, we discuss the importance of the combination of heavy sleptons and 
 large  $X_N$ parameter (generated through matter-messenger mixing). }

\begin{table}
 $$
 \begin{array}{|c|c|c|c|}
\hline
\mbox{Model No} & \mbox{Interaction} & \mbox{Lepton number} & \mbox{Remarks or 
Source}\\
\hline
\hline
\multicolumn{4}{|c|}{\mbox{Models with } N^c}\\
\hline
1. \vphantom{\frac{\frac12}{1}}  & N^c Q \bar Q_m & 1 
& \in \overline{10}\\
\hline
2. \vphantom{\frac{\frac12}{1}} & N^c U^c \bar U^c_m & 1 & \in
\overline{10} \\
\hline
3. \vphantom{\frac{\frac12}{1}}  & N^c D^c \bar D^c_m & 1 & \rplb{\in 5}\\
\hline
4. \vphantom{\frac{\frac12}{1}} & N^c L H_u^m & 0 & \in
5\\
\hline
5. \vphantom{\frac{\frac12}{1}}  & N^c E^c \bar E^c_m & 2 & \in
\overline{10}\\
\hline
6. \vphantom{\frac{\frac12}{1}} & N^c H_u H_d^m & 1 & \in
\overline{5}\\
\hline
7. \vphantom{\frac{\frac12}{1}} & N^c H_d H_u^m & 1 & \in
5\\
\hline
8. \vphantom{\frac{\frac12}{1}}  & \frac12 (N^c)^2 S_m  & 2 & \in 1
\\
\hline
9. \vphantom{\frac{\frac12}{1}} & N^c S S_m  & 0  & \in 1 \\
\hline
10. \vphantom{\frac{\frac12}{1}} & N^c H_u^m H_d^m  & -- &
\in 5\oplus\bar5 \\
\hline
\multicolumn{4}{|c|}{\mbox{Models with L}}\\
\hline
11. \vphantom{\frac{\frac12}{1}} & \rplb{L Q  D^c_m} & -1 & \in
\overline{5}\\
\hline
12. \vphantom{\frac{\frac12}{1}}  & \rplb{L D^c  Q_m} & -1 & \in
\rplb{10}\\
\hline
13. \vphantom{\frac{\frac12}{1}} & L E^c H_d^m & 0 & \in
\overline{5}\\
\hline
14. \vphantom{\frac{\frac12}{1}} & L H_d E^c_m & -1 & \in 10 \\
\hline
15. \vphantom{\frac{\frac12}{1}}  & L H_u S_m & -1 & \in 1 \\
\hline
16. \vphantom{\frac{\frac12}{1}}  & L S H_u^m &  -2 & \in 5\\
\hline
17. \vphantom{\frac{\frac12}{1}} & L H_u^m S_{m} & -- &\in 1,5\\
\hline
\end{array}
 $$
\caption{\sf Classification of the models.  Note that, for the models 1, 2, 5 
and 14, messenger fields are $10\oplus \overline{10}$ and for the rest of the 
models these are $5\oplus\bar5$. Each model contains only one term and the 
corresponding coupling is chosen to be $\lambda$. In the 
third column lepton number of the messenger fields are listed. As in the model 
10 and 17 two messenger fields are appearing in each interaction terms, one can 
assign any lepton number to the messengers keeping in mind that product of 
their lepton numbers should be 1 and -1 for the model 10 and 17 respectively. 
In the last column we 
mention the representation and source of the messenger field or fields.}
\label{t:class}
 \end{table}

\subsection{Classification of the models}
We are interested to study the effect of messenger-matter
 interaction in the inverse seesaw 
mechanism. We know that the Lagrangian given in \Eqn{e:superpot} has softly 
broken U(1) lepton number and the softly broken parameter, $\mu_s$, is 
responsible for the generation of neutrino mass through inverse seesaw. If 
messenger-matter interactions do not obey U(1) lepton number then we cannot 
guarantee that inverse seesaw is the only source of neutrino mass. 
Therefore  we impose U(1) lepton 
number on the messenger fields. To generate $A_t$ and/or $A_N$, at least one of 
the fields $Q,\, U^c,\, H_u,\, L$ and $N^c$ has to couple with the messenger 
fields. Models involving $N^c$ field in the messenger-matter interaction are 
not explored in the literature. We have listed  17 possible models of  
messenger-matter interactions involving $L$ and $N^c$ fields in \tabl{t:class}. 
In these models we allow only the third generation of the 
matter fields to couple with the messenger fields. The interaction term $L^2 
E^c_m$ is not there in the above list because this vanishes. Along with the 
interaction terms listed above, some new terms may be allowed by symmetry. For 
example, in model 5, the term $S^2 E^c_m$ is allowed. However, we are not 
considering this term as it will not generate $A_t$ or $A_N$. For the same 
reason, we do not list the models involving only $E^c$ or $S$.
Each of the model contains the shown interaction in the 
superpotential. Inter-generational mixing is considered to be absent. When more
than one messenger fields is considered, the matter-messenger coupling is 
considered universal over all the messenger fields. 
 
 Some of the models involving $L$, like model 11, 13 and 15, are not new. These 
are considered in  Ref. \cite{Byakti:2013ti} along with other interaction terms. 
As mentioned earlier, the 
suffix $m$ is used to indicate messengers. Messenger field with known symbol   
has the same quantum number under SM gauge group. Here models 10 and 17 are 
Type I models and rest of the models are of Type II. In each model we are
 allowing only one messenger-matter interaction term. 
In the next section, we are going to list the modification of the boundary 
conditions due to these messenger-matter interactions.

\section{Analysis of the models}\label{s:analysis}
From the previous section it is clear that the fields which are 
coupled with the messenger fields through Yukawa interactions have negative 
one-loop corrections as well as positive two-loop corrections to their soft 
masses.  Other matter fields which are not directly 
coupled to the messenger fields but, have  Yukawa interactions with the matter 
fields with direct interactions to the messenger fields,  always get two-loop 
negative contributions. On the top of these 
corrections, there are 
usual GMSB contributions to the soft masses which are always positive.  The messenger-matter
 coupling $\lambda$ cannot take arbitrary values as  it can lead to negative 
mass squared eigenvalues  for the scalars at the weak scale.

 Another important parameter is  $x \equiv \Lambda/M$. One-loop corrections 
diminish for smaller values of $x$. We consider two cases: (a) $x=0.5$ for which 
we cannot neglect 1-loop effects, and (b) $x=0.1$ for which 1-loop contributions 
can be neglected.

Because of non-observation of any SUSY particle, LHC bounds on the soft masses 
are very stringent.  In GGM models, the present lower limits on gluino 
is 1.6 TeV \cite{Khachatryan:2016ojf}, whereas on the chargino it is 650 GeV  
\cite{Aad:2015hea}. It can be easily seen that LHC bound on 
chargino mass is more stronger than that of the gluino mass in the universal 
gaugino mass case.  If one considers mGMSB with $N=d=1$ then the upper bound
 on the gluino mass forces the  stop masses to be of the order of  2 TeV. 
However one could be interested in light spectrum for various reasons including 
the fine-tuning issue.  
To resolve this issue in the models with $5\oplus\bar5$, we 
consider the number of messengers to be 3. In models 1, 2, 5 and 14 this problem is 
automatically solved as $10\oplus\overline{10}$ messenger field has $d=3$. We 
choose the following values in the numerical analysis: 
\begin{eqnarray}\label{e:lambda}
\begin{array}{|c|c| c|}
\hline
& 5\oplus\bar5 &10\oplus\overline{10}\\
\hline
{\rm Models} & 3, 4, 6, 7, \cdots,11, 13, 15, 16, 17  & 1, 2, 5, 12, 14\\
\hline
\Lambda & 100\, {\rm TeV} & 100\, {\rm TeV} \\
\hline
{\rm Number\, of\, messengers} & 3 & 1 \\
\hline
 {\rm Dynkin\, index} & 1 & 3\\
\hline
\end{array}\label{e:lamval}
\end{eqnarray}

\rplb{Gravitino is the LSP in these models. Its mass has the following 
expression:
\begin{equation}
 m_\frac32 = \frac{F}{\sqrt{3} M_{\rm Pl}} = \frac{\Lambda^2}{\sqrt{3} x M_{\rm 
Pl}} = \frac{10}{4.16 x} \,{\mbox{eV}},
\end{equation}
where $M_{\rm Pl}= 2.4\times 10^{18}$ GeV. Thus we get gravitino mass 4.8 eV 
and 24 eV for $x=0.5$ and $x=0.1$, respectively. Experimental bound on gravitino mass at $2\sigma$ limit 
is 16 eV \cite{Viel:2005qj}. Though $x=0.1$ case is ruled out by gravitino mass 
constraint, there is a way out to overcome this gravitino problem 
\cite{Kawasaki:2006gs}.}

Lepton number violating mass parameter $\mu_s$ is another important parameter. 
Upper limit of $y_N$ depends on it. The upper bound on $y_N$ comes from electroweak 
precision tests, which sets the ratio  $m_D/m_R <\frac{1}{20}$ 
\cite{delAguila:2008pw}. \rplb{We consider $\mu_s= 5 \times 10^{-4} m_e $, this 
fixes the $m_D/m_R$ ratio to be of the $\frac{1}{70}$ for a neutrino mass of 
${10^{-1}}$ eV.  And thus the limits from electroweak precision tests are always 
satisfied. As  $y_N \propto {\rm cosec}\beta$, it is insensitive to $\beta$ for 
higher values of $\tan\beta$}. We thus kept $\tan\beta= 10$ through out the 
analysis. For spectrum calculation, a modified version of the publicly available 
code SuSeFLAV \cite{Chowdhury:2011ss} is used.  All the low energy 
phenomenological constraints including flavour constraints dominantly from $BR(B 
\to X_s + \gamma)$ and mass constraints from LHC are imposed on the spectrum. We 
will now  discuss each model in  detail. 

\begin{figure}
\centering
\subfigure[]{
\includegraphics[height=5cm, width=7cm]{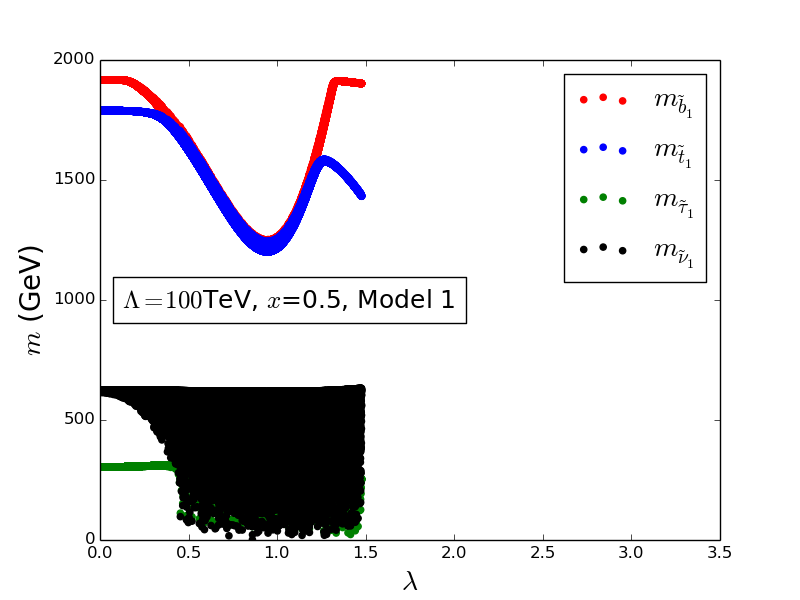}}
\subfigure[]{
\includegraphics[height=5cm, width=7cm]{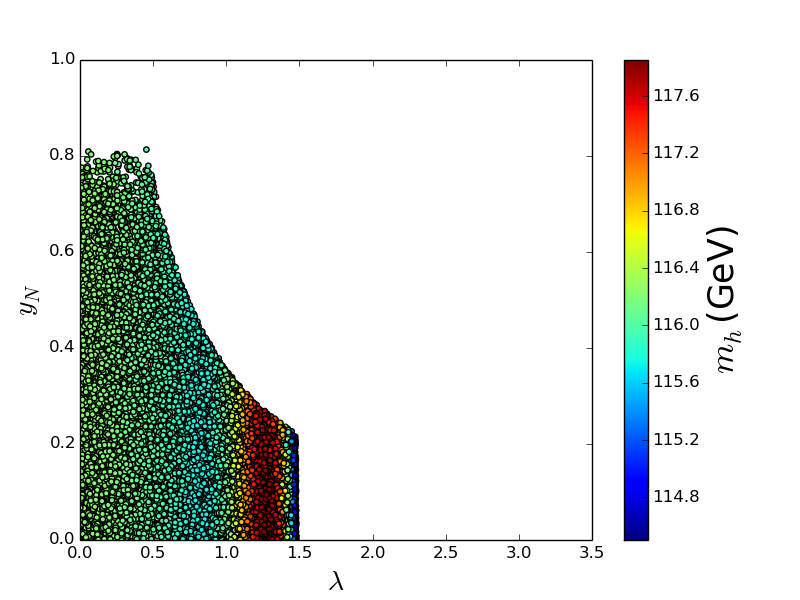}}
\subfigure[]{
\includegraphics[height=5cm, width=7cm]{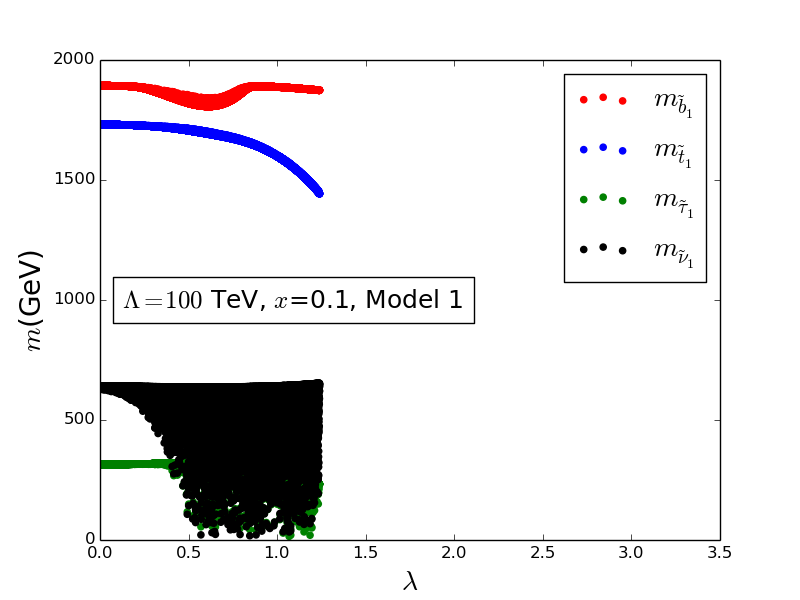}}
\subfigure[]{
\includegraphics[height=5cm, width=7cm]{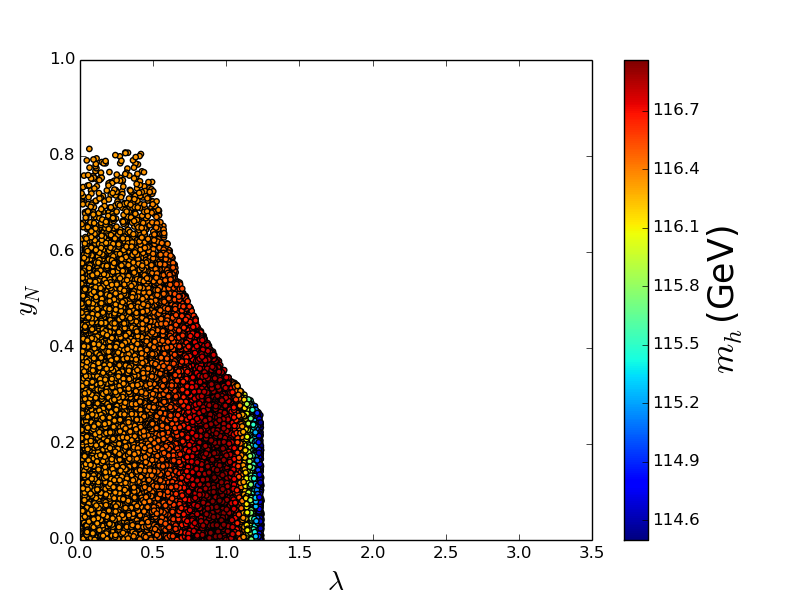}}
\caption{\sf Model 1: Variation of the third generation soft masses with 
messenger-matter interaction coupling $\lambda$ is shown in the left panel, 
and Higgs mass values in $\lambda$ and $y_N$ plane are shown in the 
right panel. The upper and lower plots correspond to $x=0.5$ and 
$x=0.1$ respectively.}\label{f:mod1}
\end{figure}

\begin{figure}
\centering
\subfigure[]{
\includegraphics[height=5cm, width=7cm]{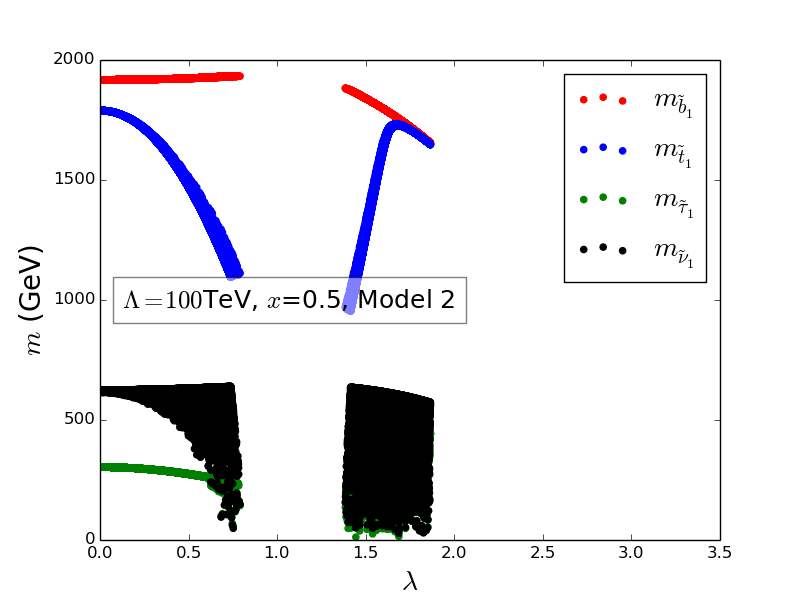}}
\subfigure[]{
\includegraphics[height=5cm, width=7cm]{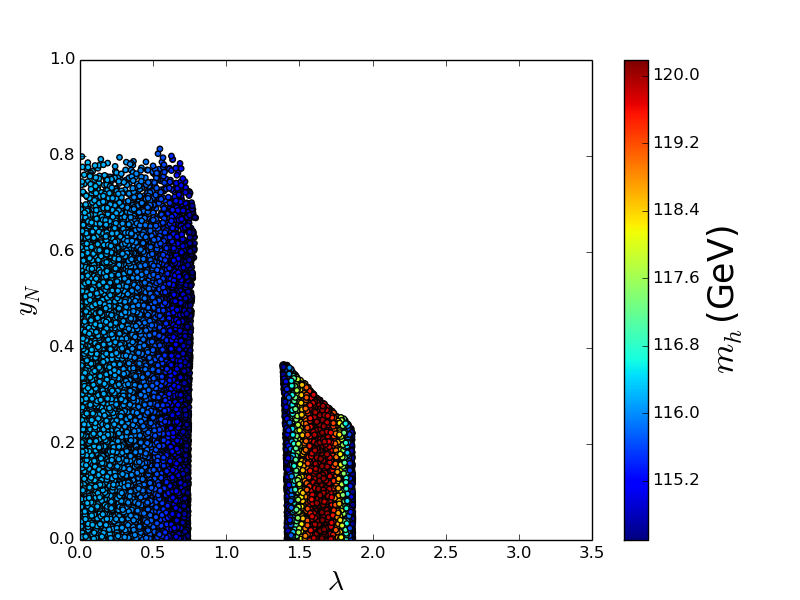}}
\subfigure[]{
\includegraphics[height=5cm, width=7cm]{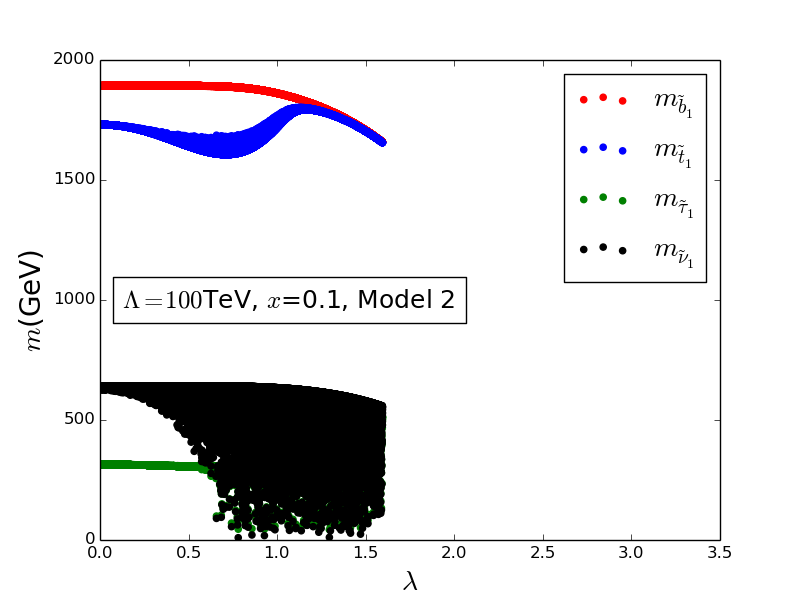}}
\subfigure[]{
\includegraphics[height=5cm, width=7cm]{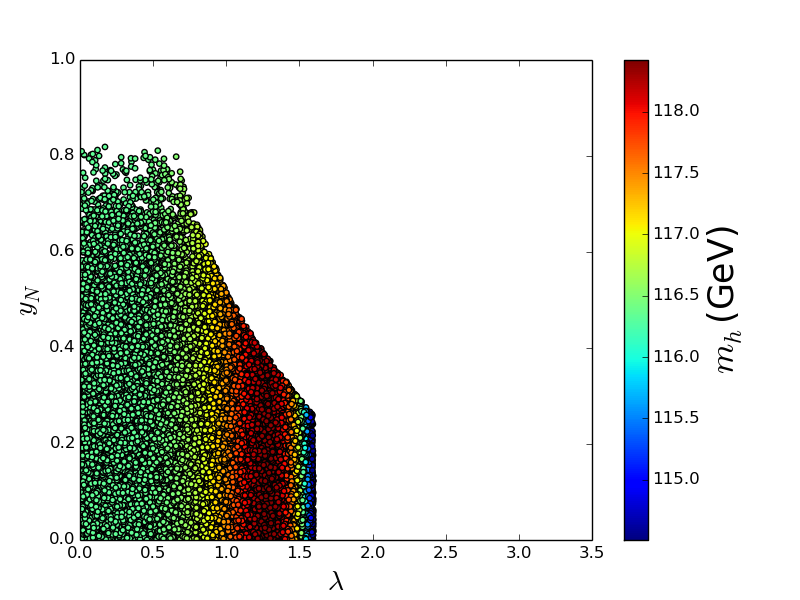}}
\caption{\sf Model 2: Upper two plots correspond to $x=0.5$ and the lower two 
are for $x=0.1$. Note that depth of blue line is higher for this model compared 
to model 1. Please see caption of \fig{f:mod1} for  details of 
notation.}\label{f:mod2}
\end{figure}

\subsection{Model 1}
Model 1 is a Type II model of our classification which interacts with the quark 
doublet fields
 and  the singlet. $W_{\rm mix} = \lambda N^c Q \bar Q_m$, the resultant 
one-loop and two-loop corrections  to the various soft masses are shown below
\beqa
 \delta M^2_{\tilde Q} &=& \Big[-\frac{\alpha_\lambda}{24\pi}x^2 h(x) + 
\frac{\alpha 
_{\lambda } \left(-\alpha _1+5 \left(-9 
\alpha
_2-16 \alpha _3+6 \alpha _N+24 \alpha _{\lambda }\right)\right)}{240 \pi ^2} 
\Big]\Lambda^2  \nonumber \\
 \delta M^2_{\tilde U^c} &=& \Big[ -\frac{\alpha _t \alpha _{\lambda }}{8 \pi 
^2} 
\Big]\Lambda^2 
\nonumber \\
 \delta M^2_{\tilde D^c} &=& \Big[ -\frac{\alpha _b \alpha _{\lambda }}{8 \pi 
^2} 
\Big]\Lambda^2 
\nonumber \\
 \delta M^2_{\tilde L} & =& \Big[-\frac{3 \alpha _N \alpha _{\lambda }}{8 \pi 
^2} \Big]\Lambda^2 \nonumber 
\eeqa
\beqa
 \delta M^2_{\tilde N^c} &=& \Big[\SOLo + \frac{\alpha _{\lambda } 
\left(-\alpha 
_1+5 
\left(-9
\alpha _2-16 \alpha _3+3 \left(\alpha _b+\alpha _t+8 \alpha _{\lambda
}\right)\right)\right)}{40 \pi ^2} \Big]\Lambda^2 \nonumber \\
\delta M^2_{H_u} &=& \Big[ -\frac{3 \left(\alpha _t+2 \alpha _N\right) \alpha
_{\lambda }}{16 \pi ^2} \Big]\Lambda^2 \nonumber \\
 \delta M^2_{H_d} &=& \Big[ -\frac{3 \alpha _b \alpha _{\lambda }}{16 \pi ^2} 
\Big]\Lambda^2 \nonumber \\
\delta A_t &=& \Big[ -\frac{\alpha _{\lambda }}{4 \pi } \Big]\Lambda \nonumber 
\\
\delta A_b &=& \Big[ -\frac{\alpha _{\lambda }}{4 \pi } \Big]\Lambda \nonumber 
\\
\delta A_N &=& \Big[ -\frac{3 \alpha _{\lambda }}{2 \pi } \Big]\Lambda,
\label{eqmod1}\eeqa
where 
\begin{eqnarray}
\alpha_t = \frac{y_t^2}{4\pi},\,
\alpha_b = \frac{y_b^2}{4\pi},\,
\alpha_\tau = \frac{y_\tau^2}{4\pi},\,
\alpha_\lambda = \frac{\lambda^2}{4\pi}
\mbox{ and }
\alpha_N = \frac{y_N^2}{4\pi}.
\end{eqnarray}

In this model $\tilde Q$ and $\tilde N^c$ get one-loop negative corrections to 
their masses.  For $x=0.5$, the interplay between the one-loop,  two-loop 
corrections and the standard GMSB contributions  is clearly evident in the 
lightest stop and sbottom masses shown in Fig. \ref{f:mod1} (a). As can be seen, 
 when $\lambda$ is relatively small, the negative one-loop contributions 
significantly cancel with the standard GMSB contributions,  lowering the 
lightest eigenvalue to smaller values. The cancellation is maximum around  
$\lambda \sim1$. Beyond those values of $\lambda$, positive two-loop 
contributions start dominating over the one-loop contributions resulting in 
positive and  larger spectra.  For $x=0.1$ the one-loop 
effects are no longer important and the cancellation regions disappear as can be
seen from the Fig \ref{f:mod1}(c).  For both the cases, as $\lambda$ increases,
staus and stops start becoming tachyonic for values $\lambda \gtrsim 1.5$.  Remember
that the staus receive negative contributions at two-loops from 
matter-messenger mixing terms  
at the messenger scale (Eq. \ref{eqmod1}). 


The Higgs mass values for the allowed parameter space are presented 
for $x = 0.5$ in Fig. \ref{f:mod1}(b) and for $x = 0.1$
in Fig. \ref{f:mod1}(d). 
This model can not produce the correct Higgs mass because although both $A_t$ and $A_N$ are generated 
but sleptons are not that heavy. However $A_N$ is dominant over $A_t$ by a factor three
as can be seen from  Eq. (\ref{eqmod1}).  To disentangle the effects from each contribution, 
we use three different notations to illustrate the  corrections to the Higgs mass. We used 
$m_h^{0}$ parameter for pure MSSM Higgs  mass, 
$m_h^{0+\Delta}$ denotes Higgs mass, calculated including the matter messenger
 mixing terms in the boundary conditions but without considering
 the corrections from the neutrino-sneutrino sector, on the top of the MSSM and 
$m_h$ is the  actual Higgs mass when all the corrections are considered. 


Four benchmark points (two each with $x=0.5$ and $x=0.1$) are presented in 
Table \ref{t:mod1}.  As can be seen from the benchmark points, even both 
$A_t$ and $A_N$ are not sufficient to provide correct Higgs mass. The stop spectrum is relatively
light $\sim 1.7$ TeV. The sneutrino mixing parameter $A_N$ is relatively large compared
to the $A_t$ generated. 

\begin{table}
 $$
 \begin{array}{|c|c|c|c|c||c|c|c|c|c|}
 \hline{\mbox{Parameter}}&{\rm{ x=0.5}}
  &{\rm  x=0.5} & {\rm x=0.1} & {\rm x=0.1} &
 {\mbox{Parameter}}&{\rm{ x=0.5}}&{\rm  x=0.5}
  & {\rm x=0.1} & {\rm x=0.1}\\  \hline
\lambda &1.34 &0.53 &0.55 &            
  0.55   &
   y_N & 0.24 &0.51 &0.51 &            
  0.21  \\
  \hline
    A_N &           -6650.7 &           -1021.8 &           -1074.6 &           
-1136.3   &
   A_t &            -1477.9 &           -727.3 &           -829.4 &           
-843.4  \\
    m_R &            5823.8 &            6195.3 &            6469.5 &           
 3732.3   &
   \mu &1400. &790. &935. &            
  823.  \\
 m_{\tilde\nu_1} &133. &           
   441. &426. &610.   &
 m_{\tilde\nu_{2,3}}/m_{\nu_{2,3}} &             5824. &          
   6196. &             6470. &             3732.  \\
  \hline \hline
    m_h &            117.05 &            115.62 &            116.06 &           
 117.21   &
 m_{H} &1523. &             1003. &             1129. &           
   999.  \\
  m_h^{0+\Delta} &             117.05 &          
   115.60 &            116.04 &            117.21   &
        m_{A^0}  &             1539. &           
  1013. &             1141. &             1001.  \\
        m_h^{0} & 115.84 &         
  115.84 &            116.18 &             116.18   &
      m_{H^\pm} &1541. &           
  1016. &             1144. &             1003.  \\
  \hline 
 M_{S}  &             1728. &             1706. &             1768. &           
  1759.   &
   m_{\tilde g} &2187. &           
  2180. &             2108. &             2108.  \\
      \tilde t_1 &             1561. &           
  1611. &             1707. &             1703.   &
     \tilde t_2 &1913. &           
  1806. &             1832. &             1816.  \\
      \tilde b_1 &             1912. &           
  1620. &             1831. &             1809.   &
     \tilde b_2 &1926. &           
  1928. &             1899. &             1888.  \\
   \tilde \tau_1 &119. &           
   317. &316. &311.   &
  \tilde \tau_2 & 299. &           
   321. &323. &324.  \\
      \tilde u_1 &             1935. &           
  1916. &             1897. &             1892.   &
     \tilde u_2 &2003. &           
  1936. &             1909. &             1910.  \\
      \tilde d_1 &             1927. &           
  1931. &             1901. &             1902.   &
     \tilde d_2 &1993. &           
  1931. &             1902. &             1902.  \\
      \tilde e_1 &299. &           
   321. &323. &324.   &
     \tilde e_2 & 311. &           
   446. &432. &614.  \\
   N_1  &431. &430. &412. &           
   412.   &
   N_2 & 822. &749. &772. &           
   750.  \\
    N_3 &             1408. &795. &942. &           
   829.   &
     N_4&             1413. &878. &965. &           
   873.  \\
   C_1  &806. &740. &758. &           
   739.   &
   C_2 &1404. &863. &955. &           
   862.  \\
 \hline\end{array}
 $$
 \caption{\sf{Model 1: Matter-messenger mixing parameter, neutrino sector 
parameters, Higgs mass and SUSY spectrum for the benchmark points from $x=0.5$ 
and $x=0.1$ cases. In each case the considered  benchmark represents
small and large allowed $\lambda$ range. Here $m_h^{0}$ means pure MSSM Higgs 
mass calculated at the two-loop level, $m_h^{0+\Delta}$ is the Higgs mass when 
all the mixing terms 
except the neutrino mixing are considered on the top of the MSSM 
and $m_h$ is the actual Higgs mass when all the corrections were 
considered. All the masses are given in GeV. The $\Lambda$ values
 are taken as in \Eqn{e:lamval}.
}
\label{t:mod1}
}\end{table}

\subsection{Model 2}
This model has similar structure as Model 1 with quark doublets
 replaced by the up type quark singlet. The resultant one-loop
 and two-loop corrections are listed below in Eq.\ref{eq.mod2}
\beqa
 \delta M^2_{\tilde Q} &=& \Big[ -\frac{\alpha _t \alpha _{\lambda }}{16 \pi 
^2} 
\Big]\Lambda^2  
\nonumber  \\
 \delta M^2_{\tilde U^c} &=& \Big[\OOLo+ \frac{\alpha _{\lambda } \left(-16 
\alpha 
_1-80 
\alpha
_3+30 \alpha _N+75 \alpha _{\lambda }\right)}{240 \pi ^2} \Big]\Lambda^2 
\nonumber  
\\
 \delta M^2_{\tilde L} &=& \Big[ -\frac{3 \alpha _N \alpha _{\lambda }}{16 \pi 
^2} 
\Big]\Lambda^2 
\nonumber  \\
 \delta M^2_{\tilde N^c} &=& \Big[\ThOLo+ \frac{\alpha _{\lambda } \left(-16 
\alpha 
_1-80 
\alpha
_3+30 \alpha _t+75 \alpha _{\lambda }\right)}{80 \pi ^2} \Big]\Lambda^2 
\nonumber  \\
 \delta M^2_{H_u} &=& \Big[ -\frac{3 \left(\alpha _t+\alpha _N\right) \alpha 
_{\lambda
}}{16 \pi ^2} \Big]\Lambda^2 \nonumber  \\
\delta A_t &=& \Big[ -\frac{\alpha _{\lambda }}{4 \pi } \Big]\Lambda  \nonumber 
\\
\delta A_N &=& \Big[ -\frac{3 \alpha _{\lambda }}{4 \pi } \Big]\Lambda.
\label{eq.mod2}
\eeqa
In this case $\tilde U^c$ and $N^c$ fields get one-loop negative and two-loop
positive contributions to 
their soft masses.  From Figs. \ref{f:mod2}(a) and \ref{f:mod2}(c), we can see 
the variation
of the lightest third generation mass eigenvalues with respect to $\lambda$. 
For $x = 0.5$, cancellations only appear for the stop  sector and not in the
bottom sector as the messenger matter interactions are only active for the 
up-type
singlet sector. The cancellations are however much deeper here as the standard
GMSB contributions for the singlet up squarks is lesser compared to the doublet
squarks.  The cancellation is milder as expected for $x=0.1$.  

In this case $A_N$ dominates $A_t$ by a factor three as in Model 1. For large $\lambda 
\gtrsim ~2$,
sneutrinos, stops and sbottoms become tachyonic making the model unviable. 
The Higgs mass values are presented in Figs. \ref{f:mod2}(b) and 
\ref{f:mod2}(d).

Four benchmark points are given in \tabl{t:mod2}, as before two for 
$x=0.5$ and two for $x=0.1$. As can be seen from the points, neutrino/sneutrino
contribution is not significant in this model. 

\begin{table}
 $$
 \begin{array}{|c|c|c|c|c||c|c|c|c|c|}
 \hline{\mbox{Parameter}}&{\rm{ x=0.5}}
  &{\rm  x=0.5} & {\rm x=0.1} & {\rm x=0.1} &
 {\mbox{Parameter}}&{\rm{ x=0.5}}&{\rm  x=0.5}
  & {\rm x=0.1} & {\rm x=0.1}\\  \hline
\lambda &1.58 &0.66 &1.32 &           
   0.72   &
   y_N & 0.27 &0.62 &0.36 &           
   0.65  \\
  \hline
    A_N &           -4599.7 &            -785.0 &           -3137.3 &           
 -882.2   &
   A_t &            -1845.6 &            -814.3 &           -1495.5 &           
 -925.6  \\
    m_R &            6162.5 &            8773.4 &            7394.6 &           
 9977.0   &
   \mu &1551. &852. &             1527. &           
  1050.  \\
 m_{\tilde\nu_1} &306. &           
   425. &115. &322.   &
 m_{\tilde\nu_{2,3}}/m_{\nu_{2,3}} &             6163. &          
   8774. &             7395. &             9978.  \\
  \hline \hline
    m_h &            119.24 &            114.84 &            117.80 &           
 116.24   &
 m_{H} &1671. &             1053. &             1643. &           
  1217.  \\
  m_h^{0+\Delta} &             119.22 &          
  114.82 &            117.78 &            116.22   &
        m_{A^0}  &             1686. &           
  1071. &             1662. &             1241.  \\
        m_h^{0} &119.22 &          
  114.82 &            115.95 &            115.94   &
      m_{H^\pm} &1688. &           
  1074. &             1664. &             1244.  \\
  \hline 
 M_{S}  &             1722. &             1569. &             1820. &           
  1761.   &
   m_{\tilde g} &2182. &           
  2184. &             2114. &             2110.  \\
      \tilde t_1 &             1600. &           
  1291. &             1771. &             1666.   &
     \tilde t_2 &1853. &           
  1906. &             1871. &             1862.  \\
      \tilde b_1 &             1808. &           
  1928. &             1783. &             1888.   &
     \tilde b_2 &1922. &           
  1934. &             1889. &             1901.  \\
   \tilde \tau_1 &269. &           
   256. &108. &288.   &
  \tilde \tau_2 & 321. &           
   262. &418. &315.  \\
      \tilde u_1 &             1922. &           
  1948. &             1885. &             1910.   &
     \tilde u_2 &1935. &           
  1974. &             1990. &             1927.  \\
      \tilde d_1 &             1930. &           
  1934. &             1896. &             1901.   &
     \tilde d_2 &1930. &           
  1959. &             1897. &             1909.  \\
      \tilde e_1 &321. &           
   262. &418. &315.   &
     \tilde e_2 & 356. &           
   430. &422. &349.  \\
   N_1  &431. &430. &413. &           
   412.   &
   N_2 & 822. &783. &782. &           
   780.  \\
    N_3 &             1559. &858. &             1536. &           
  1058.   &
     N_4&             1563. &908. &             1540. &           
  1071.  \\
   C_1  &806. &769. &773. &           
   766.   &
   C_2 &1554. &895. &             1529. &           
  1062.  \\
 \hline\end{array}
 $$
 \caption{\sf{Benchmark   points for Model 2. As can be seen,
 the neutrino corrections can be significant in some regions
 of the parameter space compared to the matter messenger mixing corrections. 
See 
caption of \tabl{t:mod1} for details of notation.}}
 \label{t:mod2}
 \end{table}

\begin{figure}
\centering
\subfigure[]{
\includegraphics[height=5cm, width=7cm]{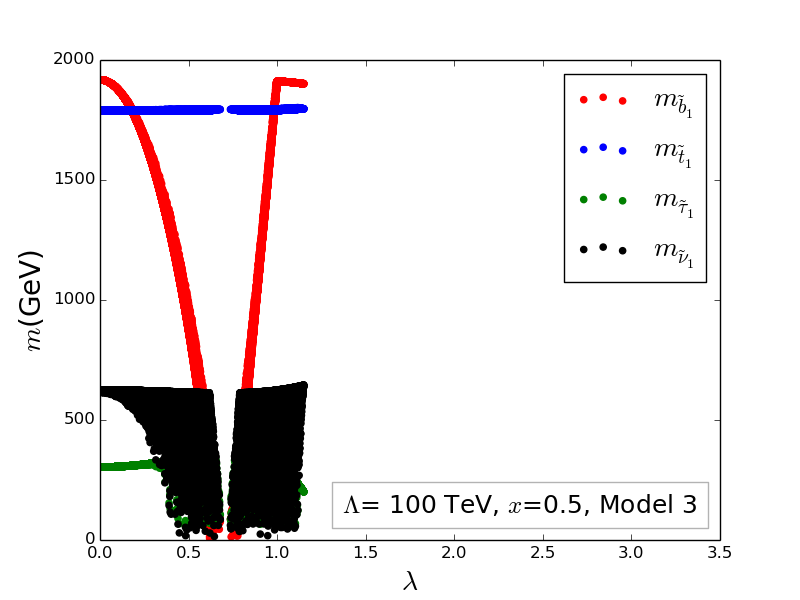}}
\subfigure[]{
\includegraphics[height=5cm, width=7cm]{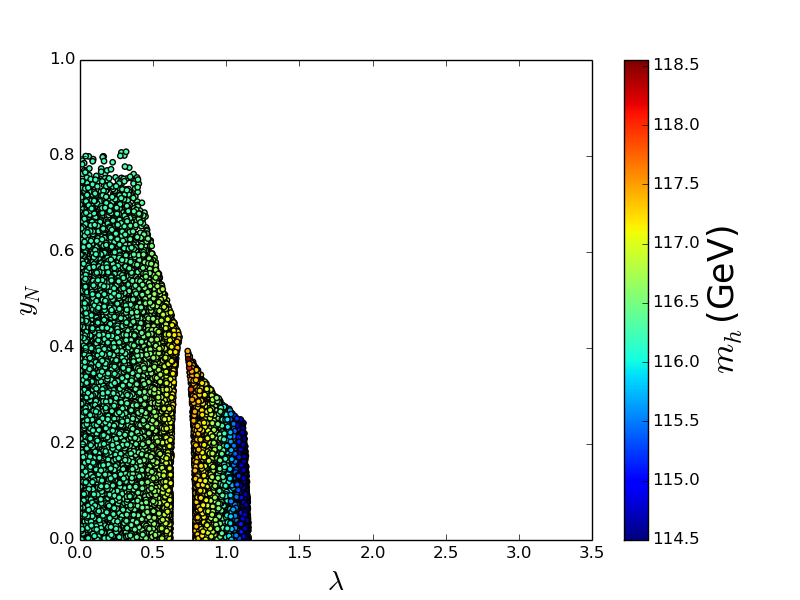}}
\subfigure[]{
\includegraphics[height=5cm, width=7cm]{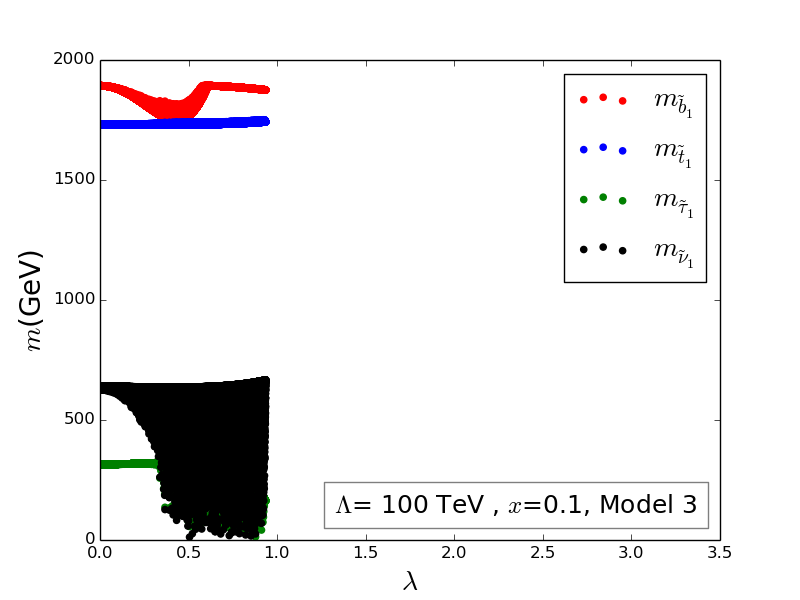}}
\subfigure[]{
\includegraphics[height=5cm, width=7cm]{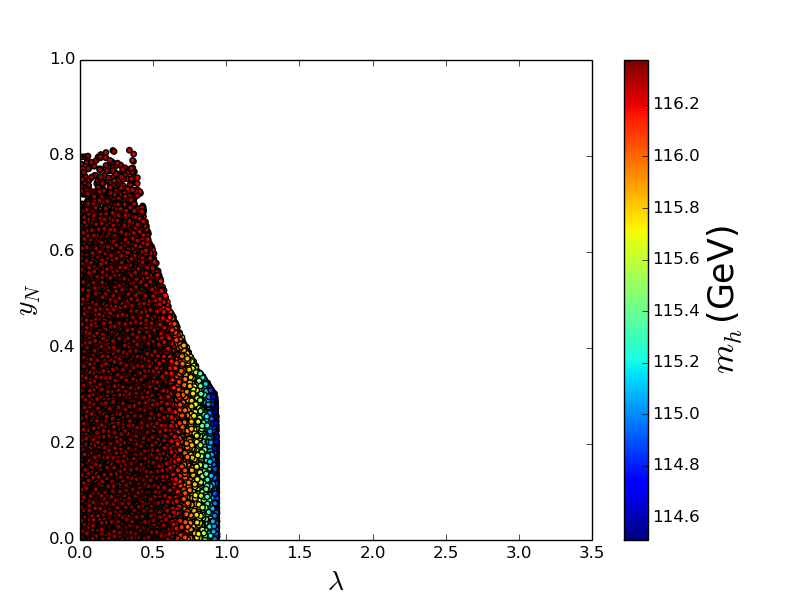}}
\caption{\sf Model 3: Spectrum variation with $\lambda$ and Higgs 
mass data points in $y_N$ and $\lambda$ plane. Note that in (b) 
origin is not at zero but at 0.7 for $\lambda$. 
Please see caption of \fig{f:mod1} for  details of notations. }\label{f:mod3}
\end{figure}


\subsection{Model 3}
Model 3 is a Type II model of our classification in which down quark interacts 
with a messenger field transforming as a conjugate representation of $D^c$: 
$W_{\rm mix} = \lambda N^c D^c \bar D^c_m$. The resultant one-loop and two-loop 
corrections to the various soft masses are shown below 
\beqa\label{e:mod3}
 \delta  M^2_{\tilde Q } &=& \Big[ -\frac{3 \alpha _b \alpha _{\lambda }}{16 
\pi 
^2} \Big] \Lambda^2 \nonumber\\*
 \delta  M^2_{\tilde {D^c} } &=& \Big[\OOL+ \frac{\alpha _{\lambda } \left(-4 
\alpha 
_1-80 \alpha 
_3+30 \alpha _N+195 \alpha _{\lambda }\right)}{80 \pi ^2} \Big] \Lambda^2 
\nonumber\\*
 \delta  M^2_{\tilde L } &=& \Big[ -\frac{9 \alpha _N \alpha _{\lambda }}{16 
\pi 
^2} \Big] \Lambda^2 \nonumber \\*
 \delta  M^2_{ {H_u} } &=& \Big[ -\frac{9 \alpha _N \alpha _{\lambda 
}}{16 
\pi ^2} \Big] \Lambda^2 \nonumber \\*
 \delta  M^2_{ {H_d} } &=& \Big[ -\frac{9 \alpha _b \alpha _{\lambda 
}}{16 
\pi ^2} \Big] \Lambda^2 \nonumber \\*
 \delta  M^2_{\tilde {N^c} } &=& \Big[\ThOL + \frac{3 \alpha _{\lambda } 
\left(-4 
\alpha _1-80 \alpha 
_3+30 \alpha _b+195 \alpha _{\lambda }\right)}{80 \pi ^2} \Big] \Lambda^2 
\nonumber \\*
 \delta  A_b  &=& \Big[ -\frac{3 \alpha _{\lambda }}{4 \pi } \Big] \Lambda
\nonumber\\*
 \delta  A_N  &=& \Big[ -\frac{9 \alpha _{\lambda }}{4 \pi } \Big] \Lambda.
\label{eq:mod3}
\eeqa
Here $M_{\tilde D^c}$ and $M_{\tilde N^c}$  get one-loop negative 
contributions. Cancellation between the one-loop contribution and the two-loop 
contributions for $M_{\tilde D^c}$ is so severe in $x=0.5$ case that 
the $m_{\tilde b_1}$ becomes tachyonic for  $\lambda \approx ~0.6 - 0.7$ 
(\fig{f:mod3}(a)). One 
may wonder why $m_{\tilde b_1}$ goes to negative here while  $m_{\tilde t_1}$ 
of 
previous model does not touch the zero line. Note that $\tilde U^c$ mass 
at the boundary is slightly higher than that of $\tilde D^c$ because former has 
hypercharge double of the later one. Hence the cancellation is less severe in 
model 2. This cancellation is less for $x=0.1$ as expected (\fig{f:mod3}(c)).

In this model, no $A_t$ term is generated at the boundary. We see that 
parameter space of $\lambda$ splits into two parts for $x=0.5$ (\fig{f:mod3}(b)) 
and one can not obtain the correct Higgs mass in this model. For the case of 
$x=0.1$, however, parameter space for $\lambda$ is continuous (\fig{f:mod3}(d)) 
and the spectrum becomes tachyonic for $\lambda$ $>$ 1.0. The benchmark points 
are given in \tabl{t:mod3}.

 \begin{table}
 $$
 \begin{array}{|c|c|c|c|c||c|c|c|c|c|}
 \hline{\mbox{Parameter}}&{\rm{ x=0.5}}
  &{\rm  x=0.5} & {\rm x=0.1} & {\rm x=0.1} &
 {\mbox{Parameter}}&{\rm{ x=0.5}}&{\rm  x=0.5}
  & {\rm x=0.1} & {\rm x=0.1}\\  \hline
   \lambda &              0.83 &              0.41 &              0.81 &         
     0.45   &
y_N &               0.28 &              0.54 &              0.32 &              
0.59  \\
  \hline
 A_N &           -3921.8 &            -955.1 &           -3716.3 &           
-1072.6   &
A_t &            -572.2 &           -575.5 &           -679.5 &           
-679.6  \\
 m_R &            4753.9 &            8559.6 &            9412.0 &            
7235.1   &
\mu &               803. &              770. &              937. &              
905.  \\
                                 m_{\tilde\nu_1} &              330. &          
    421. &              273. &              325.   &
               m_{\tilde\nu_{2,3}}/m_{\nu_{2,3}} &             4754. &          
   8560. &             9412. &             7236.  \\
  \hline \hline
 m_h &            117.22 &            116.43 &            115.48 &            
116.30   &
    m_{H} &              1004. &           
   983. &             1111. &             1098.  \\
                                  m_h^{0+\Delta} &             117.20 &         
   116.42 &            115.46 &            116.27   &
  m_{A^0}  &             1011. &           
   997. &             1135. &             1115.  \\
  m_h^{0} &              116.23 &          
  116.22 &            116.33 &            116.33   &
m_{H^\pm} &              1014. &           
  1000. &             1137. &             1117.  \\
  \hline 
    M_{S}  &             1842. &           
  1842. &             1803. &             1798.   &
                                   m_{\tilde g} &              2183. &          
   2182. &             2117. &             2110.  \\
\tilde t_1 &             1792. &           
  1792. &             1744. &             1738.   &
                                     \tilde t_2 &              1893. &          
   1893. &             1864. &             1861.  \\
\tilde b_1 &              694. &           
  1266. &             1885. &             1800.   &
                                     \tilde b_2 &              1921. &          
   1923. &             1894. &             1897.  \\
                                   \tilde \tau_1 &              317. &          
    326. &              233. &              299.   &
                                  \tilde \tau_2 &               344. &          
    332. &              259. &              324.  \\
\tilde u_1 &             1927. &           
  1927. &             1912. &             1906.   &
                                     \tilde u_2 &              1971. &          
   1971. &             1938. &             1942.  \\
\tilde d_1 &             1926. &           
  1924. &             1927. &             1899.   &
                                     \tilde d_2 &              1960. &          
   1956. &             1994. &             1925.  \\
\tilde e_1 &              344. &           
   332. &              259. &              324.   &
                                     \tilde e_2 &               361. &          
    427. &              299. &              350.  \\
N_1  &              430. &              429. &              413. &              
412.   &
N_2 &               756. &              736. &              772. &              
768.  \\
 N_3 &              808. &              775. &              943. &              
912.   &
  N_4&              881. &              870. &              966. &              
938.  \\
C_1  &              747. &              727. &              759. &              
755.   &
C_2 &               868. &              855. &              956. &              
928.  \\
 \hline\end{array}
 $$
 \caption{\sf{Benchmark   points for Model 3. See caption of \tabl{t:mod1} for 
details of notation.}}
 \label{t:mod3}
 \end{table}
 
\begin{figure}
\centering
\subfigure[]{
\includegraphics[height=4.9cm, width=7cm]{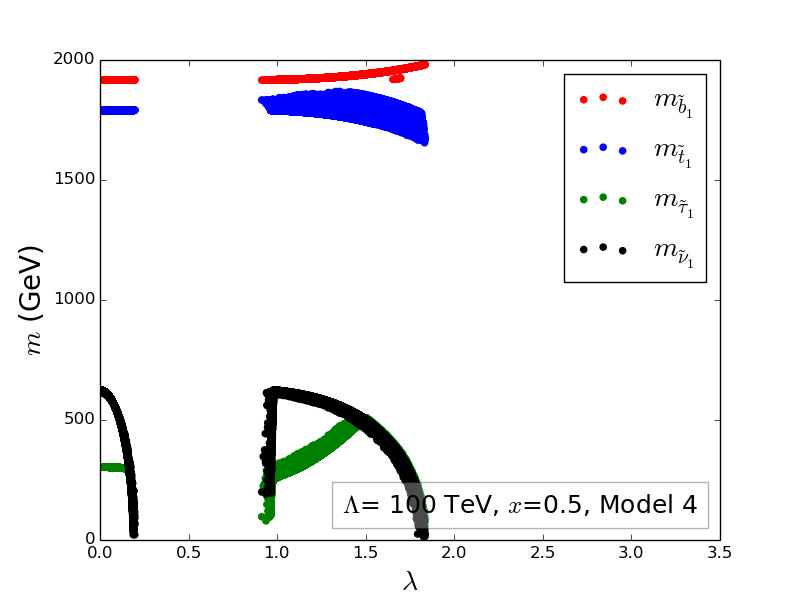}}
\subfigure[]{
\includegraphics[height=4.9cm, width=7cm]{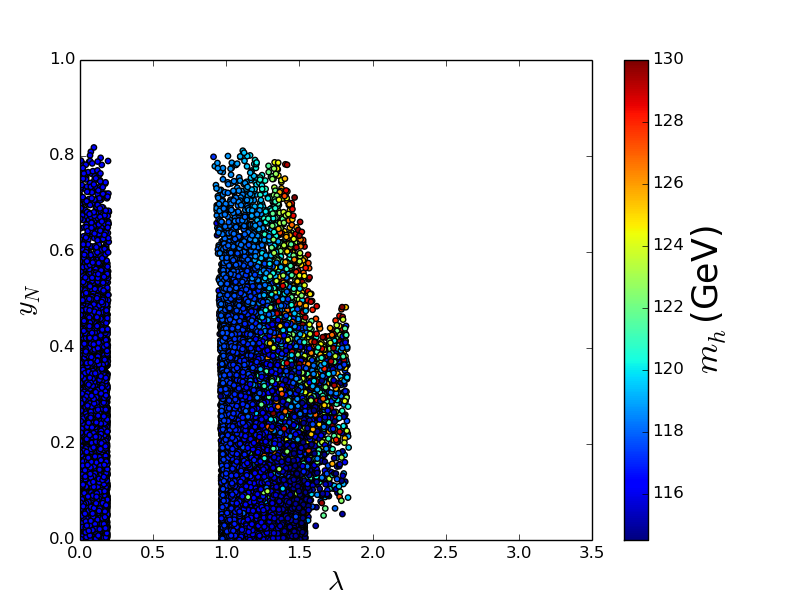}}
\subfigure[]{
\includegraphics[height=4.9cm, width=7cm]{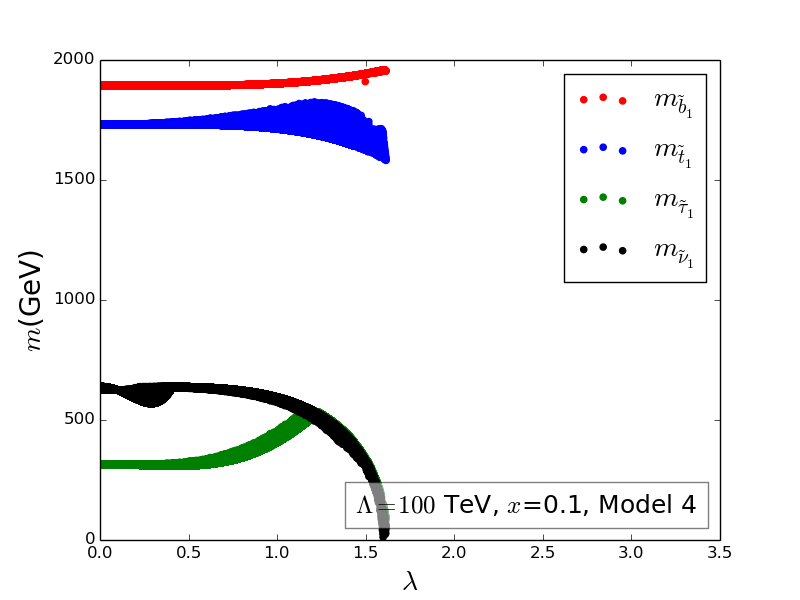}}
\subfigure[]{
\includegraphics[height=4.9cm, width=7cm]{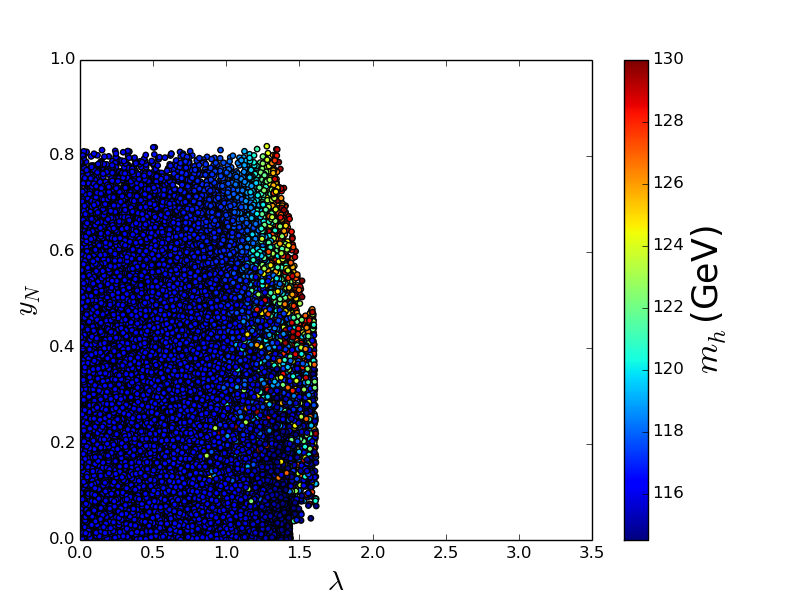}}
\caption{\sf Model 4: Spectrum variation with $\lambda$ and Higgs 
mass values in $y_N$ and $\lambda$ plane. Please see caption of 
\fig{f:mod1} 
for  details of notation.}\label{f:mod4}
\end{figure}

\begin{figure}
\centering
\subfigure[]{
\includegraphics[height=5cm, width=7cm]{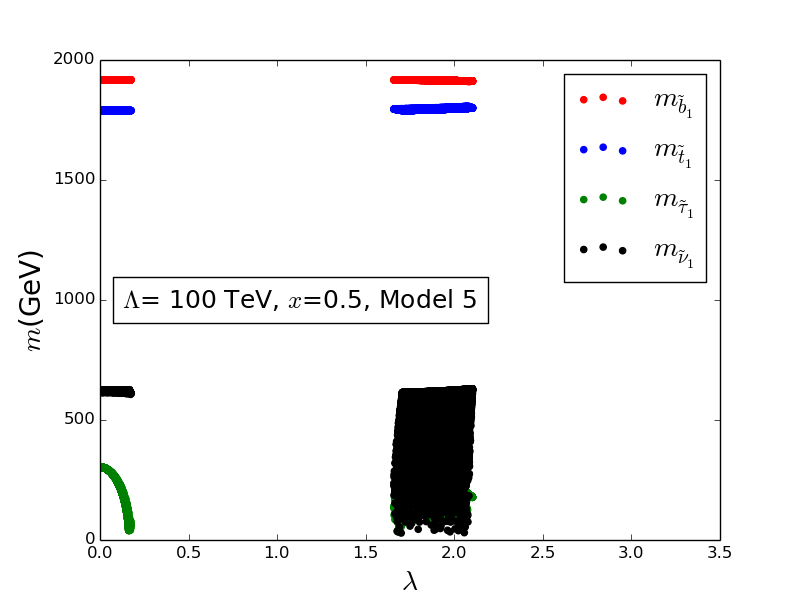}}
\subfigure[]{
\includegraphics[height=5cm, width=7cm]{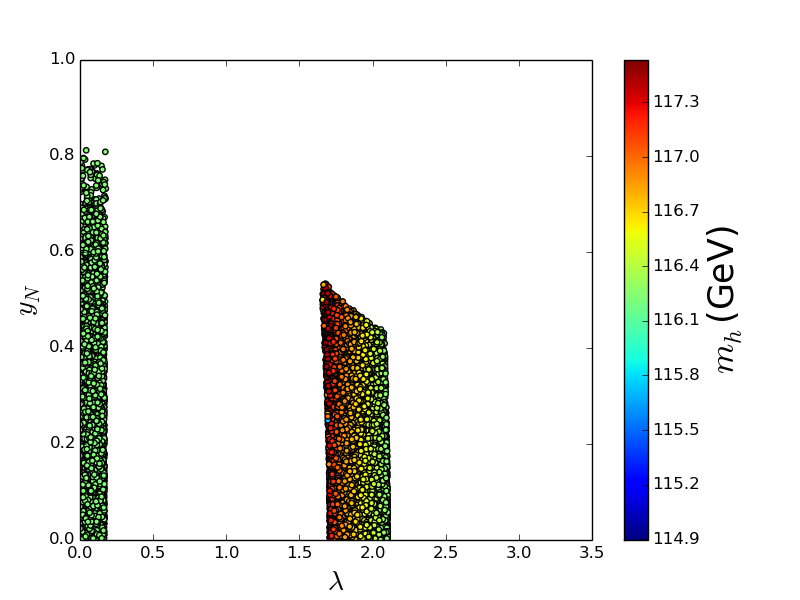}}
\subfigure[]{
\includegraphics[height=5cm, width=7cm]{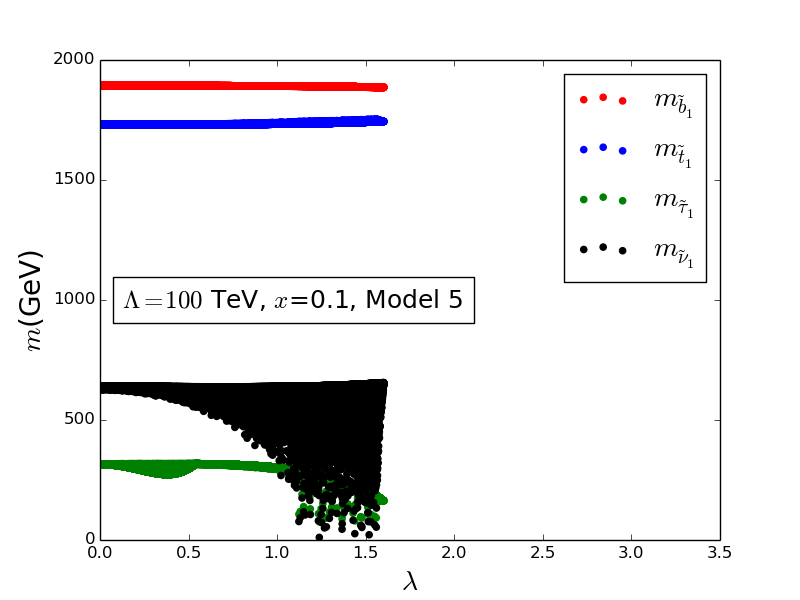}}
\subfigure[]{
\includegraphics[height=5cm, width=7cm]{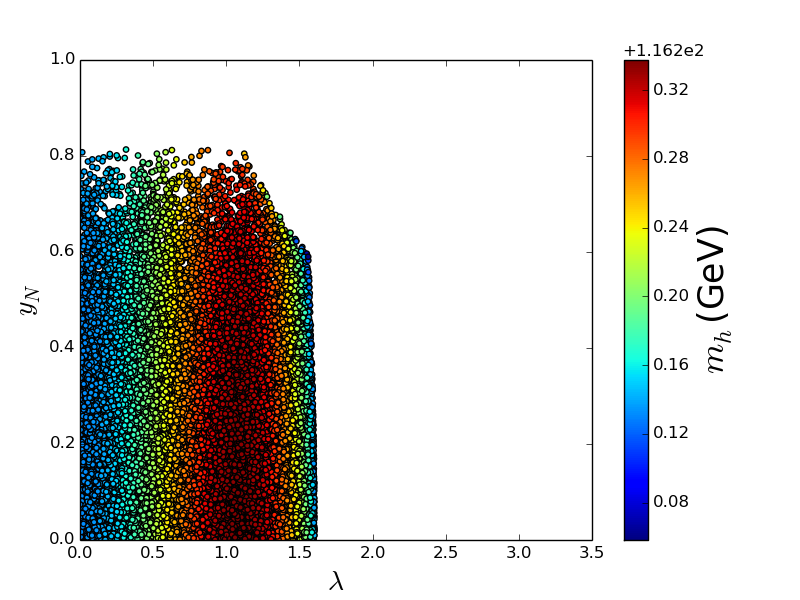}}
\caption{\sf Model 5: Spectrum variation with $\lambda$ and Higgs 
mass values in $y_N$ and $\lambda$ plane. Please see caption of 
\fig{f:mod1} 
for  details of notation.}\label{f:mod5}
\end{figure}

\subsection{Model 4}
This is a Type II  model with messenger matter interaction $ W_{\rm mix} =  
\lambda N^c L H_u^m$. Like the previous model, this model has $5\oplus\bar 5$ 
messengers. One-loop and two-loop contribution to the soft masses are shown in 
\Eqn{e:mod4}. 
\beqa\label{e:mod4}
 \delta  M^2_{\tilde L } &=& \Big[\OOL+ \frac{3 \alpha _{\lambda } \left(-3 
\alpha 
_1-15 \alpha _2+10 
\left(\alpha _N+5 \alpha _{\lambda }\right)\right)}{80 \pi ^2} \Big] \Lambda^2 
\nonumber\\*
 \delta  M^2_{\tilde {E^c} } &=& \Big[ -\frac{3 \alpha _{\lambda } \alpha 
_{\tau }}{8 \pi ^2} \Big] \Lambda^2 \nonumber\\*
 \delta  M^2_{{H_u} } &=& \Big[ -\frac{9 \alpha _N \alpha _{\lambda 
}}{16 \pi ^2} \Big] \Lambda^2 \nonumber\\*
 \delta  M^2_{{H_d} } &=& \Big[ -\frac{3 \alpha _{\lambda } \alpha 
_{\tau }}{16 \pi ^2} 
\Big] \Lambda^2 \nonumber\\*
 \delta  M^2_{\tilde {N^c} } &=& \Big[\TwOL+ \frac{3 \alpha _{\lambda } \left(5 
\left(-3 \alpha _2+2 
\alpha _N+10 \alpha _{\lambda }+\alpha _{\tau }\right)-3 \alpha _1\right)}{40 
\pi ^2} \Big] \Lambda^2 \nonumber\\*
\delta  A_{\tau }  &=& \Big[ -\frac{3 \alpha _{\lambda }}{4 \pi } \Big] 
\Lambda \nonumber\\*
 \delta  A_N  &=& \Big[ -\frac{9 \alpha _{\lambda }}{4 \pi } \Big] \Lambda.
\eeqa
Here $M^2_{\tilde L }$ and $M^2_{\tilde {N^c} }$ get one-loop 
negative contributions and two-loop positive contributions. We know 
that usual GMSB contributions to the $M_{\tilde L}^2$ and $M_{\tilde 
E^c}^2$ are small. Thus in the models where these masses get one-loop negative 
contributions, $M_{\tilde L}^2$ will be large (comparable to $m_R$). Therefore 
$d_{1,2}$ (Eq. \ref{d1e} and \ref{d2e}) will be small which appear in the denominator of the 
Higgs mass 
expressions and enhances the contribution. Therefore one will correct Higgs 
mass.

In \fig{f:mod4}(a) one can see that $m_{\tilde \tau_1}$ becomes tachyonic for 
vary small values of $\lambda\sim 0.18$ and for larger $\lambda$ one can get 
positive spectrum. The larger band $(\lambda > 
1.0)$ has correct Higgs mass. For $x=0.1$, strength of one-loop contribution 
diminishes 
and hence larger values of $\lambda$ are allowed.  
Beyond $\lambda=1.6$, $m_{\tilde \tau_1}$ is tachyonic as can be seen in 
\fig{f:mod4}(c). For these values of $\lambda$ two-loop negative contribution 
to $M_{\tilde E^c}$ dominates over usual GMSB contribution. In the Higgs mass 
scatter plot (\fig{f:mod4}(b) and (d)), one can see that this model is capable 
of 
giving correct Higgs mass. From these figures, we can see the importance of 
$m_R$ 
parameter. The Higgs mass is sensitive to the $m_R$ value for a given point in 
the $y_N$-$\lambda$ plane.
Like the model 3, in this model too $A_t$ term
 is not generated at the boundary. 
Hence, $A_N$ and heavy $M^2_{\tilde L}$ are mainly responsible for raising the 
Higgs mass up to the 
required value. For the $x=0.5$ case, one need the value of $- A_N$ is 
$\sim 12$ to 4  TeV (\tabl{t:mod4}) to raise the Higgs mass from its pure MSSM 
value  as shown in \tabl{t:mod4}.

 \begin{table}
 $$
 \begin{array}{|c|c|c|c|c||c|c|c|c|c|}
 \hline{\mbox{Parameter}}&{\rm{ x=0.5}}
  &{\rm  x=0.5} & {\rm x=0.1} & {\rm x=0.1} &
 {\mbox{Parameter}}&{\rm{ x=0.5}}&{\rm  x=0.5}
  & {\rm x=0.1} & {\rm x=0.1}\\  \hline
\lambda &1.49 &1.18 &1.37 &           
   0.82   &
   y_N & 0.39 &0.14 &0.46 &           
   0.17  \\
  \hline
    A_N &          -12389.9 &           -7925.4 &          -10240.4 &           
-3869.9   &
   A_t &           -513. &           -578.0 &          -581.2 &           
-693.0  \\
    m_R &            7564.6 &            2891.9 &            7985.1 &           
 2096.7   &
   \mu &1624. &743. &             1893. &           
   798.  \\
 m_{\tilde\nu_1} &             5801. &           
  2732. &             6190. &             2161.   &
 m_{\tilde\nu_{2,3}}/m_{\nu_{2,3}} &             7567. &          
   2900. &             7987. &             2079.  \\
  \hline \hline
    m_h &            124.93 &            122.69 &            125.04 &           
 124.53   &
 m_{H} &1558. &949. &             1767. &           
  1011.  \\
  m_h^{0+\Delta} &             112.79 &          
  116.08 &             112.59 &           116.15   &
        m_{A^0}  &             1709. &           
   953. &             1953. &             1013.  \\
        m_h^{0} &116.20 &          
   116.2 &            116.33 &             116.3   &
      m_{H^\pm} &1709. &           
   955. &             1953. &             1016.  \\
  \hline 
 M_{S}  &             1821. &             1834. &             1767. &           
  1789.   &
   m_{\tilde g} &2189. &           
  2188. &             2113. &             2111.  \\
      \tilde t_1 &             1783. &           
  1783. &             1727. &             1726.   &
     \tilde t_2 &1860. &           
  1887. &             1807. &             1854.  \\
      \tilde b_1 &             1939. &           
  1922. &             1928. &             1897.   &
     \tilde b_2 &1945. &           
  1927. &             1936. &             1903.  \\
   \tilde \tau_1 &501. &           
   324. &436. &350.   &
  \tilde \tau_2 & 5804. &           
   2741. & 6194. & 2139.  \\
      \tilde u_1 &             1879. &           
  1919. &             1827. &             1899.   &
     \tilde u_2 &1992. &           
  1971. &             1981. &             1942.  \\
      \tilde d_1 &             1945. &           
  1927. &             1936. &             1903.   &
     \tilde d_2 &1985. &           
  1957. &             1975. &             1926.  \\
      \tilde e_1 &520. &           
   389. &652. &383.   &
     \tilde e_2 & 612. &           
   605. &735. &626.  \\
   N_1  &431. &429. &413. &           
   411.   &
   N_2 & 824. &716. &791. &           
   738.  \\
    N_3 &             1633. &748. &             1893. &           
   803.   &
     N_4&             1637. &864. &             1895. &           
   861.  \\
   C_1  &805. &707. &773. &           
   726.   &
   C_2 &1626. &847. &             1893. &           
   848.  \\
 \hline\end{array}
 $$
 \caption{\sf{Benchmark   points for Model 4. See caption of \tabl{t:mod1} for 
details of notation.}}
 \label{t:mod4}\end{table}

\begin{figure}
\centering
\subfigure[]{
\includegraphics[height=5cm, width=7cm]{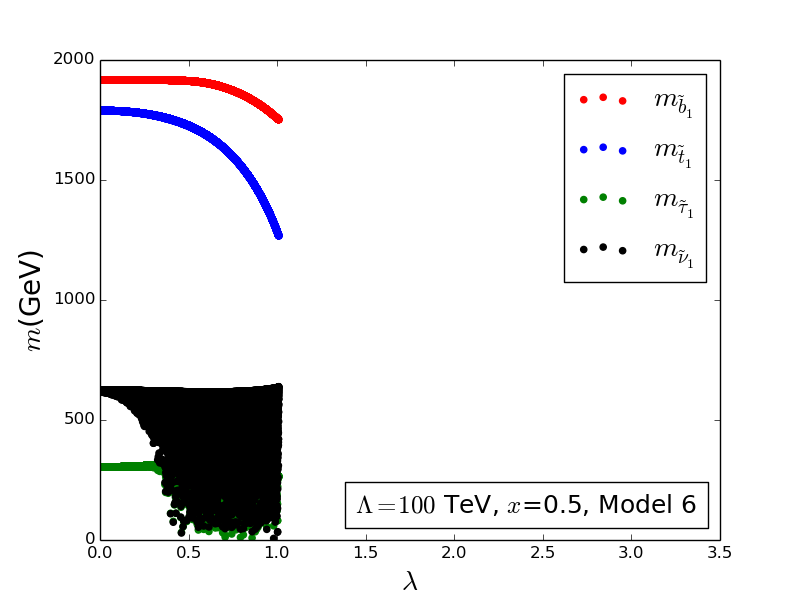}}
\subfigure[]{
\includegraphics[height=5cm, width=7cm]{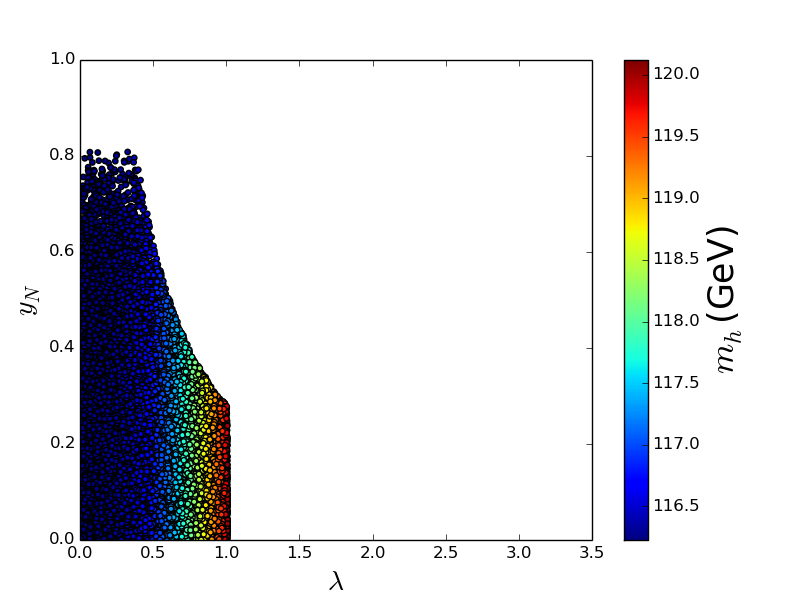}}
\subfigure[]{
\includegraphics[height=5cm, width=7cm]{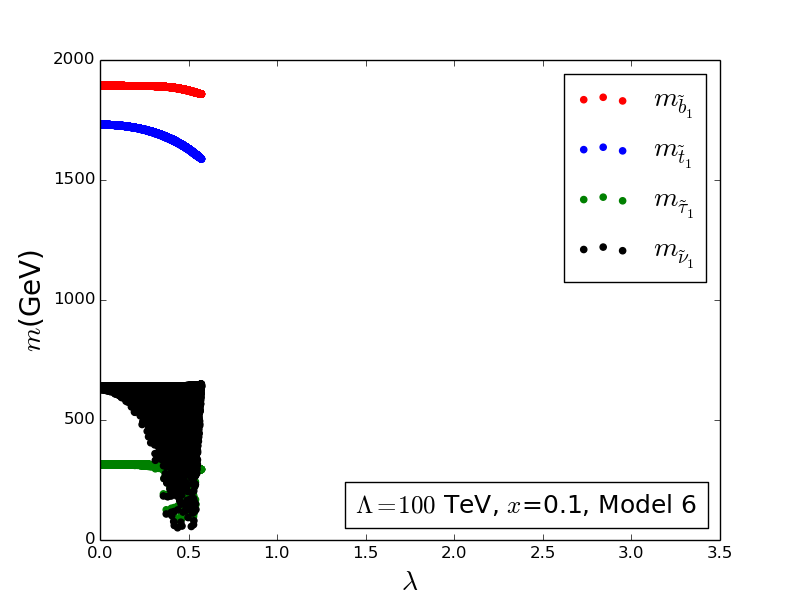}}
\subfigure[]{
\includegraphics[height=5cm, width=7cm]{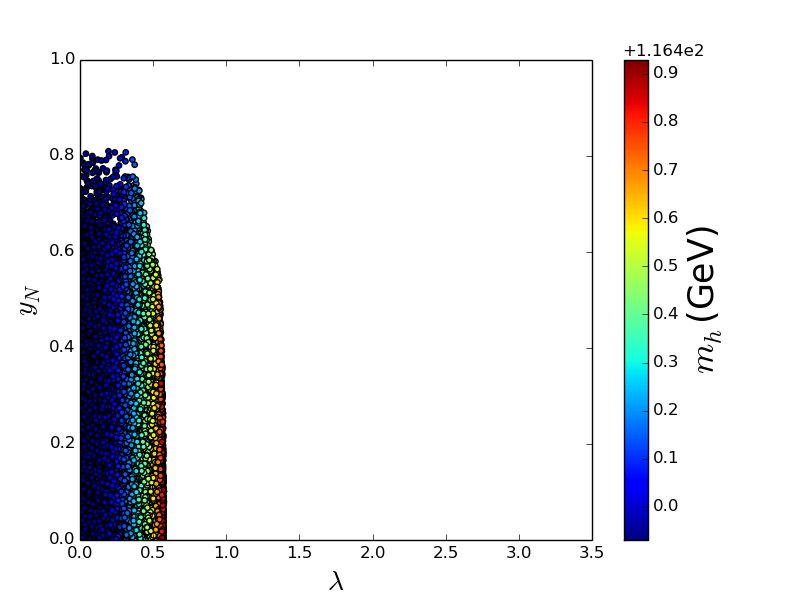}}
\caption{\sf Model 6: Spectrum variation with $\lambda$ and Higgs 
mass values in $y_N$ and $\lambda$ plane. Please see caption of 
\fig{f:mod1} 
for  details of notation.}\label{f:mod6}
\end{figure}

\subsection{Model 5}
This model represents the interaction of $N^c$ and $E^c$ field with
 the messenger field $E^c_m$ so consequently $N^c$ and $E^c$ get
 negative corrections at one-loop and positive two-loop corrections.  The other 
fields interacting with them
 receive two-loop corrections, as given below 
\beqa
 \delta M^2_{\tilde L} &=& \Big[ -\frac{\alpha _{\lambda } \left(\alpha 
_N+\alpha _{\tau
}\right)}{16 \pi ^2} \Big]\Lambda^2 \nonumber  \\
 \delta M^2_{\tilde E^c} &=& \Big[\OOLo + \frac{\alpha _{\lambda } \left(-12 
\alpha 
_1+10 
\alpha
_N+15 \alpha _{\lambda }\right)}{80 \pi ^2} \Big]\Lambda^2 \nonumber  \\
\delta M^2_{\tilde N^c} &=& \Big[\OOLo+ \frac{\alpha _{\lambda } \left(-12 
\alpha 
_1+15 
\alpha
_{\lambda }+10 \alpha _{\tau }\right)}{80 \pi ^2} \Big]\Lambda^2 \nonumber  \\
 \delta M^2_{H_u} &=& \Big[ -\frac{\alpha _N \alpha _{\lambda }}{16 \pi ^2} 
\Big]\Lambda^2 
\nonumber  \\
 \delta M^2_{H_d} &=& \Big[ -\frac{\alpha _{\lambda } \alpha _{\tau }}{16 \pi 
^2} 
\Big]\Lambda^2 \nonumber  \\
\delta A_{\tau } &=& \Big[ -\frac{\alpha _{\lambda }}{4 \pi } \Big]\Lambda 
\nonumber
\\
\delta A_N &=& \Big[ -\frac{\alpha _{\lambda }}{4 \pi } \Big]\Lambda.
\label{eq:mod5}
\eeqa

As can be seen from the above, \Eqn{eq:mod5}, this model leaves the stop sector
including $A_t$ untouched. The variation of the  lightest sparticles 
eigenvalues 
 with $\lambda$ are presented in \fig{f:mod5}(a).  As can be seen from 
the figure,  for $x=0.5$, the cancellations are really strong
and there are  two distinct regions of $\lambda$  which are viable 
phenomenologically:
$\lambda \lesssim 0.2 $ and $1.7 \lesssim \lambda \lesssim  2.2 5$.  For lower 
values of the $\lambda$, the
stau becomes tachyonic as the one-loop contributions from matter messenger 
mixing lead
to significant cancellations in the boundary conditions. For larger values of 
$\lambda \gtrsim 2.0$, 
the $A_\tau$ becomes large leading again to  tachyonic spectrum of the 
$\tilde{\tau}$. 
The region with $\lambda \lesssim 0.2 $ is irrelevant phenomenologically and 
thus not pursued
further.  From  \fig{f:mod5}(c) we see that the cancellations are extremely 
mild and almost 
disappear for $ x=0.1$. For $\lambda \gtrsim 1.5$, the spectrum turns tachyonic 
and thus unphysical.

The Higgs mass values are shown in Figs. \ref{f:mod5}(b) and \ref{f:mod5}(d) 
for $x=0.5$ and $x=0.1$ respectively.  Unlike in Models 1 and 2, $A_N$ does 
not have a factor three in its messenger-matter contributions, 
Eq.(\ref{eq:mod5}). 
Since sleptons are not heavy so this model is also not able to produce correct 
Higgs mass.

In \tabl{t:mod5}, the spectra for 
four benchmark points are shown.  It can be seen that for all the 
points, Higgs mass is 116 GeV as $A_N$ is not sufficiently large and sleptons are 
light.  

\begin{table}
 $$
 \begin{array}{|c|c|c|c|c||c|c|c|c|c|}
 \hline{\mbox{Parameter}}&{\rm{ x=0.5}}
  &{\rm  x=0.5} & {\rm x=0.1} & {\rm x=0.1} &
 {\mbox{Parameter}}&{\rm{ x=0.5}}&{\rm  x=0.5}
  & {\rm x=0.1} & {\rm x=0.1}\\  \hline
\lambda &2.00 &1.79 &1.46 &           
   1.02   &
   y_N & 0.40 &0.45 &0.47 &           
   0.28  \\
  \hline
    A_N &           -2491.1 &           -1974.5 &           -1303.8 &           
 -683.0   &
   A_t &            -569.4 &           -569.3 &           -685.0 &            
-698.0  \\
    m_R &            5960.8 &            5980.6 &            6631.0 &           
 4818.2   &
   \mu & 848. &836. &884. &           
   765.  \\
 m_{\tilde\nu_1} &271. &           
   260. &435. &611.   &
 m_{\tilde\nu_{2,3}}/m_{\nu_{2,3}} &             5961. &          
   5981. &             6632. &             4818.  \\
  \hline \hline
    m_h &            116.11 &            116.64 &            116.04 &           
 116.15   &
 m_{H} &1052. &             1038. &             1096. &           
  1004.  \\
  m_h^{0+\Delta} &             116.08 &          
  116.60 &             116.03 &           116.14   &
        m_{A^0}  &             1066. &           
  1052. &             1108. &             1005.  \\
        m_h^{0} &115.84 &          
  115.84 &            115.94 &             115.95   &
      m_{H^\pm} &1068. &           
  1055. &             1110. &             1008.  \\
  \hline 
 M_{S}  &             1851. &             1847. &             1806. &           
  1797.   &
   m_{\tilde g} &2188. &           
  2188. &             2112. &             2111.  \\
      \tilde t_1 &             1804. &           
  1798. &             1747. &             1735.   &
     \tilde t_2 &1901. &           
  1897. &             1867. &             1861.  \\
      \tilde b_1 &             1913. &           
  1916. &             1888. &             1892.   &
     \tilde b_2 &1917. &           
  1920. &             1893. &             1898.  \\
   \tilde \tau_1 &221. &           
   262. &216. &300.   &
  \tilde \tau_2 & 221. &           
   285. &216. &300.  \\
      \tilde u_1 &             1936. &           
  1931. &             1916. &             1909.   &
     \tilde u_2 &1969. &           
  1970. &             1940. &             1940.  \\
      \tilde d_1 &             1917. &           
  1920. &             1894. &             1898.   &
     \tilde d_2 &1955. &           
  1956. &             1923. &             1923.  \\
      \tilde e_1 &275. &           
   285. &438. &612.   &
     \tilde e_2 & 647. &           
   634. &665. &648.  \\
   N_1  &431. &430. &412. &           
   411.   &
   N_2 & 779. &774. &764. &           
   720.  \\
    N_3 &854. &842. &890. &           
   770.   &
     N_4&904. &898. &920. &           
   845.  \\
   C_1  &768. &763. &751. &           
   711.   &
   C_2 & 892. &886. &910. &           
   832.  \\
 \hline\end{array}
 $$
 \caption{\sf{Benchmark   points for  Model 5. See caption of \tabl{t:mod1} for 
details of notation.}
 \label{t:mod5}}
 \end{table}

\subsection{Model 6}
This model represents the interaction $W_{\rm mix } = \lambda N^c H_u H^m_d$. 
This is also a Type II model. One-loop and two-loop corrections at the boundary 
are given below:
 \beqa\label{e:mod6}
 \delta  M^2_{\tilde Q } &=& \Big[ -\frac{3 \alpha _t \alpha _{\lambda }}{16 
\pi ^2} \Big] \Lambda^2 \nonumber\\*
 \delta  M^2_{\tilde {U^c} } &=& \Big[ -\frac{3 \alpha _t \alpha _{\lambda }}{8 
\pi ^2} \Big] \Lambda^2 \nonumber\\*
 \delta  M^2_{\tilde L } &=& \Big[ -\frac{9 \alpha _N \alpha _{\lambda }}{16 
\pi ^2} \Big] \Lambda^2 \nonumber\\*
 \delta  M^2_{ {H_u} } &=& \Big[\OOL+ \frac{3 \alpha _{\lambda } \left(-3 
\alpha _1-15 \alpha 
_2+10 \left(\alpha _N+5 \alpha _{\lambda }\right)\right)}{80 \pi ^2} \Big] 
\Lambda^2 \nonumber\\*
 \delta  M^2_{\tilde {N^c} } &=& \Big[\TwOL+ \frac{3 \alpha _{\lambda } \left(5 
\left(-3 \alpha _2+2 
\alpha _N+3 \alpha _t+10 \alpha _{\lambda }\right)-3 \alpha _1\right)}{40 \pi 
^2} \Big] \Lambda^2 \nonumber\\*
 \delta  A_t  &=& \Big[ -\frac{3 \alpha _{\lambda }}{4 \pi } \Big] \Lambda 
\nonumber\\*
 \delta  A_N  &=& \Big[ -\frac{9 \alpha _{\lambda }}{4 \pi } \Big] \Lambda.
\eeqa
In this model, two-loop messenger-matter contributions to $\tilde Q, \tilde 
U^c$ and $\tilde L$ masses are negative whereas $\delta M_{H_u}^2$ gets both 
negative and positive contribution. For $x=0.5$ case the maximum allowed value 
for $\lambda$ is $\sim 1.0$ as shown in \fig{f:mod6}(a). In  \fig{f:mod6}(b), 
Higgs mass values are shown in the $\lambda$ vs $y_N$ 
plane. One can not achieve correct Higgs mass in this model.

For $x=0.1$, one-loop effect is small and thus $M_{H_u}^2$ becomes more massive 
for increasing 
$\lambda$. But  we cannot consider very large $\lambda$ value because above 
certain value electro-weak symmetry breaking (EWSB) will be impossible. The 
allowed value of $\lambda$ for successful  EWSB is $\lesssim 0.6$ as shown in  
\fig{f:mod6}(c). As allowed values of $\lambda$ are small, $A_N$ is not very 
large, consequently one can not get correct Higgs mass.

In \tabl{t:mod6}, we show four benchmark points, two for $x=0.5$ and two for 
$x=0.1$. 
Here $A_t$ is less than $A_N$. Only for the first benchmark point, Higgs mass 
is 
raised by 2 GeV (from MSSM contribution) and others have small $A_N$, so there 
is no correction at all.
\begin{table}
 $$
 \begin{array}{|c|c|c|c|c||c|c|c|c|c|}
 \hline{\mbox{Parameter}}&{\rm{ x=0.5}}
  &{\rm  x=0.5} & {\rm x=0.1} & {\rm x=0.1} &
 {\mbox{Parameter}}&{\rm{ x=0.5}}&{\rm  x=0.5}
  & {\rm x=0.1} & {\rm x=0.1}\\  \hline
\lambda &0.89 &0.53 &0.54 &           
   0.30   &
   y_N & 0.28 &0.54 &0.48 &           
   0.21  \\
  \hline
    A_N &           -4301.5 &           -1496.9 &           -1563.2 &           
 -524.4   &
   A_t &            -1768.2 &           -992.2 &           -1100.1 &            
-828.1  \\
    m_R &            3925.9 &            6678.3 &            9942.6 &           
 7868.5   &
   \mu &1395. &             1510. &249. &           
   781.  \\
 m_{\tilde\nu_1} &297. &           
   257. &333. &629.   &
 m_{\tilde\nu_{2,3}}/m_{\nu_{2,3}} &             3927. &          
   6679. &             9943. &             7869.  \\
  \hline \hline
    m_h &            118.93 &            116.84 &            117.10 &           
 116.53   &
 m_{H} &1535. &             1629. &666. &           
  1016.  \\
  m_h^{0+\Delta} &118.92 &         
   116.80 &            117.08 &           116.53   &
        m_{A^0}  &             1543. &           
  1649. &676. &             1018.  \\
        m_h^{0} &116.23 &          
  116.22 &             116.33 &            116.33   &
      m_{H^\pm} &1545. &           
  1651. &681. &             1020.  \\
  \hline 
 M_{S}  &             1643. &             1803. &             1724. &           
  1778.   &
   m_{\tilde g} &2181. &           
  2186. &             2108. &             2111.  \\
      \tilde t_1 &             1462. &           
  1715. &             1604. &             1701.   &
     \tilde t_2 &1846. &           
  1896. &             1853. &             1859.  \\
      \tilde b_1 &             1821. &           
  1910. &             1863. &             1891.   &
     \tilde b_2 &1925. &           
  1925. &             1897. &             1900.  \\
   \tilde \tau_1 &259. &           
   233. &289. &313.   &
  \tilde \tau_2 & 301. &           
   320. &293. &316.  \\
      \tilde u_1 &             1928. &           
  1930. &             1908. &             1908.   &
     \tilde u_2 &1942. &           
  1948. &             1914. &             1930.  \\
      \tilde d_1 &             1933. &           
  1925. &             1902. &             1900.   &
     \tilde d_2 &1934. &           
  1933. &             1902. &             1911.  \\
      \tilde e_1 &301. &           
   320. &293. &316.   &
     \tilde e_2 & 341. &           
   341. &338. &631.  \\
   N_1  &431. &431. &236. &           
   411.   &
   N_2 & 823. &821. &253. &           
   729.  \\
    N_3 &             1403. &             1518. &422. &           
   787.   &
     N_4&             1408. &             1523. &807. &           
   853.  \\
   C_1  &805. &807. &243. &           
   719.   &
   C_2 &1399. &             1512. &790. &           
   840.  \\
 \hline\end{array}
 $$
\caption{\sf{Benchmark   points for Model 6. See caption of \tabl{t:mod1} for 
details of notation.} \label{t:mod6}}\end{table}

\begin{figure}
\centering
\subfigure[]{
\includegraphics[height=5cm, width=7cm]{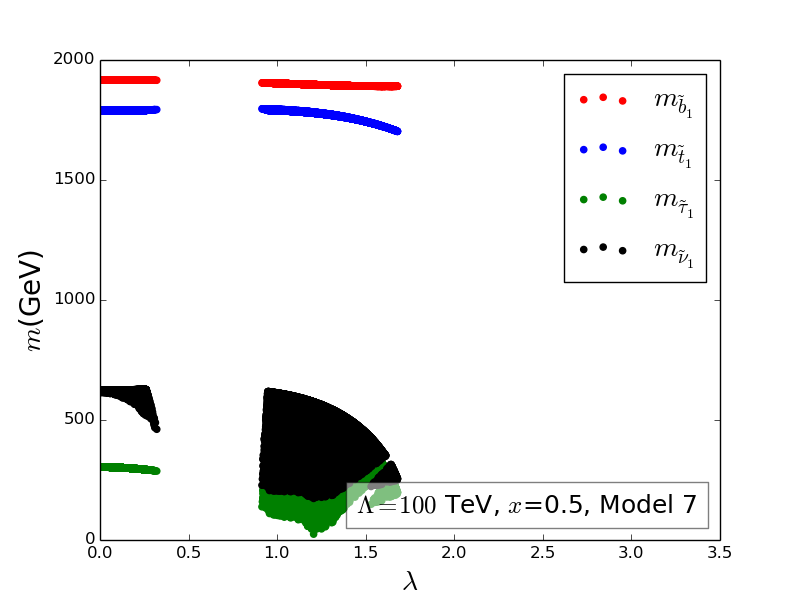}}
\subfigure[]{
\includegraphics[height=5cm, width=7cm]{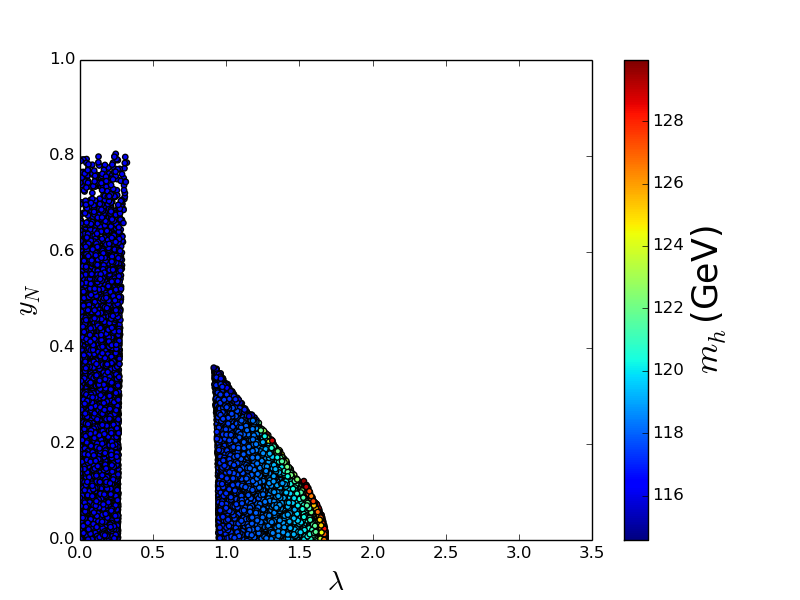}}
\subfigure[]{
\includegraphics[height=5cm, width=7cm]{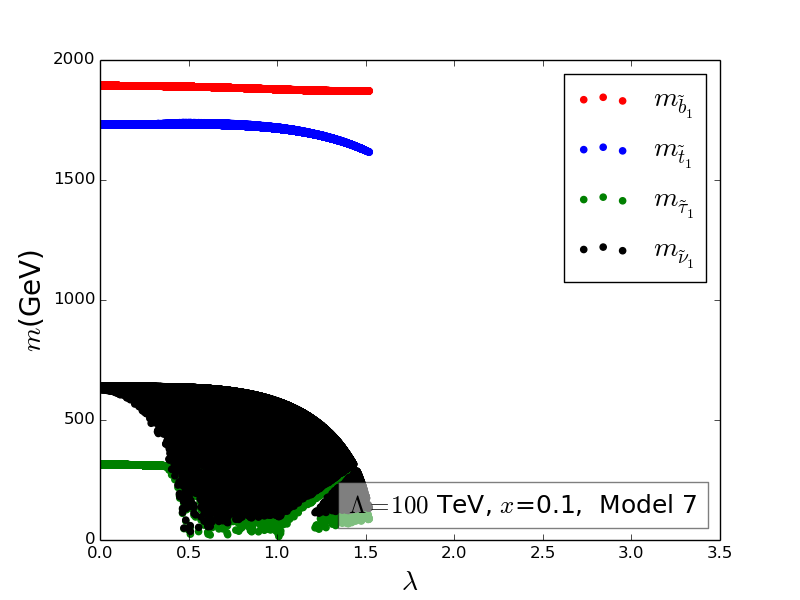}}
\subfigure[]{
\includegraphics[height=5cm, width=7cm]{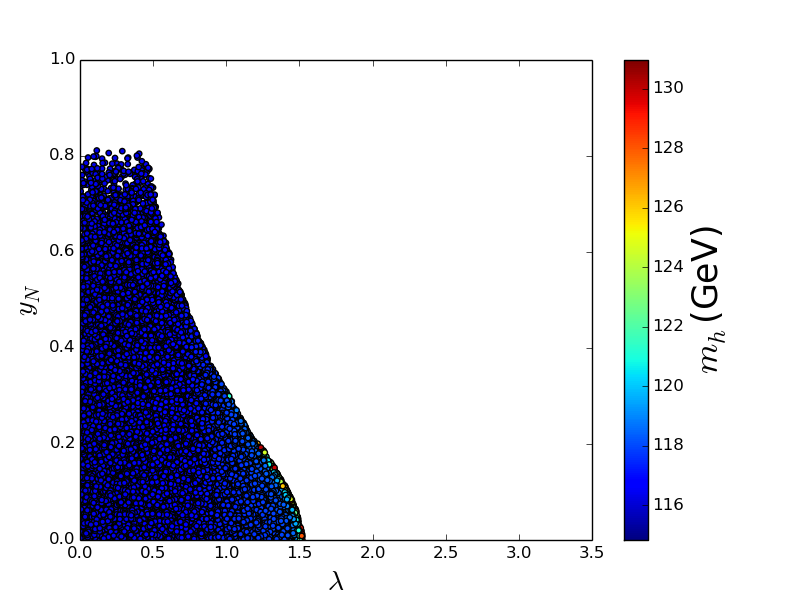}}
\caption{\sf Model 7: Spectrum variation with $\lambda$ and Higgs 
mass values in $y_N$ and $\lambda$ plane. Please see caption of 
\fig{f:mod1} 
for  details of notation.}\label{f:mod7}
\end{figure}

  
\subsection{Model 7}
This model is a Type II model. Here messenger-matter interaction 
superpotential is $W_{\rm mix} = \lambda N^c H_d H_u^m$. Non-zero corrections 
to the different soft parameters are shown in \Eqn{e:mod7}. $\delta  M^2_{ 
{H_d} }$ and $\delta  M^2_{\tilde {N^c} }$ get both negative one-loop and 
positive two-loop corrections and other soft masses gets two-loop negative 
corrections only.
\beqa\label{e:mod7}
 \delta  M^2_{\tilde Q } &=& \Big[ -\frac{3 \alpha _b \alpha _{\lambda }}{16 
\pi ^2} \Big]\Lambda^2 \nonumber\\*
 \delta  M^2_{\tilde {D^c} } &=& \Big[ -\frac{3 \alpha _b \alpha _{\lambda }}{8 
\pi ^2} \Big]\Lambda^2 \nonumber\\*
 \delta  M^2_{\tilde L } &=& \Big[ -\frac{3 \alpha _{\lambda } \left(2 \alpha 
_N+\alpha _{\tau }\right)}{16 \pi ^2} \Big]\Lambda^2 \nonumber\\*
 \delta  M^2_{\tilde {E^c} } &=& \Big[ -\frac{3 \alpha _{\lambda } \alpha 
_{\tau }}{8 \pi ^2} \Big]\Lambda^2 \nonumber\\*
 \delta  M^2_{ {H_u} } &=& \Big[ -\frac{3 \alpha _N \alpha _{\lambda }}{8 
\pi ^2} \Big]\Lambda^2 \nonumber\\*
 \delta  M^2_{ {H_d} } &=& \Big[\OOL + \frac{3 \alpha _{\lambda } 
\left(-3 
\alpha _1-15 \alpha _2+10 \left(\alpha _N+5 \alpha _{\lambda 
}\right)\right)}{80 \pi ^2} \Big]\Lambda^2 \nonumber\\*
 \delta  M^2_{\tilde {N^c} } &=& \Big[\TwOL + \frac{3 \alpha _{\lambda } 
\left(5 
\left(-3 \alpha _2+3 \alpha _b+10 \alpha _{\lambda }+\alpha _{\tau }\right)-3 
\alpha _1\right)}{40 \pi ^2} \Big]\Lambda^2 \nonumber\\*
 \delta  A_b  &=& \Big[ -\frac{3 \alpha_{\lambda }}{4 \pi } \Big]\Lambda 
\nonumber\\*
 \delta  A_{\tau }  &=& \Big[ -\frac{3 \alpha _{\lambda }}{4 \pi } \Big]\Lambda 
\nonumber\\*
 \delta  A_N  &=& \Big[ -\frac{3 \alpha _{\lambda }}{2 \pi } \Big]\Lambda. 
\eeqa
\rpla{{{\sf\bf To be deleted} $\Rightarrow$}The supersymmetric mass of ${\tilde 
N^c}$ is $m_R$ and this is much 
larger 
than $\delta M_{\tilde N^c}^2$. Thus, the messenger-matter correction has no 
effect on 
 ${\tilde N^c}$ mass. On the other hand} There is no constraint on $H_d$ mass 
from tachyonic mass condition. Thus one may expect large $\lambda$ here. 
\rplb{For both the value of $x$, maximum allowed values of $\lambda$ is 
$\sim 1.5$ because beyond this point lightest stau becomes tachyonic. However, 
there exists two band of $\lambda$ values for $x=0.5$
because of ``non-convergent Higgs spectrum'' as can be seen in \fig{f:mod7}(a). 
One gets a 125 GeV Higgs (\fig{f:mod7}(b)\& (d)) for $\lambda\sim 1.5$ in both 
cases.}

\rpla{{{\sf\bf To be deleted} $\Rightarrow$}For the case of $x=0.1$, the upper 
value of $\lambda$ is $\sim1.5$
because beyond this point lightest stau becomes tachyonic (\fig{f:mod7}(b)). 
This value of $\lambda$ is sufficiently large to generate 125 GeV Higgs. 
From the Higgs mass scatter plot (\fig{f:mod7}(d)), one can conclude that 
$\lambda\gtrsim0.5$ is sufficient to obtain this.}

In this model $A_t$ term is not generated at the boundary.  $A_N$ is 
responsible for generation of correct Higgs mass. \rplb{One can see  from the 
benchmark points in \tabl{t:mod7} that absolute values of $A_N$ can be $\sim 
10$ TeV for $x=0.5$. Typical correction to the Higgs mass from its MSSM 
value ($m_h^0$) is 8 GeV. In all the benchmark points, $A_t(M_{\rm SUSY})$ 
is always less than $A_N$.}

\begin{table}
 $$
 \begin{array}{|c|c|c|c|c||c|c|c|c|c|}
 \hline{\mbox{Parameter}}&{\rm{ x=0.5}}
  &{\rm  x=0.5} & {\rm x=0.1} & {\rm x=0.1} &
 {\mbox{Parameter}}&{\rm{ x=0.5}}&{\rm  x=0.5}
  & {\rm x=0.1} & {\rm x=0.1}\\  \hline
 \lambda &              1.60 &              1.34 &              1.26 &          
 
   1.43   &
    y_N &               0.07 &              0.17 &              0.17 &          
 
   0.09  \\
  \hline
     A_N &           -9785.4 &           -6806.3 &           -5988.2 &          
 
-7756.1   &
    A_t &            -583.6 &           -574.8 &           -692.3 &           
-698.4  \\
     m_R &            6480.5 &            3029.5 &            7172.6 &          
 
 7859.8   &
    \mu &               929. &              858. &              932. &          
 
   957.  \\
                                 m_{\tilde\nu_1} &              275. &          
 
   281. &              238. &              164.   &
               m_{\tilde\nu_{2,3}}/m_{\nu_{2,3}} &             6481. &          
 
  3030. &             7173. &             7860.  \\
  \hline \hline
     m_h &            125.73 &            120.38 &            118.71 &          
 
 123.27   &
  m_{H} &              7172. &             4394. &             5315. &          
 
  6890.  \\
                                  m_h^{0+\Delta} &            125.73 &          
 
  120.47 &           118.71 &            123.27   &
m_{A^0}  &             7165. &             4392. &             5316. &          
 
  6885.  \\
m_h^{0} &              116.23 &           116.23 &             116.33 &         
 
  116.33   &
m_{H^\pm} &              7165. &          
 
  4393. &             5317. &             6885.  \\
  \hline 
  M_{S}  &             1779. &             1821. &             1752. &          
 
  1717.   &
                                   m_{\tilde g} &              2187. &          
 
  2187. &             2111. &             2110.  \\
\tilde t_1 &             1723. &          
 
  1769. &             1685. &             1646.   &
                                     \tilde t_2 &              1838. &          
 
  1875. &             1821. &             1791.  \\
\tilde b_1 &             1891. &          
 
  1894. &             1873. &             1870.   &
                                     \tilde b_2 &              1933. &          
 
  1922. &             1906. &             1914.  \\
                                   \tilde \tau_1 &              210. &          
 
   201. &              202. &              112.   &
                                  \tilde \tau_2 &               448. &          
 
   462. &              522. &              414.  \\
\tilde u_1 &             1862. &          
 
  1905. &             1858. &             1822.   &
                                     \tilde u_2 &              1968. &          
 
  1966. &             1941. &             1942.  \\
\tilde d_1 &             1957. &          
 
  1934. &             1925. &             1943.   &
                                     \tilde d_2 &              2005. &          
 
  1980. &             1967. &             1988.  \\
\tilde e_1 &              448. &          
 
   485. &              522. &              414.   &
                                     \tilde e_2 &               625. &          
 
   485. &              568. &              690.  \\
    N_1  &              430. &              430. &              411. &          
 
   411.   &
    N_2 &               804. &              785. &              774. &          
 
   777.  \\
     N_3 &              938. &              866. &              942. &          
 
   967.   &
      N_4&              967. &              913. &              964. &          
 
   987.  \\
    C_1  &              790. &              773. &              760. &          
 
   763.   &
    C_2 &               955. &              901. &              953. &          
 
   976.  \\
 \hline\end{array}
 $$
 \caption{\sf{Benchmark  points for Model 7. See caption of \tabl{t:mod1} for 
details of notation.}}
 \label{t:mod7}
 \end{table}

\begin{figure}
\centering
\subfigure[]{
\includegraphics[height=5cm, width=7cm]{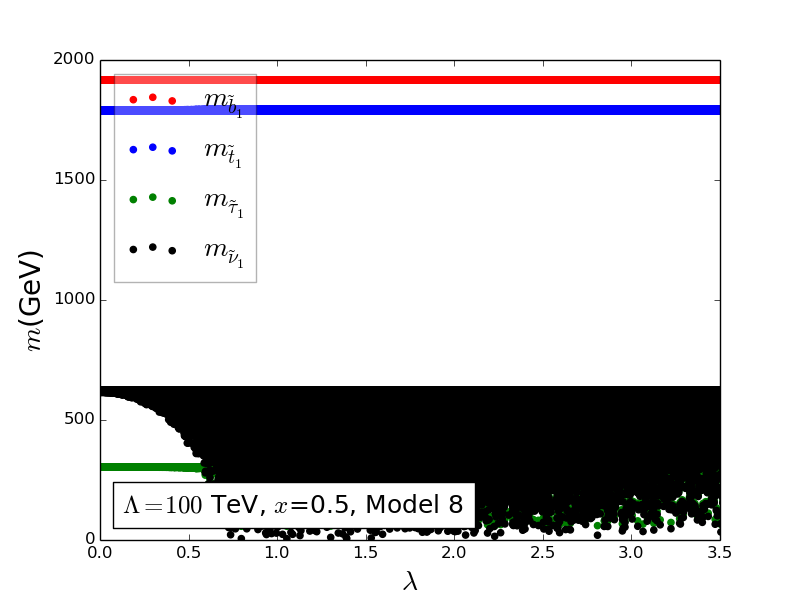}}
\subfigure[]{
\includegraphics[height=5cm, width=7cm]{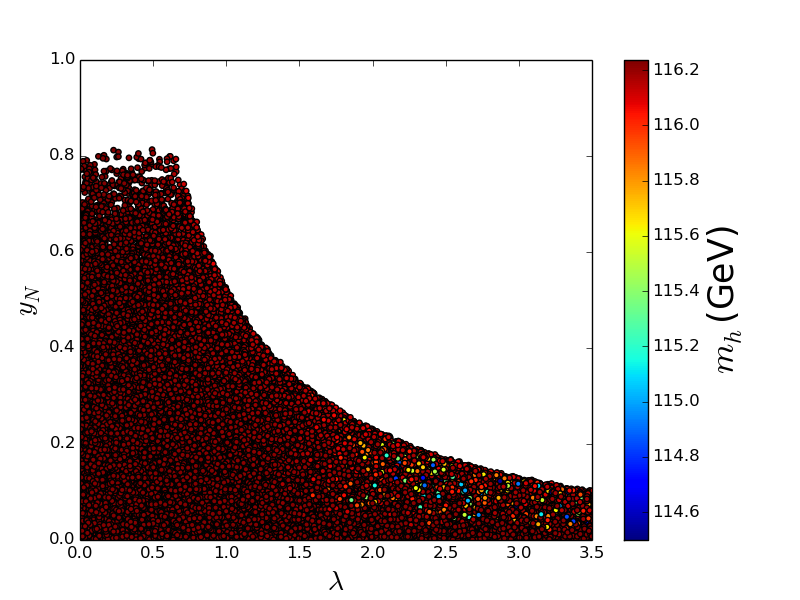}}
\subfigure[]{
\includegraphics[height=5cm, width=7cm]{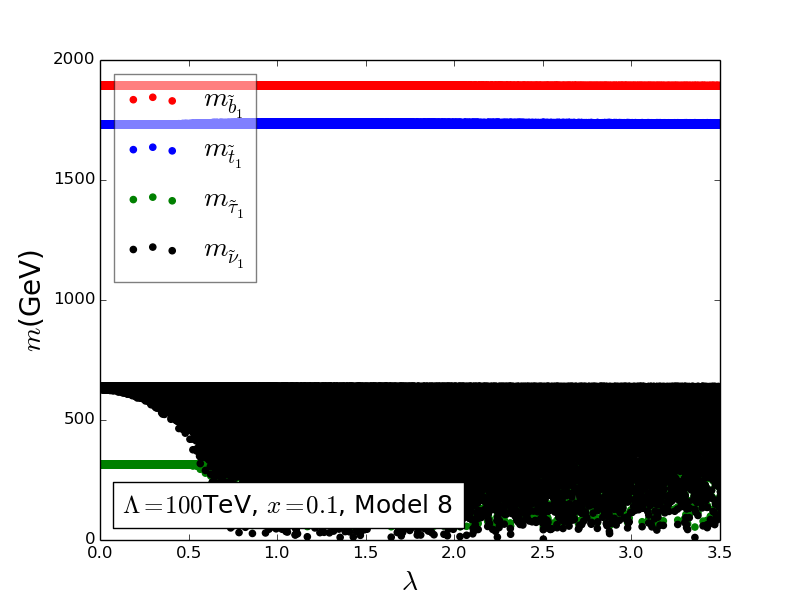}}
\subfigure[]{
\includegraphics[height=5cm, width=7cm]{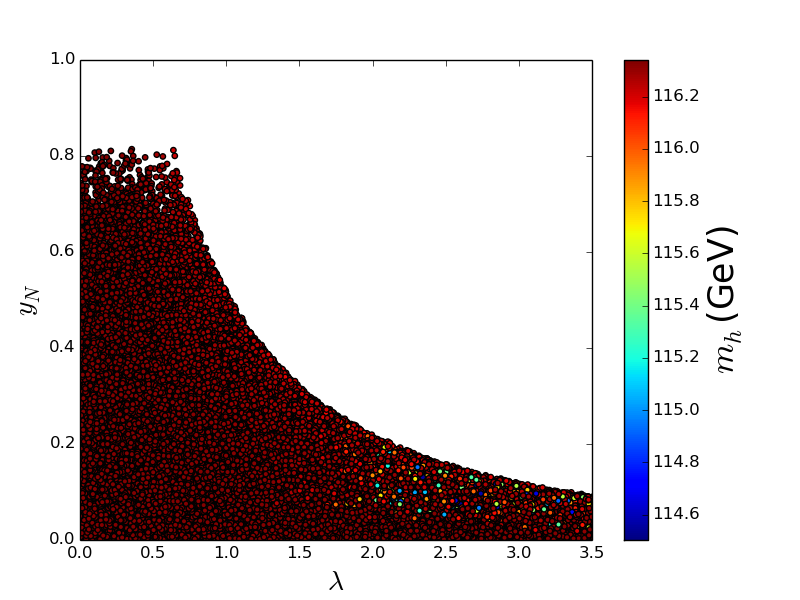}}
\caption{\sf Model 8: Spectrum variation with $\lambda$ and Higgs 
mass values in $y_N$ and $\lambda$ plane. Please see caption of 
\fig{f:mod1} 
for  details of notation.}\label{f:mod8}
\end{figure}

\subsection{Model 8}
This is a Type II model having $W_{mix}$ $=$ $\frac12 \lambda N^c N^c S^m$. 
Note that $S_m$ is a gauge singlet. Thus if we consider the messenger sector to 
be consists of only this field then no usual  soft masses will be generated. 
That is why we consider two sets of messengers: (a) singlet ($S_m$), which 
appears in the $W_{\rm mix}$ and $5\oplus\bar5$, which gives usual GMSB 
boundary conditions. None of the MSSM  fields get one-loop corrections here. 
However they get two-loop negative corrections as shown in the equation below. 
The field $\tilde N^c$ gets both the one-loop and two-loop corrections.
 \beqa
 \delta  M^2_{\tilde L } &=& \Big[ -\frac{3 \alpha _N \alpha _{\lambda }}{16 
\pi 
^2} \Big]\Lambda^2 \nonumber\\*
 \delta  M^2_{{H_u} } &=& \Big[ -\frac{3 \alpha _N \alpha _{\lambda }}{16 
\pi ^2} \Big]\Lambda^2 \nonumber\\*
 \delta  M^2_{\tilde {N^c} } &=& \Big[\OOL + \frac{3 \alpha _{\lambda } \left(4 
\alpha _N+13 \alpha _{\lambda }\right)}{32 \pi ^2} \Big]\Lambda^2 \nonumber\\*
 \delta  A_N  &=& \Big[ -\frac{3 \alpha _{\lambda }}{4 \pi } \Big]\Lambda.
\eeqa
\rplb{As $\delta  M^2_{\tilde L }$ is negative, $d_{1, 2}$ will not be small 
enough to raise the Higgs mass. That is why we are not getting a 125 GeV Higgs 
even though very large values of  $\lambda$ are allowed (\fig{f:mod8}(a) \& 
(c)) and  $\sim 10$ TeV $A_N$ is generated the $M_{\rm SUSY}$ scale 
(\tabl{t:mod8}).}

\rpla{{{\sf\bf To be deleted} $\Rightarrow$}As mentioned earlier, 
messenger-matter corrections to the $\tilde N^c$ is much 
smaller than its supersymmetric mass and thus upper limit of $\lambda$ is 
determined by tachyonic mass for lightest stau. Correction to the $\delta  
M^2_{\tilde L }$ is from two-loop only. That is why this model has similar 
spectrum for  $x=0.1$ and $x=0.5$ case as shown in \fig{f:mod8}(a) and  
\fig{f:mod8}(c). This model has wide range of allowed values of $\lambda$. As 
the squarks do not get any correction at the boundary, so the corresponding 
eigenvalues do not change with $\lambda$ (\fig{f:mod8}(a) and  
\fig{f:mod8}(c)). One can not achieve  125 GeV Higgs mass in this model  as 
can be seen from the in Higgs mass values, \fig{f:mod8}(b) and 
\fig{f:mod8}(d). Note that these two scatter plots are also almost identical to 
each other.}

In this model no other $A$-terms 
except the $A_N$ is generated at the boundary. As we have 
discussed earlier only $A_N$ is not sufficient. One also need direct coupling 
of $\tilde L$ with the messenger field to get correct Higgs mass.

 \begin{table}
 $$
 \begin{array}{|c|c|c|c|c||c|c|c|c|c|}
 \hline{\mbox{Parameter}}&{\rm{ x=0.5}}
  &{\rm  x=0.5} & {\rm x=0.1} & {\rm x=0.1} &
 {\mbox{Parameter}}&{\rm{ x=0.5}}&{\rm  x=0.5}
  & {\rm x=0.1} & {\rm x=0.1}\\  \hline
\lambda &2.53 &0.98 &2.79 &           
   1.52   &
   y_N & 0.12 &0.37 &0.12 &           
   0.29  \\
  \hline
    A_N &          -12137.9 &           -1814.3 &          -14733.5 &           
-4307.4   &
   A_t &           -576.05 &           -575.24 &          -689.4 &           
-682.4  \\
    m_R &            2374.2 &            5253.4 &            4854.8 &           
 7984.7   &
   \mu & 800. &762. &926. &           
   921.  \\
 m_{\tilde\nu_1} &367. &           
   456. &261. &272.   &
 m_{\tilde\nu_{2,3}}/m_{\nu_{2,3}} &             2377. &          
   5254. &             4855. &             7985.  \\
  \hline \hline
    m_h &            115.47 &            116.20 &            116.23 &           
 116.27   &
 m_{H} &1028. &989. &             1131. &           
  1114.  \\
  m_h^{0+\Delta} &             116.21 &          
   116.19 &           116.27 &            116.25   &
        m_{A^0}  &             1026. &           
   995. &             1136. &             1132.  \\
        m_h^{0} &116.23 &          
   116.22 &           116.33 &            116.33   &
      m_{H^\pm} &1029. &           
   998. &             1138. &             1135.  \\
  \hline 
 M_{S}  &             1843. &             1843. &             1798. &           
  1799.   &
   m_{\tilde g} &2188. &           
  2188. &             2111. &             2111.  \\
      \tilde t_1 &             1793. &           
  1793. &             1738. &             1739.   &
     \tilde t_2 &1894. &           
  1894. &             1861. &             1861.  \\
      \tilde b_1 &             1917. &           
  1917. &             1892. &             1892.   &
     \tilde b_2 &1921. &           
  1921. &             1898. &             1898.  \\
   \tilde \tau_1 &299. &           
   302. &256. &261.   &
  \tilde \tau_2 & 306. &           
   306. &318. &318.  \\
      \tilde u_1 &             1929. &           
  1929. &             1907. &             1907.   &
     \tilde u_2 &1969. &           
  1969. &             1942. &             1942.  \\
      \tilde d_1 &             1921. &           
  1921. &             1899. &             1898.   &
     \tilde d_2 &1955. &           
  1955. &             1926. &             1926.  \\
      \tilde e_1 &306. &           
   306. &318. &318.   &
     \tilde e_2 & 390. &           
   461. &330. &330.  \\
   N_1  &430. &430. &412. &           
   412.   &
   N_2 & 755. &730. &771. &           
   770.  \\
    N_3 &806. &767. &932. &           
   928.   &
     N_4&880. &867. &956. &           
   952.  \\
   C_1  &746. &721. &757. &           
   757.   &
   C_2 & 867. &853. &946. &           
   942.  \\
 \hline\end{array}
 $$
 \caption{\sf{Benchmark   points for Model 8. See caption of \tabl{t:mod1} for 
details of notation.}
 \label{t:mod8}
 }\end{table}

\begin{figure}
\centering
\subfigure[]{
\includegraphics[height=5cm, width=7cm]{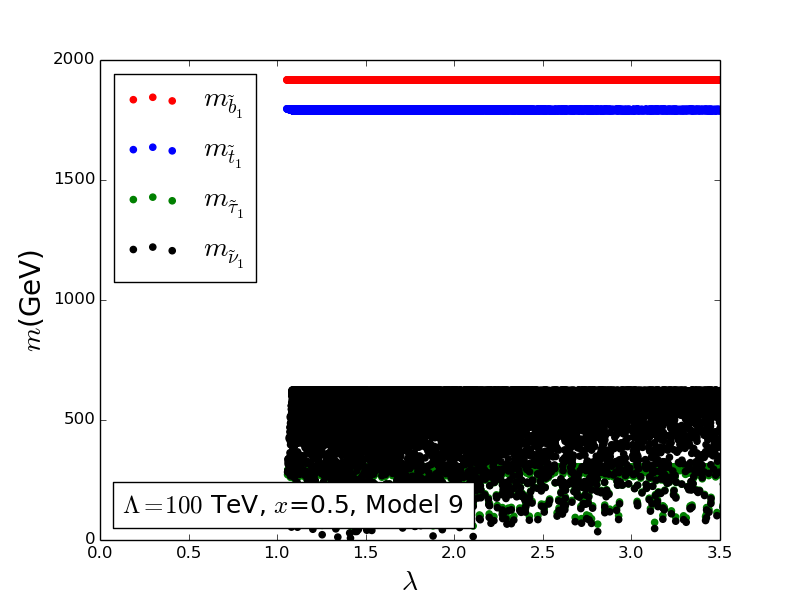}}
\subfigure[]{
\includegraphics[height=5cm, width=7cm]{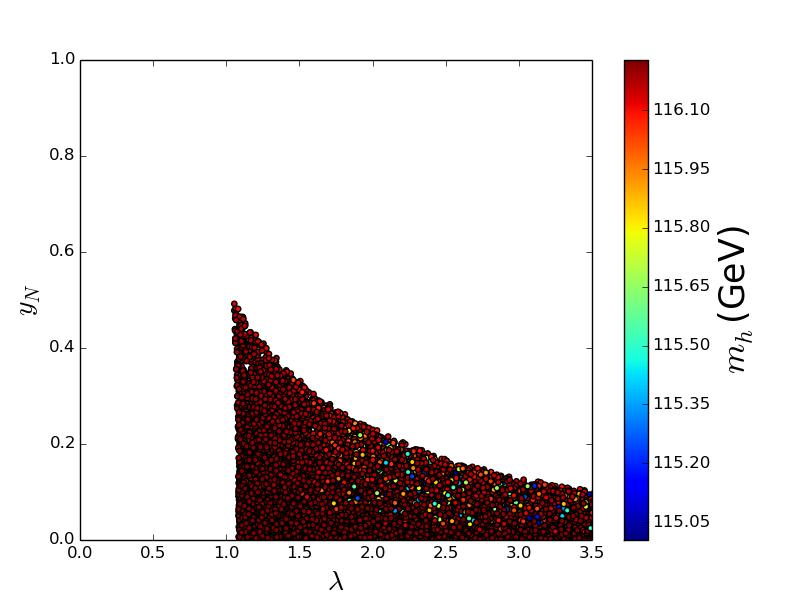}}
\subfigure[]{
\includegraphics[height=5cm, width=7cm]{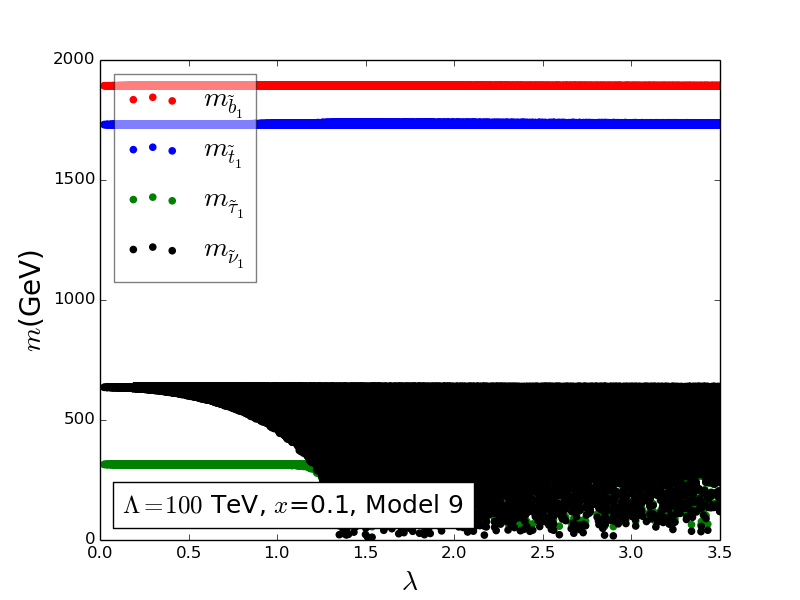}}
\subfigure[]{
\includegraphics[height=5cm, width=7cm]{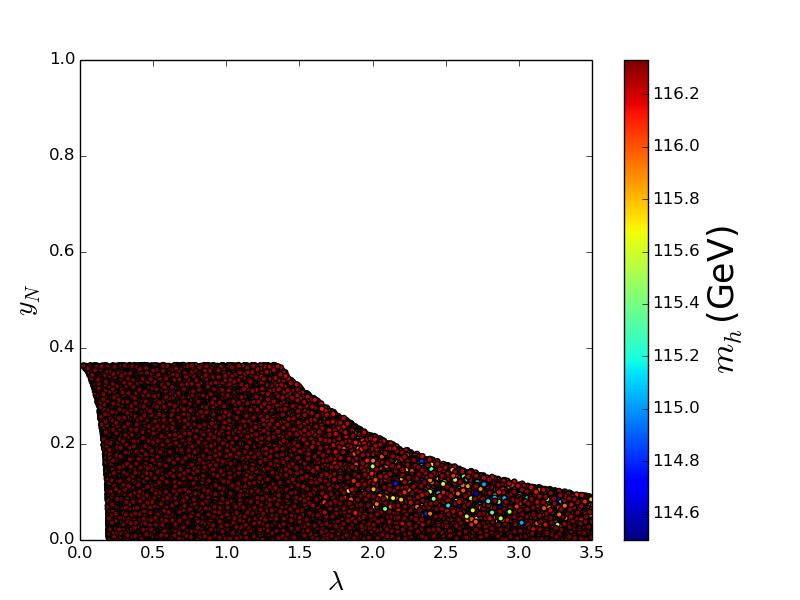}}
\caption{\sf Model 9: Spectrum variation with $\lambda$ and Higgs 
mass values in $y_N$ and $\lambda$ plane. Please see caption of 
\fig{f:mod1} 
for  details of notation.}\label{f:mod9}
\end{figure}

\subsection{Model 9}
This model is based on the $W_{mix} = \lambda N^c S S_m$ interaction. Like the 
previous model, messenger sector consists of two types of fields for the 
same reasons: (a) singlet ($S_m$), which appears in the $W_{\rm mix}$ and 
$5\oplus\bar5$, which gives usual GMSB boundary conditions.  Soft mass 
corrections from $W_{\rm mix}$  are given below:
\beqa
 \delta  M^2_{\tilde L } &=& \Big[ -\frac{3 \alpha _N \alpha _{\lambda }}{16 
\pi 
^2} \Big]\Lambda^2 \nonumber\\*
 \delta  M^2_{ {H_u} } &=& \Big[ -\frac{3 \alpha _N \alpha _{\lambda }}{16 
\pi ^2} \Big]\Lambda^2 \nonumber\\*
 \delta  M^2_{\tilde {N^c} } &=& \Big[\OOL + \frac{21 \alpha _{\lambda }^2}{16 
\pi ^2} \Big]\Lambda^2 \nonumber\\*
 \delta  M^2_{\tilde S } &=& \Big[\OOL + \frac{3 \alpha _{\lambda } \left(2 
\alpha _N+7 \alpha _{\lambda }\right)}{16 \pi ^2} \Big]\Lambda^2 \nonumber\\*
 \delta  A_N  &=& \Big[ -\frac{3 \alpha _{\lambda }}{4 \pi } \Big]\Lambda.
\eeqa

Here $\delta M^2_{\tilde S}$ has both the one-loop and two-loop contributions. 
Note 
that SUSY preserving mass of $\tilde S$ is $\mu_s$ which is  too small. 
Secondly, $\tilde S$ do not get any mass from gauge mediation as it is SM gauge 
singlet. Thus lower values of $\lambda$ are ruled out by requirement of 
non-negative soft mass of $\tilde S$:
\begin{equation}
\lambda\geq \sqrt{\left(\frac23 x^2 h(x) \pi -2 \alpha_N\right) \frac47\pi}.
\end{equation}
Another point to be noted that $\delta M^2_{\tilde S}$ is RG invariant below 
the messenger scale as there is no Yukawa or 
trilinear coupling involving $\tilde S$ below that.

Spectrum of this model is similar to the previous model except for the fact 
that $\lambda$ has now lower limit as shown in \fig{f:mod9}(a) and 
\fig{f:mod9}(c). 
These values are $\sim1$ and $\sim0.2$ for $x=0.5$ and $x=0.1$ respectively.  
The squark  masses  are independent of $\lambda$ as can be seen in 
\fig{f:mod9}(a) and \fig{f:mod9}(c). 

Higgs mass values are given in $\lambda$ vs $y_N$ plane. Note 
that we are getting Higgs mass around 116 GeV. Failure of this depicts the 
importance of heavy slepton. In this model, sleptons are not heavy as they are 
not getting any messenger-matter correction. Hence $d_{1, 2}$ are large and 
there is no significant increase in the Higgs mass. In this model also only 
$A_N$ is 
generated at the boundary. In \tabl{t:mod9},  
four benchmark points are shown.

 \begin{table}
 $$
 \begin{array}{|c|c|c|c|c||c|c|c|c|c|}
 \hline{\mbox{Parameter}}&{\rm{ x=0.5}}
  &{\rm  x=0.5} & {\rm x=0.1} & {\rm x=0.1} &
 {\mbox{Parameter}}&{\rm{ x=0.5}}&{\rm  x=0.5}
  & {\rm x=0.1} & {\rm x=0.1}\\  \hline
 \lambda &              2.40 &              1.49 &              2.26 &          
 
   1.08   &
    y_N &               0.10 &              0.22 &              0.17 &          
 
   0.36  \\
  \hline
     A_N &          -10950.2 &           -4218.6 &           -9702.2 &          
 
-2178.6   &
    A_t &           -580.8 &           -575.3 &           -684.6 &           
-684.3  \\
     m_R &            5772.3 &            3053.2 &            3746.7 &          
 
 4631.7   &
    \mu &               732. &              763. &              939. &          
 
   872.  \\
                                 m_{\tilde\nu_1} &              508. &          
 
   452. &              195. &              418.   &
               m_{\tilde\nu_{2,3}}/m_{\nu_{2,3}} &             5772. &          
 
  3054. &             3748. &             4632.  \\
  \hline \hline
     m_h &            116.21 &            116.17 &            116.10 &          
 
 116.29   &
  m_{H} &               968. &              996. &             1144. &          
 
  1085.  \\
                                  m_h^{0+\Delta} &             116.21 &         
 
   116.19 &          116.25 &             116.27   &
m_{A^0}  &              972. &              996. &             1147. &          
 
  1091.  \\
m_h^{0} &              116.23 &            116.23 &            116.33 &         
 
 116.33   &
m_{H^\pm} &               975. &          
 
   999. &             1149. &             1094.  \\
  \hline 
  M_{S}  &             1842. &             1843. &             1799. &          
 
  1798.   &
                                   m_{\tilde g} &              2188. &          
 
  2188. &             2111. &             2111.  \\
\tilde t_1 &             1792. &          
 
  1793. &             1739. &             1737.   &
                                     \tilde t_2 &              1894. &          
 
  1894. &             1861. &             1861.  \\
\tilde b_1 &             1917. &          
 
  1917. &             1892. &             1892.   &
                                     \tilde b_2 &              1921. &          
 
  1921. &             1898. &             1898.  \\
                                   \tilde \tau_1 &              303. &          
 
   302. &              202. &              311.   &
                                  \tilde \tau_2 &               306. &          
 
   306. &              318. &              318.  \\
\tilde u_1 &             1929. &          
 
  1929. &             1907. &             1907.   &
                                     \tilde u_2 &              1969. &          
 
  1969. &             1942. &             1941.  \\
\tilde d_1 &             1921. &          
 
  1921. &             1898. &             1899.   &
                                     \tilde d_2 &              1954. &          
 
  1955. &             1926. &             1925.  \\
\tilde e_1 &              306. &          
 
   306. &              318. &              318.   &
                                     \tilde e_2 &               512. &          
 
   459. &              324. &              425.  \\
    N_1  &              429. &              430. &              412. &          
 
   412.   &
    N_2 &               707. &              731. &              772. &          
 
   761.  \\
     N_3 &              737. &              768. &              945. &          
 
   878.   &
      N_4&              860. &              868. &              967. &          
 
   911.  \\
    C_1  &              699. &              722. &              759. &          
 
   749.   &
    C_2 &               845. &              853. &              958. &          
 
   901.  \\
 \hline\end{array}
 $$
 \caption{\sf{Benchmark 
  points for Model 9. See caption of \tabl{t:mod1} for details of notation.}
\label{t:mod9}
}\end{table}

\begin{figure}
\centering
\subfigure[]{
\includegraphics[height=5cm, width=7cm]{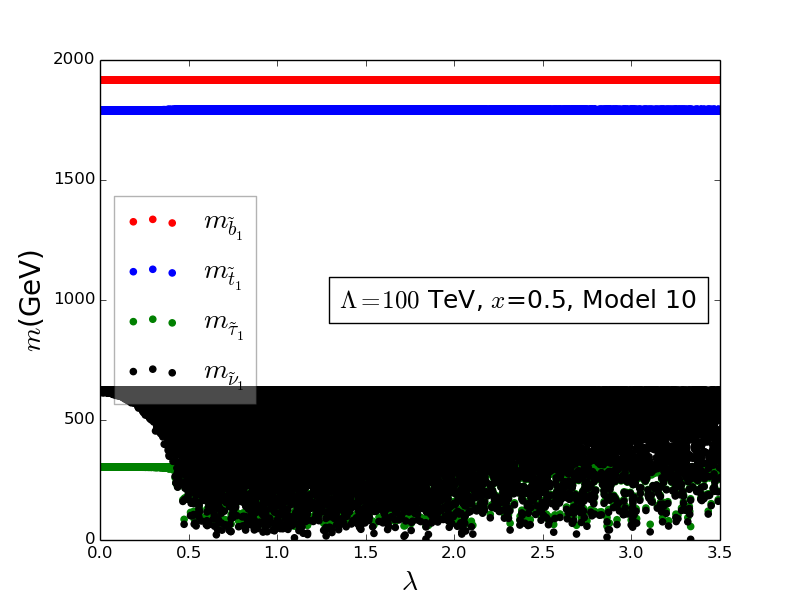}}
\subfigure[]{
\includegraphics[height=5cm, width=7cm]{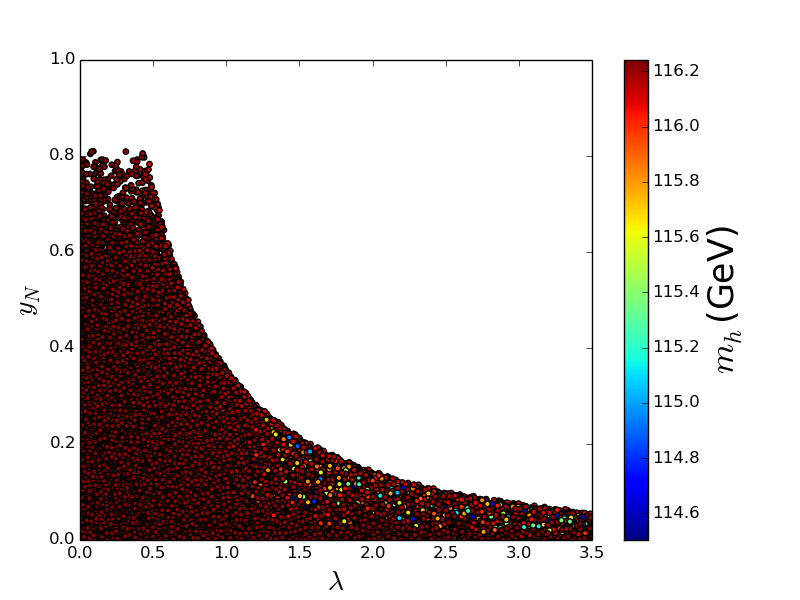}}
\subfigure[]{
\includegraphics[height=5cm, width=7cm]{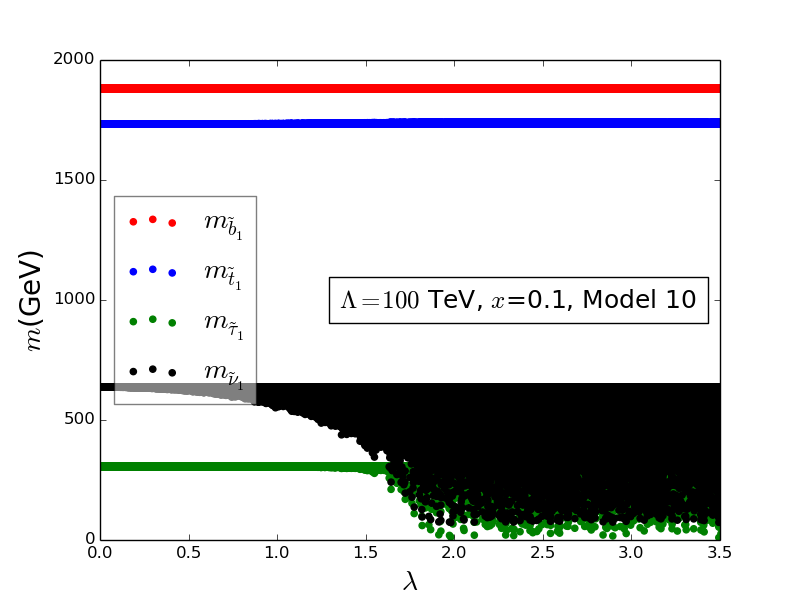}}
\subfigure[]{
\includegraphics[height=5cm, width=7cm]{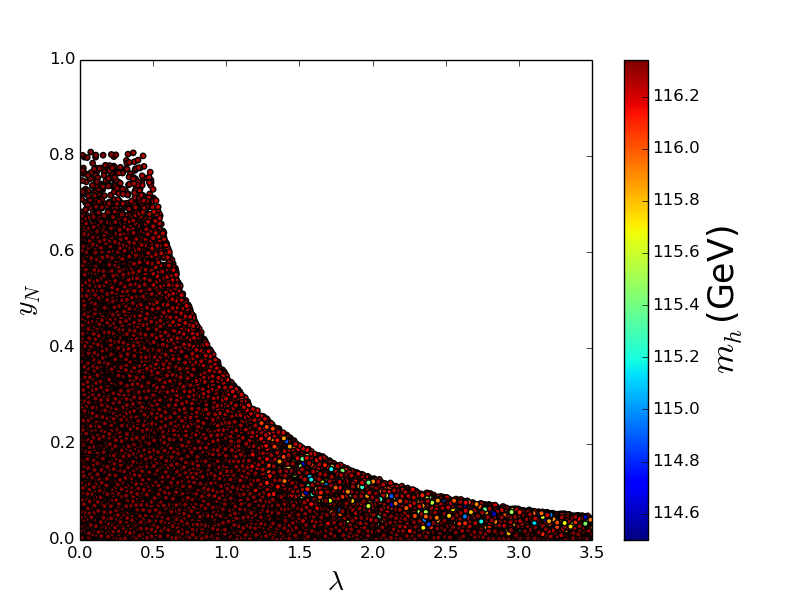}}
\caption{\sf Model 10: Spectrum variation with $\lambda$ and Higgs 
mass values in $y_N$ and $\lambda$ plane. Please see caption of 
\fig{f:mod1} 
for  details of notation.}\label{f:mod10}
\end{figure}

\subsection{Model 10}
This is a Type I model with $ W_{mix}$ $=$ $ \lambda N^c H^m_u H^m_d$. Note 
that, 
in this interaction there is only one matter field. Thus choices of the 
messenger fields are arbitrary. For example, one can have (a) $N^c 1_m 
\bar1_m$, (b) $N^c 5_m \bar5_m$, and (c) $N^c 10_m \overline{10}_m$. Here we 
consider the messenger having representation $5\oplus\bar5$. The corrections to 
the 
soft masses for this messenger-matter interaction at the boundary are given in 
\Eqn{e:mod10}. The  $\delta  M^2_{\tilde {N^c} }$ gets both one-loop negative 
and 
two-loop positive corrections. The $\delta  M^2_{\tilde L }$ and $\delta  
M^2_{{H_u} }$ are generated at two-loop level.
\beqa\label{e:mod10}
 \delta  M^2_{\tilde L } &=& \Big[  -\frac{3 \alpha _N \alpha _{\lambda }}{8 
\pi 
^2} \Big]\Lambda^2 \nonumber\\*
 \delta  M^2_{{H_u} } &=& \Big[  -\frac{3 \alpha _N \alpha _{\lambda }}{8 
\pi ^2} \Big]\Lambda^2 \nonumber\\*
 \delta  M^2_{\tilde {N^c} } &=& \Big[ -\frac{\alpha_\lambda}{2\pi}x^2 h(x) +  
\frac{3 \alpha _{\lambda } \left(-3 \alpha _1-15 \alpha _2+40 \alpha _{\lambda 
}\right)}{40 \pi ^2} \Big]\Lambda^2 \nonumber\\*
 \delta  A_N  &=& \Big[  -\frac{3 \alpha _{\lambda }}{2 \pi } \Big]\Lambda.
\eeqa
Spectral variation with $\lambda$ of this model is similar to models 8 and 9 
which is expected from the nature of messenger-matter corrections: (a) all of 
them get corrections to $M_{\tilde L}^2, M_{\tilde N^c}^2, M_{H_u}^2$ and 
$A_N$, and (b) right handed neutrinos are much heavier compared to the 
messenger-matter corrections at the boundary. Though $\lambda$ can be very 
large (\fig{f:mod10}(a) \& (c)) and a very large $A_N$ is generated at the 
$M_{\rm SUSY}$ scale (\tabl{t:mod10}), Higgs mass obtained can be atmost 
117 GeV (\fig{f:mod10}(b) \& (d)). This model also fails like the previous two 
models for the same reason. The benchmark points are given in \tabl{t:mod10}.

\begin{table}
 $$
 \begin{array}{|c|c|c|c|c||c|c|c|c|c|}
 \hline{\mbox{Parameter}}&{\rm{ x=0.5}}
  &{\rm  x=0.5} & {\rm x=0.1} & {\rm x=0.1} &
 {\mbox{Parameter}}&{\rm{ x=0.5}}&{\rm  x=0.5}
  & {\rm x=0.1} & {\rm x=0.1}\\  \hline
\lambda &2.28 &0.99 &2.25 &           
   1.04   &
   y_N & 0.10 &0.36 &0.10 &           
   0.27  \\
  \hline
    A_N &          -19775.1 &           -3696.8 &          -19295.2 &           
-4050.9   &
   A_t &           -575.6 &           -568.4 &          -690.3 &           
-684.8  \\
    m_R &            3577.1 &            8253.0 &            6582.5 &           
 5477.0   &
   \mu & 861. &873. &945. &           
   889.  \\
 m_{\tilde\nu_1} &164. &           
    81. &169. &374.   &
 m_{\tilde\nu_{2,3}}/m_{\nu_{2,3}} &             3579. &          
   8253. &             6583. &             5477.  \\
  \hline \hline
    m_h &            115.56 &            116.09 &            116.24 &           
 116.28   &
 m_{H} &1074. &             1062. &             1144. &           
  1096.  \\
  m_h^{0+\Delta} &             116.17 &          
  116.07 &            116.26 &            116.27   &
        m_{A^0}  &             1076. &           
  1086. &             1153. &             1105.  \\
        m_h^{0} &116.23 &          
  116.23 &            116.33 &            116.33   &
      m_{H^\pm} &1079. &           
  1089. &             1155. &             1108.  \\
  \hline 
 M_{S}  &             1844. &             1845. &             1798. &           
  1798.   &
   m_{\tilde g} &2188. &           
  2188. &             2111. &             2111.  \\
      \tilde t_1 &             1795. &           
  1796. &             1738. &             1737.   &
     \tilde t_2 &1894. &           
  1895. &             1861. &             1861.  \\
      \tilde b_1 &             1917. &           
  1917. &             1892. &             1892.   &
     \tilde b_2 &1921. &           
  1921. &             1898. &             1898.  \\
   \tilde \tau_1 &186. &           
    81. &170. &308.   &
  \tilde \tau_2 & 306. &           
   306. &318. &318.  \\
      \tilde u_1 &             1929. &           
  1929. &             1907. &             1907.   &
     \tilde u_2 &1970. &           
  1970. &             1942. &             1942.  \\
      \tilde d_1 &             1921. &           
  1921. &             1898. &             1899.   &
     \tilde d_2 &1956. &           
  1957. &             1926. &             1925.  \\
      \tilde e_1 &306. &           
   306. &318. &318.   &
     \tilde e_2 & 312. &           
   310. &323. &386.  \\
   N_1  &430. &431. &412. &           
   412.   &
   N_2 & 784. &788. &773. &           
   765.  \\
    N_3 &867. &879. &952. &           
   896.   &
     N_4&913. &921. &973. &           
   925.  \\
   C_1  &772. &776. &760. &           
   752.   &
   C_2 & 901. &909. &964. &           
   915.  \\
 \hline\end{array}
 $$
 \caption{\sf{Benchmark   points for Model 10. See caption of \tabl{t:mod1} for 
details of notation.}}
\label{t:mod10}
\end{table}

\begin{figure}
\centering
\subfigure[]{
\includegraphics[height=5cm, width=7cm]{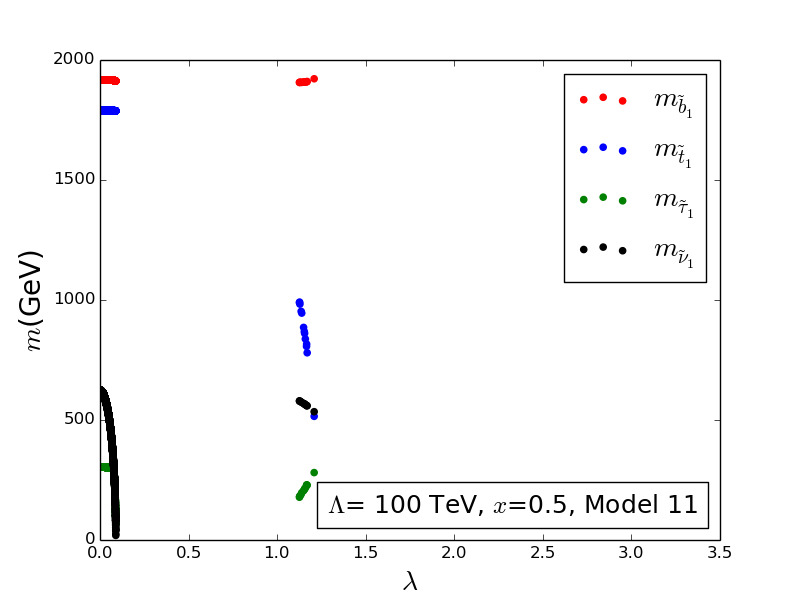}}
\subfigure[]{
\includegraphics[height=5cm, width=7cm]{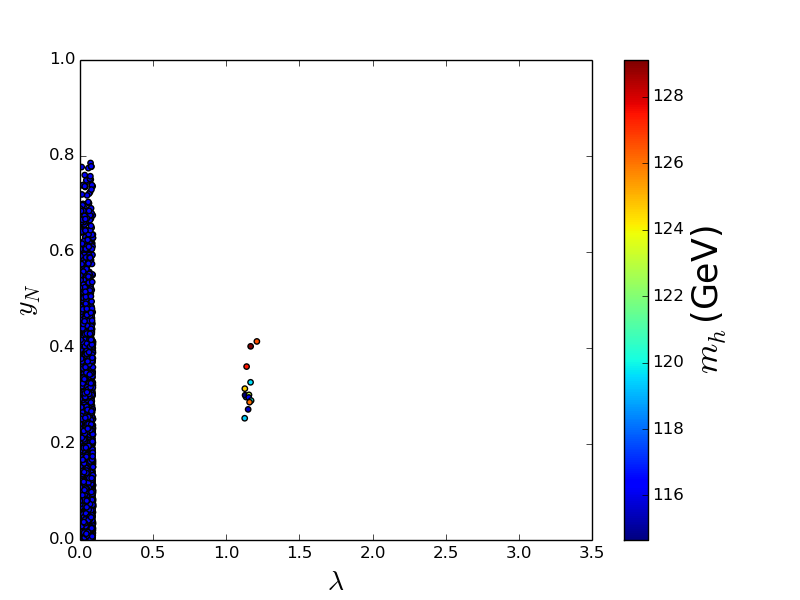}}
\subfigure[]{
\includegraphics[height=5cm, width=7cm]{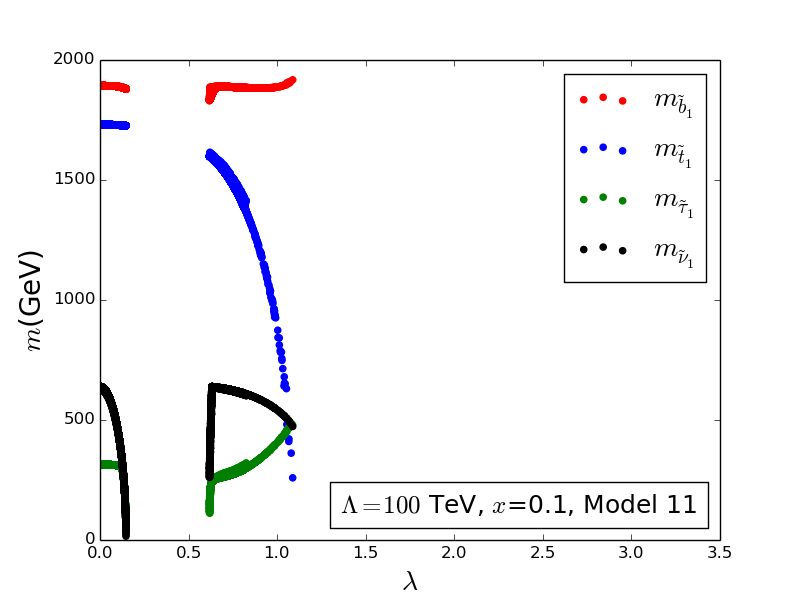}}
\subfigure[]{
\includegraphics[height=5cm, width=7cm]{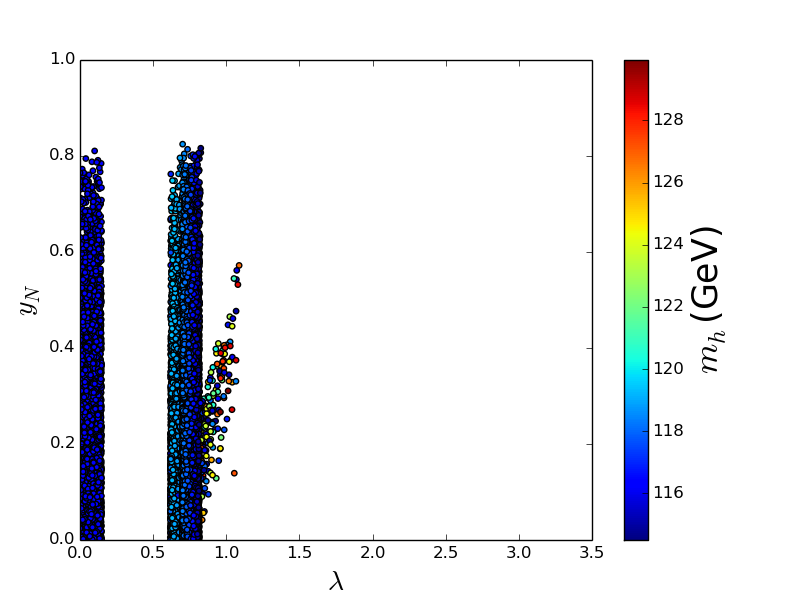}}
\caption{\sf Model 11: Spectrum variation with $\lambda$ and Higgs 
mass values in $y_N$ and $\lambda$ plane. Please see caption of 
\fig{f:mod1} 
for  details of notations. }\label{f:mod11}
\end{figure}

\subsection{Model 11}
This is a Type II model with the messenger-matter interaction term 
\rplb{$W_{\rm mix} 
= \lambda L Q D_m$.} Here messenger fields belong to $5\oplus\bar5$ 
representation. As both 
the matter fields are SU(2) doublets, all the MSSM fields get 
corrections to their soft masses as well as non-zero trilinear 
couplings.
 \beqa
 \delta  M^2_{\tilde Q } &=& \Big[\OOL + \frac{\alpha _{\lambda } \left(5 
\left(-9 \alpha _2-16 \alpha _3+3 \left(\alpha _N+14 \alpha _{\lambda }+\alpha 
_{\tau }\right)\right)-7 \alpha _1\right)}{80 \pi ^2} \Big]\Lambda^2 
\nonumber\\*
 \delta  M^2_{\tilde {U^c} } &=& \Big[ -\frac{3 \alpha _t \alpha _{\lambda }}{8 
\pi ^2} \Big]\Lambda^2 \nonumber\\*
 \delta  M^2_{\tilde {D^c} } &=& \Big[ -\frac{3 \alpha _b \alpha _{\lambda }}{8 
\pi ^2} \Big]\Lambda^2 \nonumber\\*
 \delta  M^2_{\tilde L } &=& \Big[\ThOL + \frac{3 \alpha _{\lambda } \left(5 
\left(-9 \alpha _2-16 \alpha _3+3 \left(\alpha _b+\alpha _t+14 \alpha _{\lambda 
}\right)\right)-7 \alpha _1\right)}{80 \pi ^2} \Big]\Lambda^2 \nonumber
\eeqa
\beqa
 \delta  M^2_{\tilde {E^c} } &=& \Big[ -\frac{9 \alpha _{\lambda } \alpha 
_{\tau }}{8 \pi ^2} \Big]\Lambda^2 \nonumber\\*
 \delta  M^2_{ {H_u} } &=& \Big[ -\frac{9 \left(\alpha _N+\alpha _t\right) 
\alpha _{\lambda }}{16 \pi ^2} \Big]\Lambda^2 \nonumber\\*
 \delta  M^2_{{H_d} } &=& \Big[ -\frac{9 \alpha _{\lambda } \left(\alpha 
_b+\alpha _{\tau }\right)}{16 \pi ^2} \Big]\Lambda^2 \nonumber\\*
 \delta  M^2_{\tilde {N^c} } &=& \Big[ -\frac{9 \alpha _N \alpha _{\lambda }}{8 
\pi ^2} \Big]\Lambda^2 \nonumber \\*
\delta  A_t  &=& \Big[ -\frac{3 \alpha _{\lambda }}{4 \pi } \Big]\Lambda 
\nonumber\\*
 \delta  A_b  &=& \Big[ -\frac{3 \alpha _{\lambda }}{4 \pi } \Big]\Lambda 
\nonumber\\*
 \delta  A_{\tau }  &=& \Big[ -\frac{9 \alpha _{\lambda }}{4 \pi } \Big]\Lambda 
\nonumber\\*
 \delta  A_N  &=& \Big[ -\frac{9 \alpha _{\lambda }}{4 \pi } \Big]\Lambda.
\eeqa
Here $\delta  M^2_{\tilde L }$ gets one-loop negative and two-loop positive 
contributions. For $x=0.5$, one-loop contribution dominates over two-loop 
contributions in this model. For very small value of $\lambda (\gtrsim 
0.1$), lightest stau and sneutrino become tachyonic as shown in 
\fig{f:mod11}(a) and again become positive for $\lambda$ $>$ 1 . On the other 
hand, for $x=0.1$, two-loop negative 
contribution to $M^2_{\tilde {E^c} }$ is responsible for smallness of 
$\lambda$. Here lightest stau and sneutrino become tachyonic for $\lambda 
\gtrsim 0.15$ and large $\lambda$ region (0.6-1.1) provides positive 
spectrum. 
In this model, all the A-terms are generated at the boundary. 
We can see from  the (\fig{f:mod11}(b) and \fig{f:mod11}(d), the upper allowed 
band of 
$\lambda$ in both $x=0.5$ and $x=0.1$ case can provide correct Higgs mass.  
In \tabl{t:mod11}, four benchmark points for this models are shown.  One can 
see 
from \tabl{t:mod11} that the combination of both matter-messenger and 
neutrino-sneutrino 
corrections will reproduce correct Higgs mass and one may get tachyonic 
spectrum 
including only the matter-messenger boundary conditions.

 \begin{table}
 $$
 \begin{array}{|c|c|c|c|c||c|c|c|c|c|}
 \hline{\mbox{Parameter}}&{\rm{ x=0.5}}
  &{\rm  x=0.5} & {\rm x=0.1} & {\rm x=0.1} &
 {\mbox{Parameter}}&{\rm{ x=0.5}}&{\rm  x=0.5}
  & {\rm x=0.1} & {\rm x=0.1}\\  \hline
   \lambda &              1.16 &           
   1.13 &              0.81 &              0.74   &
y_N &               0.29 &              0.32 &              0.11 &              
0.14  \\
  \hline
 A_N &           -7366.9 &           -6930.3 &           -3616.7 &           
-3008.0   &
A_t &            -2601.8 &           -2484.2 &           -1624.7 &           
-1472.0  \\
 m_R &            5154.4 &            4568.6 &            3015.6 &            
2158.0   &
\mu &              2157. &             2095. &             1465. &             
1331.  \\
                                 m_{\tilde\nu_1} &             4947. &          
   4290. &             2982. &             2113.   &
               m_{\tilde\nu_{2,3}}/m_{\nu_{2,3}} &             5161. &          
   4576. &             3022. &             2169.  \\
  \hline \hline
 m_h &            126.14 &            124.72 &            125.66 &            
122.15   &
    m_{H} &              2153. &           
  2105. &             1576. &             1463.  \\
                                  m_h^{0+\Delta} &              -- &            
 -- &            114.25 &            117.34   &
  m_{A^0}  &             2200. &           
  2148. &             1579. &             1464.  \\
  m_h^{0} &              116.23 &          
  116.23 &            116.33 &            116.33   &
m_{H^\pm} &              2197. &           
  2146. &             1580. &             1466.  \\
  \hline 
    M_{S}  &             1264. &           
  1376. &             1613. &             1673.   &
                                   m_{\tilde g} &              2203. &          
   2200. &             2117. &             2114.  \\
\tilde t_1 &              838. &           
   991. &             1390. &             1491.   &
                                     \tilde t_2 &              1908. &          
   1909. &             1872. &             1876.  \\
\tilde b_1 &             1909. &           
  1907. &             1885. &             1888.   &
                                     \tilde b_2 &              1943. &          
   1936. &             1903. &             1902.  \\
                                   \tilde \tau_1 &              217. &          
    182. &              294. &              272.   &
                                  \tilde \tau_2 &               4958. &         
     4300. &            2993. &             2131.  \\
\tilde u_1 &             1916. &           
  1918. &             1895. &             1903.   &
                                     \tilde u_2 &              2027. &          
   2022. &             1996. &             1997.  \\
\tilde d_1 &             1943. &           
  1937. &             1903. &             1903.   &
                                     \tilde d_2 &              2028. &          
   2024. &             1997. &             1999.  \\
\tilde e_1 &              477. &           
   441. &              402. &              361.   &
                                     \tilde e_2 &               570. &          
    584. &              617. &              630.  \\
N_1  &              431. &              431. &              412. &              
412.   &
N_2 &               834. &              832. &              788. &              
787.  \\
 N_3 &             2148. &             2087. &             1473. &             
1339.   &
  N_4&             2150. &             2090. &             1478. &             
1345.  \\
C_1  &              807. &              807. &              770. &              
770.   &
C_2 &              2156. &             2095. &             1468. &             
1335.  \\
 \hline\end{array}
 $$
 \caption{\sf{Benchmark   points for Model 11. See caption of \tabl{t:mod1} for 
details of notation.}
 \label{t:mod11}}
 \end{table}

\begin{figure}
\centering
\subfigure[]{
\includegraphics[height=5cm, width=7cm]{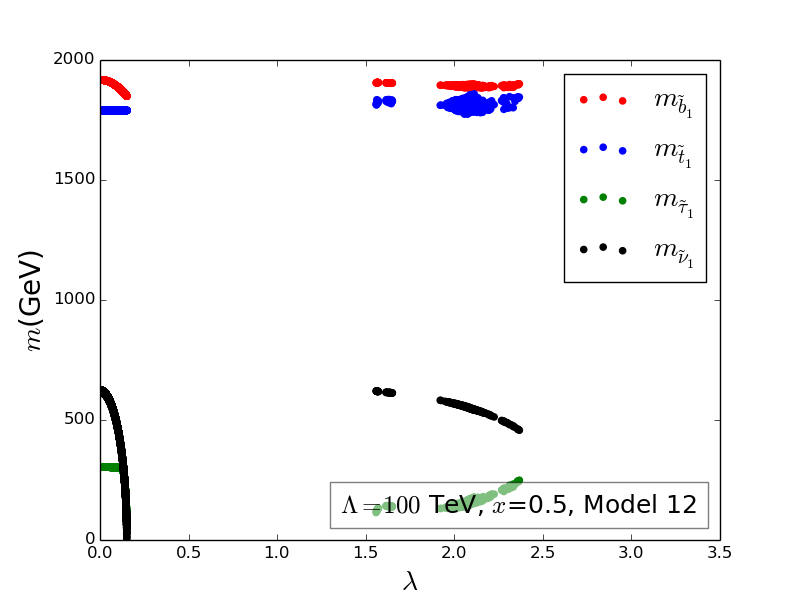}}
\subfigure[]{
\includegraphics[height=5cm, width=7cm]{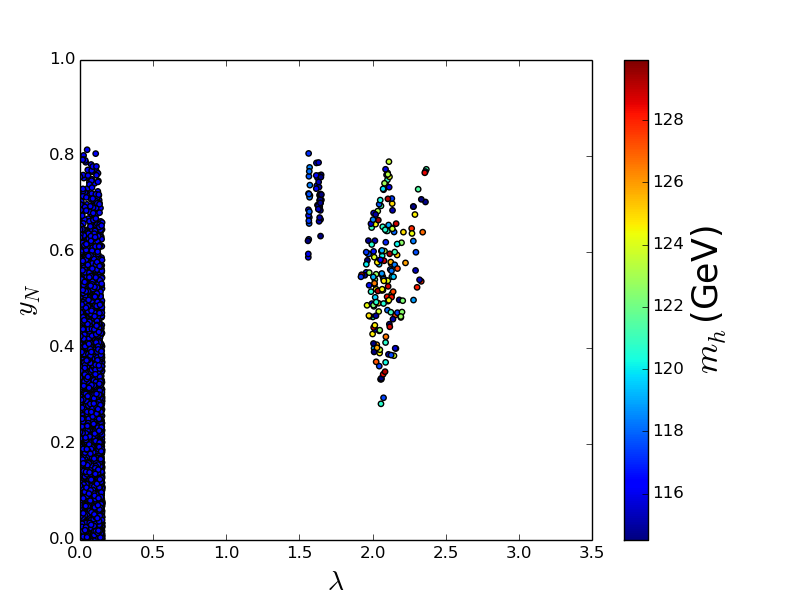}}
\subfigure[]{
\includegraphics[height=5cm, width=7cm]{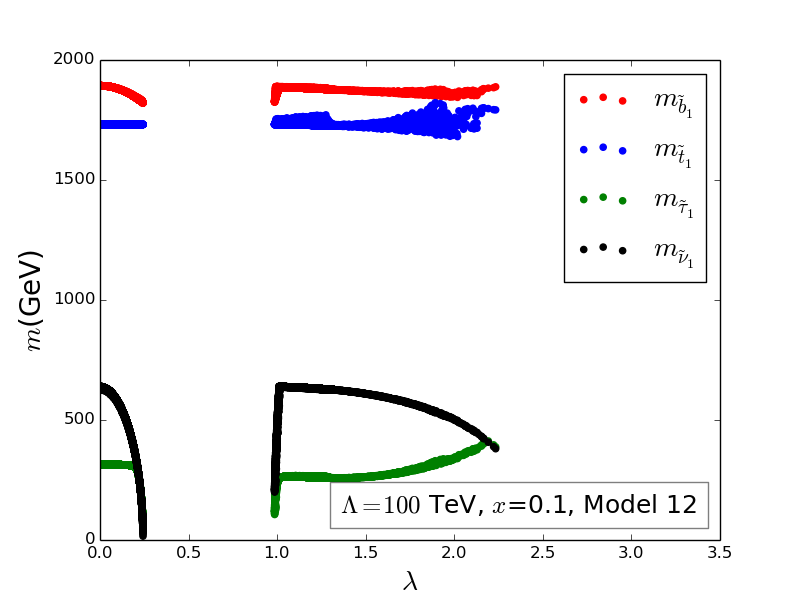}}
\subfigure[]{
\includegraphics[height=5cm, width=7cm]{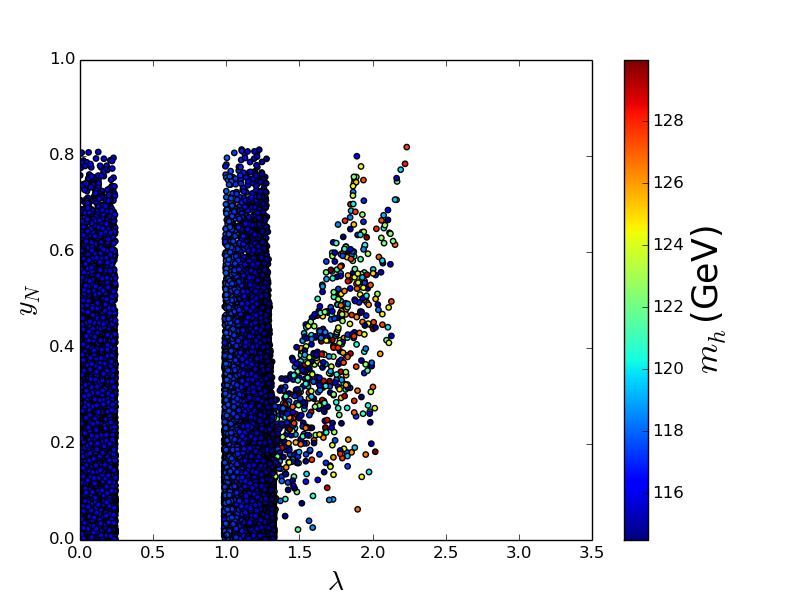}}
\caption{\sf Model 12: Spectrum variation with $\lambda$ and Higgs 
mass values in $y_N$ and $\lambda$ plane. Please see caption of 
\fig{f:mod1} 
for  details of notation.}\label{f:mod12}
\end{figure}

\subsection{Model 12}
This model represents the messenger-matter interaction superpotential 
\rplb{$W_{\rm 
mix} = \lambda L D^c  Q_m$}. Here messenger fields transform as $10\oplus 
\overline{10}$.  One-loop and two-loop corrections due to this interaction are 
shown below:
\beqa
\delta   M^2_{\tilde Q} &=& \Big[ -\frac{\alpha _b \alpha _{\lambda }}{8 \pi 
^2} 
\Big]\Lambda^2 
\nonumber  \\
\delta   M^2_{\tilde D^c} &=& \Big[\TwOLo+ \frac{\alpha _{\lambda } \left(-7 
\alpha 
_1+5 
\left(-9
\alpha _2-16 \alpha _3+3 \alpha _N+18 \alpha _{\lambda }+3 \alpha _{\tau
}\right)\right)}{120 \pi ^2} \Big]\Lambda^2 \nonumber  \\
\delta   M^2_{\tilde L} &=& \Big[\ThOLo+ \frac{\alpha _{\lambda } \left(-7 
\alpha 
_1-45 
\alpha 
_2-80
\alpha _3+30 \alpha _b+90 \alpha _{\lambda }\right)}{80 \pi ^2} 
\Big]\Lambda^2 
\nonumber  \\
\delta   M^2_{\tilde E^c} &=& \Big[ -\frac{3 \alpha _{\lambda } \alpha _{\tau 
}}{8 \pi 
^2} 
\Big]\Lambda^2 \nonumber  \\
\delta   M^2_{\tilde N^c} &=& \Big[ -\frac{3 \alpha _N \alpha _{\lambda }}{8 
\pi 
^2} 
\Big]\Lambda^2 \nonumber  \\
\delta   M^2_{H_u} &=& \Big[ -\frac{3 \alpha _N \alpha _{\lambda }}{16 \pi ^2} 
\Big]\Lambda^2 \nonumber \\
\delta   M^2_{H_d} &=& \Big[ -\frac{3 \alpha _{\lambda } \left(2 \alpha 
_b+\alpha
_{\tau }\right)}{16 \pi ^2} \Big]\Lambda^2 \nonumber  \\
\delta A_b &=& \Big[ -\frac{\alpha _{\lambda }}{2 \pi } \Big]\Lambda 
\nonumber  \\
 \delta A_{\tau } &=& \Big[ -\frac{3 \alpha _{\lambda }}{4 \pi } \Big]\Lambda 
\nonumber 
 \\
 \delta A_N &=& \Big[ -\frac{3 \alpha _{\lambda }}{4 \pi } \Big]\Lambda.
\eeqa
Here $\delta   M^2_{\tilde D^c}$ and $\delta   M^2_{\tilde L}$ get both the 
one-loop negative and two-loop positive contributions. For $x=0.5$, 
cancellation between one-loop and two-loop contribution to $ M^2_{\tilde L}$ is 
severe. We can see from \fig{f:mod12}(a) and \fig{f:mod12}(c), two band of 
$\lambda$ values are allowed in both $x=0.5$ and $x=0.1$ case.

Here all the A-terms, except $A_t$, are generated. In larger $\lambda$ region, 
we 
can achieve a 125 GeV Higgs. In the \tabl{t:mod12}, four benchmark points are 
shown 
for this model. Like model 11, in this model also, for $x=0.5$ case, only the 
combination of matter-messenger mixing and neutrino-sneutrino corrections can 
provide positive spectrum and one may get tachyonic spectrum considering only 
the matter messenger mixing boundary conditions.

 \begin{table}
 $$
 \begin{array}{|c|c|c|c|c||c|c|c|c|c|}
 \hline{\mbox{Parameter}}&{\rm{ x=0.5}}
  &{\rm  x=0.5} & {\rm x=0.1} & {\rm x=0.1} &
 {\mbox{Parameter}}&{\rm{ x=0.5}}&{\rm  x=0.5}
  & {\rm x=0.1} & {\rm x=0.1}\\  \hline
   \lambda &              2.11 &           
   1.96 &              1.55 &              1.34   &
y_N &               0.50 &              0.49 &              0.20 &              
0.24  \\
  \hline
 A_N &           -8138.8 &           -7044.0 &           -4541.8 &           
-3381.1   &
A_t &            -501.2 &           -514.9 &           -686.7 &           
-685.3  \\
 m_R &            7297.3 &            7115.6 &            5011.5 &            
3108.0   &
\mu &              1726. &             1520. &              854. &              
860.  \\
                                 m_{\tilde\nu_1} &             8098. &          
   6325. &             4904. &             3219.   &
               m_{\tilde\nu_{2,3}}/m_{\nu_{2,3}} &             7295. &          
   7119. &             5014. &             3103.  \\
  \hline \hline
 m_h &            124.62 &            124.33 &            125.54 &            
124.33   &
    m_{H} &              1536. &           
  1414. &              952. &             1009.  \\
                                  m_h^{0+\Delta} &             -- &             
-- &              109.21 &             113.62   &
  m_{A^0}  &             1726. &           
  1552. &              964. &             1017.  \\
  m_h^{0} &              116.23 &          
 116.23 &             116.33 &            116.33   &
m_{H^\pm} &              1719. &           
  1548. &              962. &             1018.  \\
  \hline 
    M_{S}  &             1836. &           
  1841. &             1781. &             1790.   &
                                   m_{\tilde g} &              2206. &          
   2203. &             2123. &             2120.  \\
\tilde t_1 &             1801. &           
  1802. &             1719. &             1729.   &
                                     \tilde t_2 &              1871. &          
   1881. &             1845. &             1853.  \\
\tilde b_1 &             1889. &           
  1892. &             1867. &             1876.   &
                                     \tilde b_2 &              1932. &          
   1927. &             1904. &             1900.  \\
                                   \tilde \tau_1 &              158. &          
    132. &              265. &              259.   &
                                  \tilde \tau_2 &               8093. &         
   6330. &            4910. &            3210.  \\
\tilde u_1 &             1893. &           
  1906. &             1890. &             1899.   &
                                     \tilde u_2 &              1965. &          
   1965. &             1932. &             1936.  \\
\tilde d_1 &             1992. &           
  1984. &             1947. &             1939.   &
                                     \tilde d_2 &              2016. &          
   2014. &             1998. &             1997.  \\
\tilde e_1 &              520. &           
   450. &              431. &              370.   &
                                     \tilde e_2 &               550. &          
    581. &              609. &              629.  \\
N_1  &              432. &              432. &              412. &              
412.   &
N_2 &               824. &              823. &              759. &              
760.  \\
 N_3 &             1733. &             1529. &              860. &              
866.   &
  N_4&             1737. &             1534. &              898. &              
902.  \\
C_1  &              805. &              805. &              744. &              
746.   &
C_2 &              1727. &             1523. &              886. &              
891.  \\
 \hline\end{array}
 $$
 \caption{\sf{Benchmark   points for Model 12. See caption of \tabl{t:mod1} for 
details of notation.}
 \label{t:mod12}}
 \end{table}

\begin{figure}
\centering
\subfigure[]{
\includegraphics[height=5cm, width=7cm]{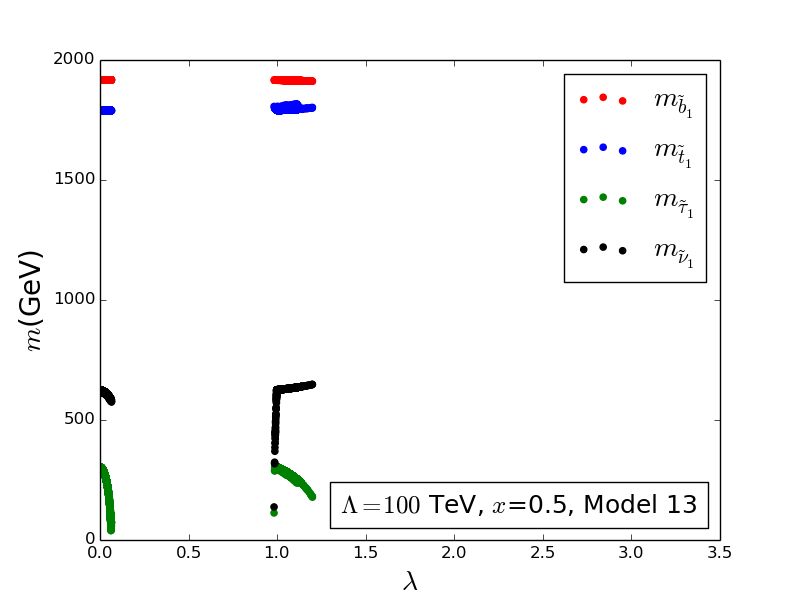}}
\subfigure[]{
\includegraphics[height=5cm, width=7cm]{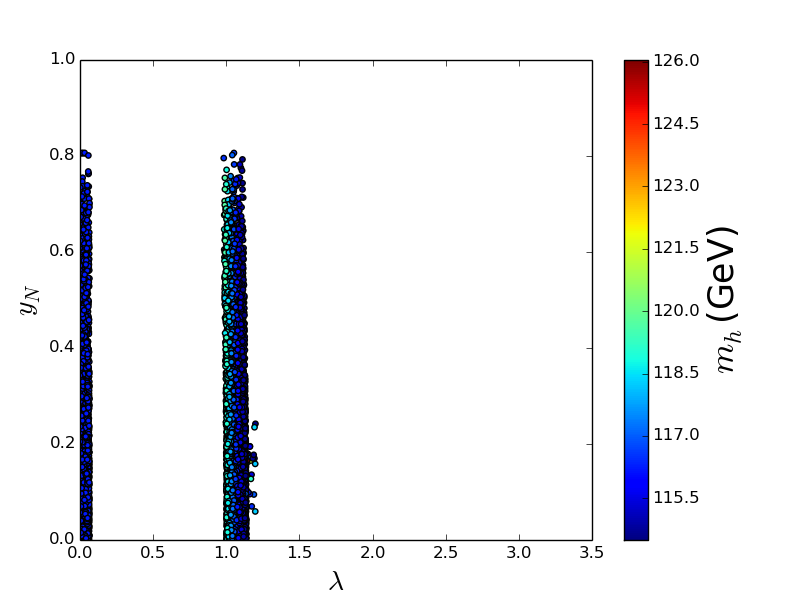}}
\subfigure[]{
\includegraphics[height=5cm, width=7cm]{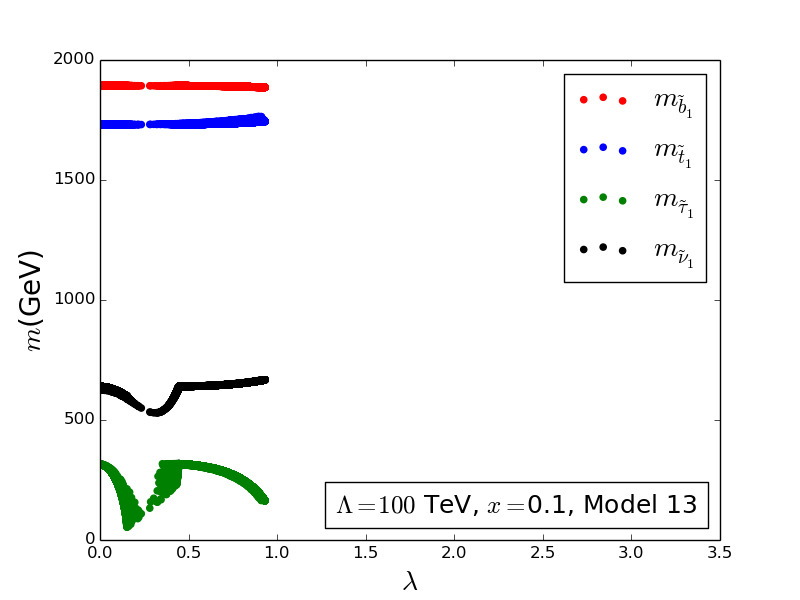}}
\subfigure[]{
\includegraphics[height=5cm, width=7cm]{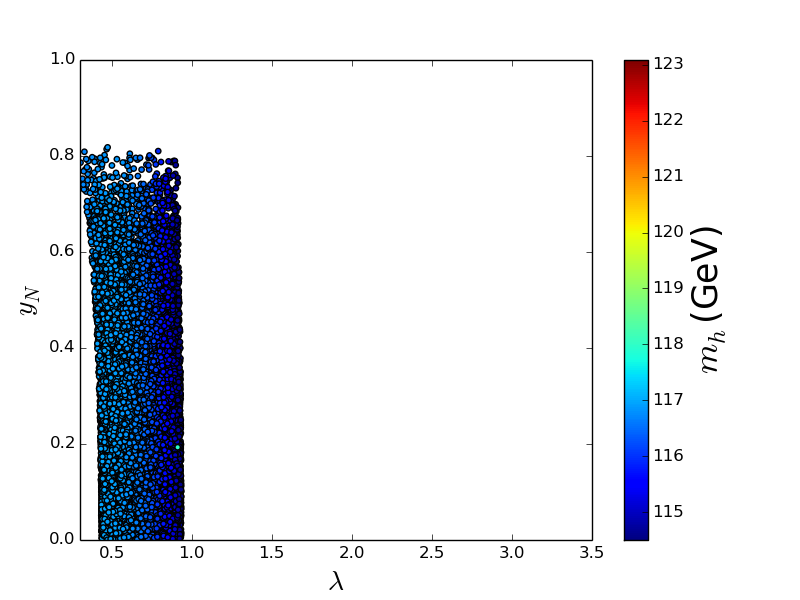}}
\caption{\sf Model 13: Spectrum variation with $\lambda$ and Higgs 
mass values in $y_N$ and $\lambda$ plane.  Note that origins 
of two right panel plots are 1.0 and 0.3 respectively.
Please see caption of \fig{f:mod1} for  details of notation.}\label{f:mod13}
\end{figure}

\subsection{Model 13}
Model 13 is Type II model having $W_{mix}$ $=$ $\lambda L E^c H_d^m$ and soft 
mass corrections are listed below :  
\beqa
 \delta  M^2_{\tilde L } &=& \Big[\OOL+ \frac{3 \alpha _{\lambda } \left(-9 
\alpha _1-15 \alpha _2+10 \left(5 \alpha _{\lambda }+\alpha _{\tau 
}\right)\right)}{80 \pi ^2} \Big]\Lambda^2 \nonumber\\*
 \delta  M^2_{\tilde {E^c} } &=& \Big[\TwOL + \frac{3 \alpha _{\lambda } 
\left(5 \left(-3 \alpha _2+\alpha _N+10 \alpha _{\lambda }+2 \alpha _{\tau 
}\right)-9 \alpha _1\right)}{40 \pi ^2} \Big]\Lambda^2 \nonumber\\*
 \delta  M^2_{ {H_u} } &=& \Big[ -\frac{3 \alpha _N \alpha _{\lambda }}{16 
\pi ^2} \Big]\Lambda^2 \nonumber\\*
 \delta  M^2_{ {H_d} } &=& \Big[ -\frac{9 \alpha _{\lambda } \alpha _{\tau 
}}{16 \pi ^2} \Big]\Lambda^2 \nonumber\\*
 \delta  M^2_{\tilde {N^c} } &=& \Big[ -\frac{3 \alpha _N \alpha _{\lambda }}{8 
\pi ^2} \Big]\Lambda^2 \nonumber\\*
 \delta  A_{\tau }  &=& \Big[ -\frac{9 \alpha _{\lambda }}{4 \pi } 
\Big]\Lambda^2 \nonumber\\*
 \delta  A_N  &=& \Big[ -\frac{3 \alpha _{\lambda }}{4 \pi } \Big]\Lambda.
\eeqa
Here both $\delta  M^2_{\tilde L }$ and $\delta  M^2_{\tilde E^c }$ get 
one-loop negative correction and two-loop positive correction. 
As earlier,  the interplay between the positive and negative contributions plays
an important role. Parameter 
space for $\lambda$ is broken down into two parts. In the region where 
$\lambda$ 
is small, one-loop contributions dominate over the two-loop contributions. Just 
above $\lambda\sim0.1$ ($\sim0.2$), lightest stau becomes tachyonic for 
$x=0.5\,(0.1)$ as shown in \fig{f:mod13}(a) and \fig{f:mod13}(c). After certain 
values of $\lambda$, the two-loop contributions dominate over the one-loop 
contributions. Note that, the  gap between two regions of parameter spaces of 
$\lambda$ is larger for $x=0.5$ which is expected. Higgs mass values 
are shown in \fig{f:mod13}(b) and \fig{f:mod13}(d). In the second region where 
$\lambda$ is higher, a 125 GeV Higgs is achievable.

\rplb{Here at the boundary, $A_\tau$ and $A_N$ are generated. In 
\tabl{t:mod13}, four 
benchmark points are shown for both $x$=0.5 and 0.1 case, the absolute values 
of $A_N$ at the $M_{\rm SUSY}$ scale is $\gtrsim 1.5$ TeV. The Higgs mass is 
raised  by 4-10 GeV in these points(\tabl{t:mod13}).}

\begin{table}
 $$
 \begin{array}{|c|c|c|c|c||c|c|c|c|c|}
 \hline{\mbox{Parameter}}&{\rm{ x=0.5}}
  &{\rm  x=0.5} & {\rm x=0.1} & {\rm x=0.1} &
 {\mbox{Parameter}}&{\rm{ x=0.5}}&{\rm  x=0.5}
  & {\rm x=0.1} & {\rm x=0.1}\\  \hline
\lambda &1.14 &0.99 &0.82 &           
   0.89   &
   y_N & 0.17 &0.57 &0.10 &           
   0.05  \\
  \hline
    A_N &           -2493.9 &           -1808.9 &           -1294.8 &           
-1517.2   &
   A_t &            -581.6 &           -564.2 &           -700.1 &           
-700.7  \\
    m_R &            2326.9 &            9630.6 &            2054.6 &           
 2437.1   &
   \mu & 700. &915. &764. &           
   763.  \\
 m_{\tilde\nu_1} &             2284. &           
   556. &             2023. &             2450.   &
 m_{\tilde\nu_{2,3}}/m_{\nu_{2,3}} &             2340. &          
   9631. &             2063. &             2434.  \\
  \hline \hline
    m_h &            126.02 &            119.55 &            123.18 &           
 119.50   &
 m_{H} & 911. &             1066. &992. &           
   991.  \\
  m_h^{0+\Delta} &             113.72 &          
  119.5 &             115.46 &             114.98   &
        m_{A^0}  &914. &           
  1104. &993. &992.  \\
        m_h^{0} & 116.23 &         
  116.22 &             116.33 &            116.33   &
      m_{H^\pm} & 916. &           
  1107. &996. &994.  \\
  \hline 
 M_{S}  &             1847. &             1847. &             1800. &           
  1802.   &
   m_{\tilde g} &2188. &           
  2188. &             2111. &             2111.  \\
      \tilde t_1 &             1797. &           
  1799. &             1738. &             1741.   &
     \tilde t_2 &1898. &           
  1896. &             1864. &             1866.  \\
      \tilde b_1 &             1914. &           
  1916. &             1889. &             1888.   &
     \tilde b_2 &1918. &           
  1921. &             1895. &             1893.  \\
   \tilde \tau_1 &2307. &           
   300. &2051. &2439.   &
  \tilde \tau_2 & 3163. &           
   937. &2763. &3327.  \\
      \tilde u_1 &             1936. &           
  1929. &             1914. &             1917.   &
     \tilde u_2 &1967. &           
  1971. &             1939. &             1938.  \\
      \tilde d_1 &             1918. &           
  1921. &             1895. &             1893.   &
     \tilde d_2 &1952. &           
  1957. &             1920. &             1919.  \\
      \tilde e_1 &644. &           
   536. &659. &666.   &
     \tilde e_2 & 644. &           
   630. &659. &666.  \\
   N_1  &430. &430. &412. &           
   412.   &
   N_2 & 681. &797. &720. &           
   719.  \\
    N_3 &705. &921. &769. &           
   769.   &
     N_4&856. &952. &845. &           
   845.  \\
   C_1  &673. &785. &710. &           
   710.   &
   C_2 & 839. &941. &831. &           
   831.  \\
 \hline\end{array}
 $$
 \caption{\sf{Benchmark   points for Model 13. See caption of \tabl{t:mod1} for 
details of notation.}
 \label{t:mod13}}
 \end{table}

\begin{figure}
\centering
\subfigure[]{
\includegraphics[height=5cm, width=7cm]{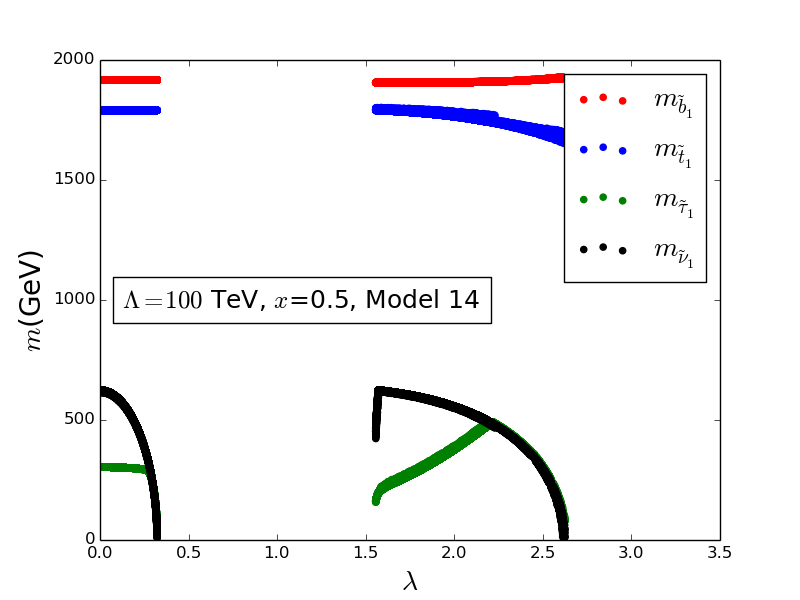}}
\subfigure[]{
\includegraphics[height=5cm, width=7cm]{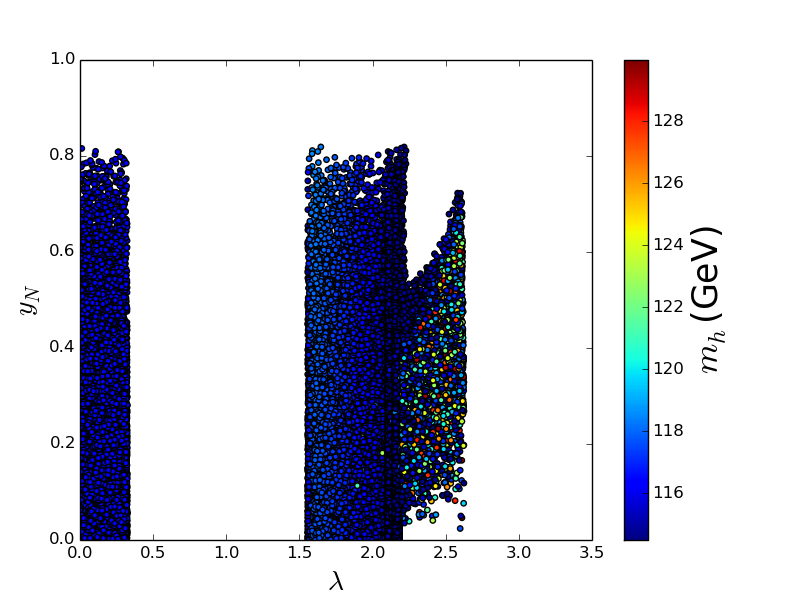}}
\subfigure[]{
\includegraphics[height=5cm, width=7cm]{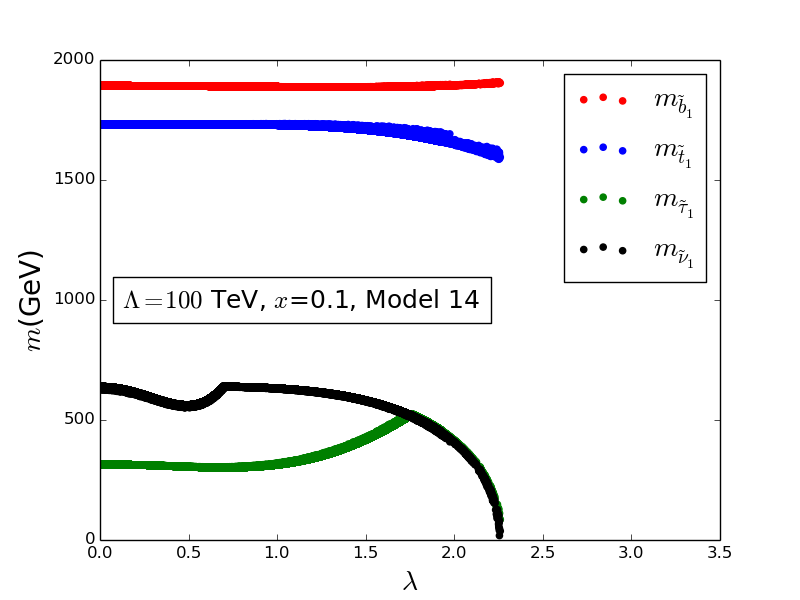}}
\subfigure[]{
\includegraphics[height=5cm, width=7cm]{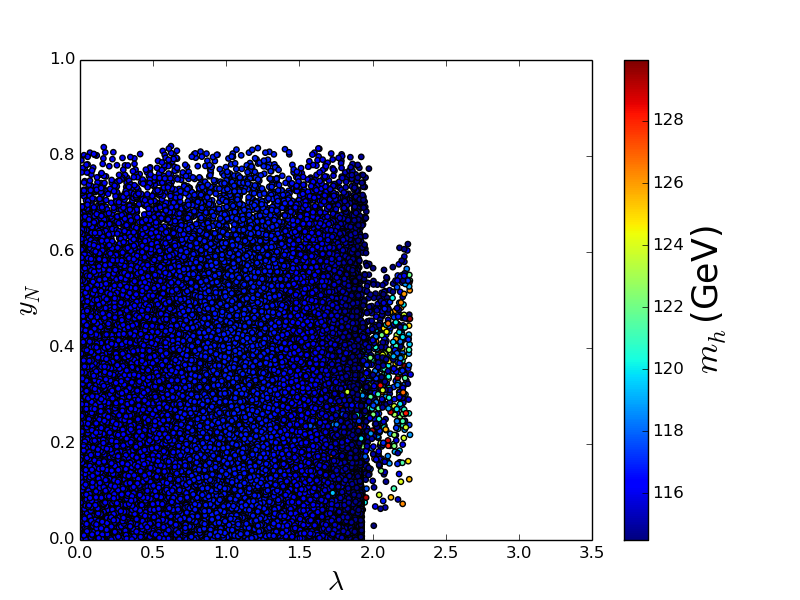}}
\caption{\sf Model 14: Spectrum variation with $\lambda$ and Higgs 
mass values in $y_N$ and $\lambda$ plane. Please see caption of 
\fig{f:mod1} 
for  details of notation.}\label{f:mod14}
\end{figure}

 \begin{table}
 $$
 \begin{array}{|c|c|c|c|c||c|c|c|c|c|}
 \hline{\mbox{Parameter}}&{\rm{ x=0.5}}
  &{\rm  x=0.5} & {\rm x=0.1} & {\rm x=0.1} &
 {\mbox{Parameter}}&{\rm{ x=0.5}}&{\rm  x=0.5}
  & {\rm x=0.1} & {\rm x=0.1}\\  \hline
\lambda &2.16 &1.82 &2.16 &           
   1.59   &
   y_N & 0.30 &0.18 &0.37 &           
   0.17  \\
  \hline
    A_N &           -2926.0 &           -2108.4 &           -2875.0 &           
-1624.1   &
   A_t &         -573.2  &     -581.1       &          -680.4  &           
-697.9  \\
    m_R &            3907.0 &            2220.9 &            5876.2 &           
 2981.9   &
   \mu & 859. &693. &             1077. &           
   777.  \\
 m_{\tilde\nu_1} &             4071. &           
  2176. &             5549. &             2935.   &
 m_{\tilde\nu_{2,3}}/m_{\nu_{2,3}} &             3905. &          
   2229. &             5877. &             2984.  \\
  \hline \hline
    m_h &            125.39 &            120.65 &            124.87 &           
 120.93   &
 m_{H} &4198. &             2308. &             5734. &           
  3058.  \\
  m_h^{0+\Delta} &             114.01 &          
  116.73 &110.76 &            115.918   &
        m_{A^0}  &             4202. &           
  2308. &             5740. &             3058.  \\
        m_h^{0} &116.23 &          
  116.23 &             116.33 &             116.33   &
      m_{H^\pm} &4202. &           
  2310. &             5740. &             3059.  \\
  \hline 
 M_{S}  &             1803. &             1831. &             1692. &           
  1767.   &
   m_{\tilde g} &2188. &           
  2188. &             2111. &             2111.  \\
      \tilde t_1 &             1750. &           
  1779. &             1623. &             1702.   &
     \tilde t_2 &1858. &           
  1884. &             1763. &             1836.  \\
      \tilde b_1 &             1908. &           
  1905. &             1900. &             1887.   &
     \tilde b_2 &1934. &           
  1922. &             1938. &             1907.  \\
   \tilde \tau_1 &461. &           
   295. &283. &451.   &
  \tilde \tau_2 & 4064. &           
   2192. &5553. &2941.  \\
      \tilde u_1 &             1884. &           
  1917. &             1789. &             1876.   &
     \tilde u_2 &1971. &           
  1966. &             1957. &             1941.  \\
      \tilde d_1 &             1945. &           
  1928. &             1958. &             1915.   &
     \tilde d_2 &1983. &           
  1966. &             1993. &             1945.  \\
      \tilde e_1 &513. &           
   406. &749. &528.   &
     \tilde e_2 & 595. &           
   600. &876. &569.  \\
   N_1  &430. &429. &412. &           
   411.   &
   N_2 & 786. &676. &787. &           
   729.  \\
    N_3 &867. &699. &             1088. &           
   785.   &
     N_4&915. &856. &             1100. &           
   854.  \\
   C_1  &772. &668. &768. &           
   718.   &
   C_2 & 901. &839. &             1089. &           
   839.  \\
 \hline\end{array}
 $$
 \caption{\sf{Benchmark   points for Model 14. See caption of \tabl{t:mod1} for 
details of notation.}}
 \label{t:mod14}
 \end{table}

\begin{figure}
\centering
\subfigure[]{
\includegraphics[height=5cm, width=7cm]{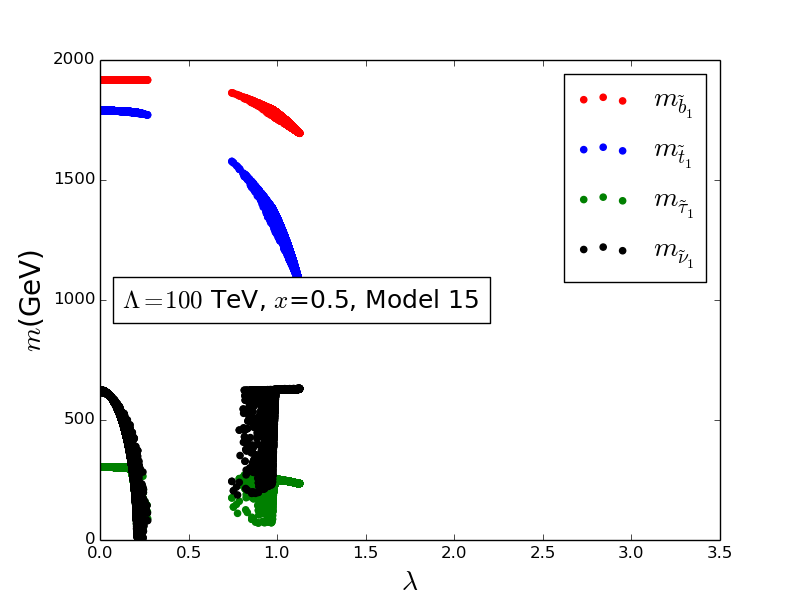}}
\subfigure[]{
\includegraphics[height=5cm, width=7cm]{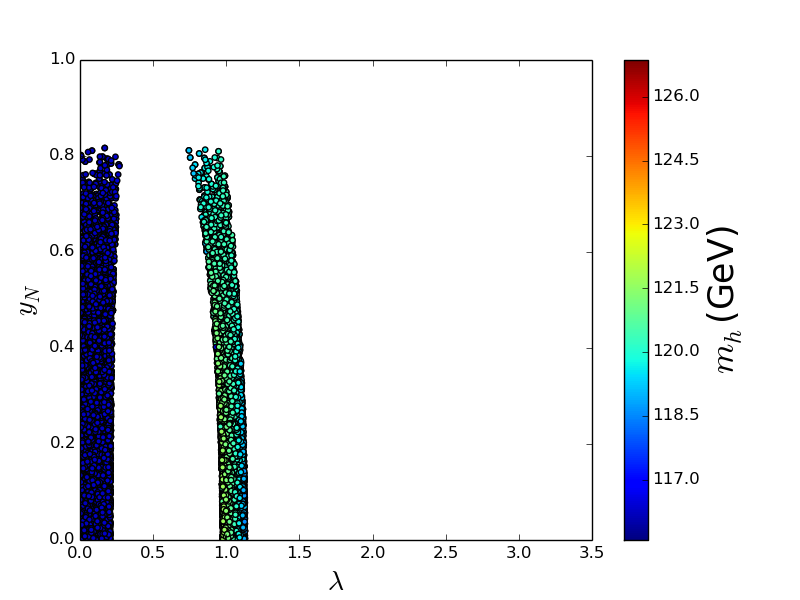}}
\subfigure[]{
\includegraphics[height=5cm, width=7cm]{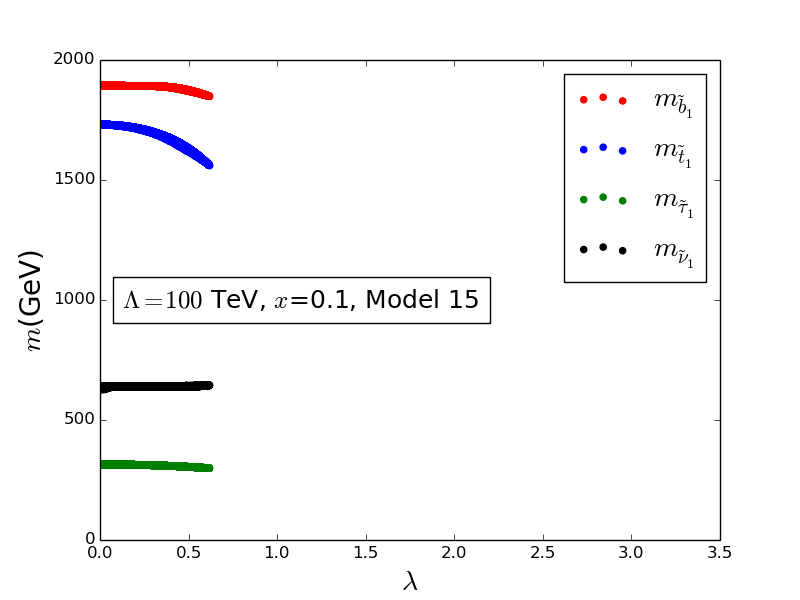}}
\subfigure[]{
\includegraphics[height=5cm, width=7cm]{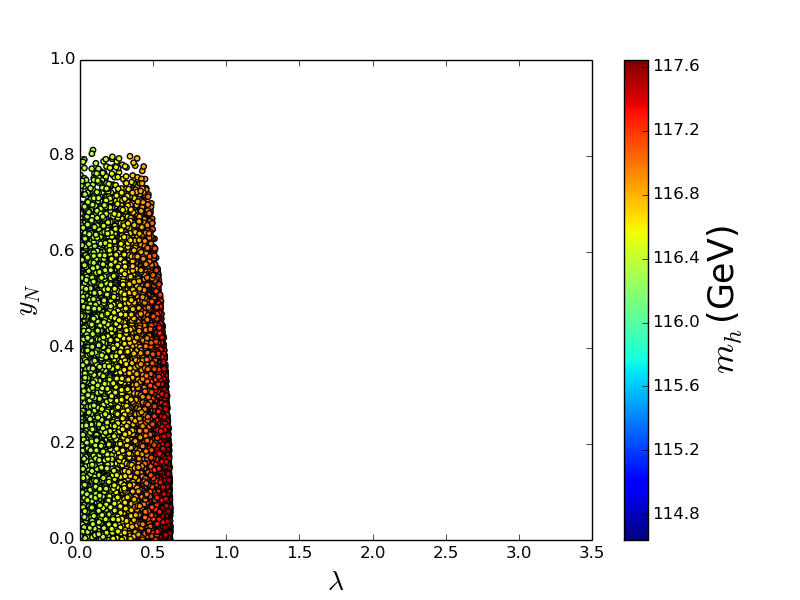}}
\caption{\sf Model 15: Spectrum variation with $\lambda$ and Higgs 
mass values in $y_N$ and $\lambda$ plane. Please see caption of 
\fig{f:mod1} 
for  details of notation.}\label{f:mod15}
\end{figure}


\subsection{Model 14}
This model has the following messenger-matter interaction superpotential, 
$ W_{\rm mix} = \lambda L H_d E^c_m$. The boundary conditions generated
 for this superpotential are shown in 
\Eqn{e:mod14}. All the scalar fields of MSSM  except  $\tilde U^c$ get these 
corrections. Both
$\delta   M^2_{\tilde L}$ and $\delta   M^2_{H_d}$ get one-loop negative and 
the two-loop positive contributions. Rest of the fields get only two-loop 
negative corrections.
\beqa\label{e:mod14}
\delta   M^2_{\tilde Q} &=& \Big[ -\frac{\alpha _b \alpha _{\lambda }}{16 \pi 
^2} 
\Big]\Lambda^2 
\nonumber \\
\delta   M^2_{\tilde D^c} &=& \Big[ -\frac{\alpha _b \alpha _{\lambda }}{8 \pi 
^2} 
\Big]\Lambda^2 
\nonumber  \\
\delta   M^2_{\tilde L} &= &\Big[\OOLo+ \frac{\alpha _{\lambda } \left(-9 
\alpha 
_1+5 
\left(-3
\alpha _2+3 \alpha _b+4 \left(\alpha _{\lambda }+\alpha _{\tau
}\right)\right)\right)}{80 \pi ^2} \Big]\Lambda^2 \nonumber  \\
\delta   M^2_{\tilde E^c} &=& \Big[ -\frac{\alpha _{\lambda } \alpha _{\tau 
}}{4 
\pi 
^2} 
\Big]\Lambda^2 \nonumber  \\
\delta   M^2_{\tilde N^c} &=& \Big[ -\frac{\alpha _N \alpha _{\lambda }}{8 \pi 
^2} 
\Big]\Lambda^2 
\nonumber \\
\delta   M^2_{H_u} &=& \Big[ -\frac{\alpha _N \alpha _{\lambda }}{16 \pi ^2} 
\Big]\Lambda^2 
\nonumber \\
\delta   M^2_{H_d} &=& \Big[\OOLo+ \frac{\alpha _{\lambda } \left(-9 \alpha 
_1+5 
\left(-3
\alpha _2+\alpha _N+4 \left(\alpha _{\lambda }+\alpha _{\tau
}\right)\right)\right)}{80 \pi ^2} \Big]\Lambda^2 \nonumber \\
A_b &=& \Big[ -\frac{\alpha _{\lambda }}{4 \pi } \Big]\Lambda \nonumber \\
 A_{\tau }&= & \Big[ -\frac{\alpha _{\lambda }}{2 \pi } \Big]\Lambda 
\nonumber 
\\
 A_N &=& \Big[ -\frac{\alpha _{\lambda }}{4 \pi } \Big]\Lambda.
\eeqa
We get two band of allowed values of $\lambda$ for $x=0.5$. Note 
that the one loop and two loop corrections are proportional to $\alpha_\lambda$ 
and $\alpha_\lambda^2$ 
respectively. For the smaller values of $\lambda$, 
one-loop effect dominates over the two loop contribution and as a consequence 
lightest stau becomes tachyonic beyond $\lambda\sim0.32$ as shown in 
\fig{f:mod14}(a). For larger values of $\lambda$, situation gets reversed. 
However, $\lambda$ cannot be arbitrary large because otherwise $M^2_{H_d}$ will 
become very heavy and radiative electroweak symmetry breaking would not be 
possible. For the other case, where $x=0.1$, one-loop effect is not so strong 
and the upper limit of $\lambda$ is around $\sim2.3$ (\fig{f:mod14}(c)).

In \fig{f:mod14}(b) and \fig{f:mod14}(d), Higgs mass values are shown in 
the $\lambda$ vs $y_N$ plane. For both the values of $x$, a 125 GeV Higgs 
is achievable. In \tabl{t:mod14} four benchmark points are shown. In this 
model, no $A_t$ 
term is generated at the boundary. Therefore heavy sleptons $(\tilde\tau_2)$ 
and $A_N$ are
responsible for giving correct Higgs mass.

\subsection{Model 15}
This is a Type II model based on $5\oplus\bar5$ and $S_m$ messenger fields like 
the model 8 and 9. The soft mass corrections based on the interaction $\lambda 
L H_u S_m$ are given by
\beqa
 \delta  M^2_{\tilde Q } &=& \Big[ -\frac{3 \alpha _t \alpha _{\lambda }}{16 
\pi 
^2} \Big]\Lambda^2 \nonumber \\
 \delta  M^2_{\tilde {U^c} } &=& \Big[ -\frac{3 \alpha _t \alpha _{\lambda }}{8 
\pi ^2} \Big]\Lambda^2 \nonumber\\*
 \delta  M^2_{\tilde L } &=& \Big[\OOL + \frac{3 \alpha _{\lambda } \left(5 
\left(-3 \alpha _2+4 \alpha _N+3 \alpha _t+8 \alpha _{\lambda }\right)-3 \alpha 
_1\right)}{80 \pi ^2} \Big]\Lambda^2 \nonumber \\
 \delta  M^2_{\tilde {e^c} } &=& \Big[ -\frac{3 \alpha _{\lambda } \alpha 
_{\tau 
}}{8 \pi ^2} \Big]\Lambda^2 \nonumber\\*
 \delta  M^2_{ {H_u} } &=& \Big[\OOL + \frac{3 \alpha _{\lambda } \left(5 
\left(-3 \alpha _2+4 \alpha _N+8 \alpha _{\lambda }+\alpha _{\tau }\right)-3 
\alpha _1\right)}{80 \pi ^2} \Big]\Lambda^2 \nonumber\\*
 \delta  M^2_{{H_d} } &=& \Big[ -\frac{3 \alpha _{\lambda } \alpha _{\tau 
}}{16 \pi ^2} \Big]\Lambda^2 \nonumber\\*
 \delta  M^2_{\tilde {N^c} } &=& \Big[ -\frac{3 \alpha _N \alpha _{\lambda }}{4 
\pi ^2} \Big]\Lambda^2 \nonumber\\*
 \delta  A_t  &=& \Big[ -\frac{3 \alpha _{\lambda }}{4 \pi } \Big]\Lambda 
\nonumber\\*
 \delta  A_{\tau }  &=& \Big[ -\frac{3 \alpha _{\lambda }}{4 \pi } \Big]\Lambda 
\nonumber\\*
 \delta  A_N  &=& \Big[ -\frac{3 \alpha _{\lambda }}{2 \pi } \Big]\Lambda.
 \eeqa
Here all the MSSM fields, except $D^c$, get these messenger-matter 
corrections. Both the $\delta  M^2_{\tilde L }$ and $\delta  M^2_{H_u }$ get 
one-loop and two-loop corrections. Like all the models where $\delta  
M^2_{\tilde L }$ gets one-loop negative contribution, this model is also a 
constrained model. One again, $\lambda$ has two allowed regions for $x=0.5$. 
Maximum value of $\lambda$ in the first band is $\sim 0.2$ as shown in 
\fig{f:mod15}(a) and that value for the upper band is $\sim 1.2$ which is set 
by radiative electroweak breaking condition. For $x=0.1$, there is only one 
band. The maximum value of $\lambda$ is $\sim 0.6$ and this 
upper limit of $\lambda$ is set by EWSB condition because  $M_{H_u}^2$ gets 
positive correction whereas squarks get negative contributions at the boundary.

In the right panel on \fig{f:mod15}, Higgs mass values are shown.
One cannot get correct Higgs mass for the case of $x=0.1$ because $\lambda$ is 
too small. In the other case, one can achieve it for the larger 
values of $\lambda$ (\fig{f:mod15}(b)).

\rplb{Four benchmark points of the model are shown in \tabl{t:mod15}. In first 
benchmark point, we have 125 GeV Higgs mass. We see that a 10 GeV rise in the 
Higgs mass from the MSSM value.}

\begin{table}
 $$
 \begin{array}{|c|c|c|c|c||c|c|c|c|c|}
 \hline{\mbox{Parameter}}&{\rm{ x=0.5}}
  &{\rm  x=0.5} & {\rm x=0.1} & {\rm x=0.1} &
 {\mbox{Parameter}}&{\rm{ x=0.5}}&{\rm  x=0.5}
  & {\rm x=0.1} & {\rm x=0.1}\\  \hline
\lambda &1.10 &0.82 &0.60 &           
   0.47   &
   y_N & 0.09 &0.73 &0.11 &           
   0.13  \\
  \hline
    A_N &           -4367.3 &           -2275.9 &           -1290.1 &           
 -802.4   &
   A_t &            -2429.9 &           -1561.6 &           -1206.5 &           
 -1012.3  \\
    m_R &            1748.7 &            9648.0 &            2062.1 &           
 1640.4   &
   \mu & 806. &             1383. &326. &           
   714.  \\
 m_{\tilde\nu_1} &             1707. &           
   468. &             1284. &945.   &
 m_{\tilde\nu_{2,3}}/m_{\nu_{2,3}} &             1767. &          
   9649. &             2062. &             1641.  \\
  \hline \hline
    m_h &            126.01 &            119.80 &            117.47 &           
 117.05   &
 m_{H} &1016. &             1440. &706. &           
   962.  \\
  m_h^{0+\Delta} &119.16 &         
    119.60 &          117.47 &            117.05   &
        m_{A^0}  &             1018. &           
  1529. &706. &962.  \\
        m_h^{0} & 116.23 &         
   116.23 &            116.33 &            116.33   &
      m_{H^\pm} &1020. &           
  1531. &710. &965.  \\
  \hline 
 M_{S}  &             1416. &             1683. &             1706. &           
  1749.   &
   m_{\tilde g} &2178. &           
  2182. &             2107. &             2109.  \\
      \tilde t_1 &             1144. &           
  1521. &             1576. &             1647.   &
     \tilde t_2 &1752. &           
  1863. &             1848. &             1857.  \\
      \tilde b_1 &             1716. &           
  1842. &             1854. &             1880.   &
     \tilde b_2 &1928. &           
  1925. &             1899. &             1900.  \\
   \tilde \tau_1 &239. &           
   252. &300. &306.   &
  \tilde \tau_2 & 1729. &           
   628. &1285. &947.  \\
      \tilde u_1 &             1942. &           
  1927. &             1903. &             1910.   &
     \tilde u_2 &1955. &           
  1939. &             1913. &             1916.  \\
      \tilde d_1 &             1948. &           
  1931. &             1904. &             1902.   &
     \tilde d_2 &1948. &           
  1932. &             1904. &             1902.  \\
      \tilde e_1 &307. &           
   306. &318. &318.   &
     \tilde e_2 & 631. &           
   485. &646. &644.  \\
   N_1  &429. &431. &308. &           
   410.   &
   N_2 & 761. &822. &331. &           
   685.  \\
    N_3 &812. &             1391. &426. &           
   719.   &
     N_4&887. &             1396. &807. &           
   829.  \\
   C_1  &748. &805. &319. &           
   677.   &
   C_2 & 869. &             1387. &790. &           
   814.  \\
 \hline\end{array}
 $$
 \caption{\sf{Benchmark points for Model 15. See caption of \tabl{t:mod1} for 
details of notation.}}
 \label{t:mod15}\end{table}

\begin{figure}
\centering
\subfigure[]{
\includegraphics[height=5cm, width=7cm]{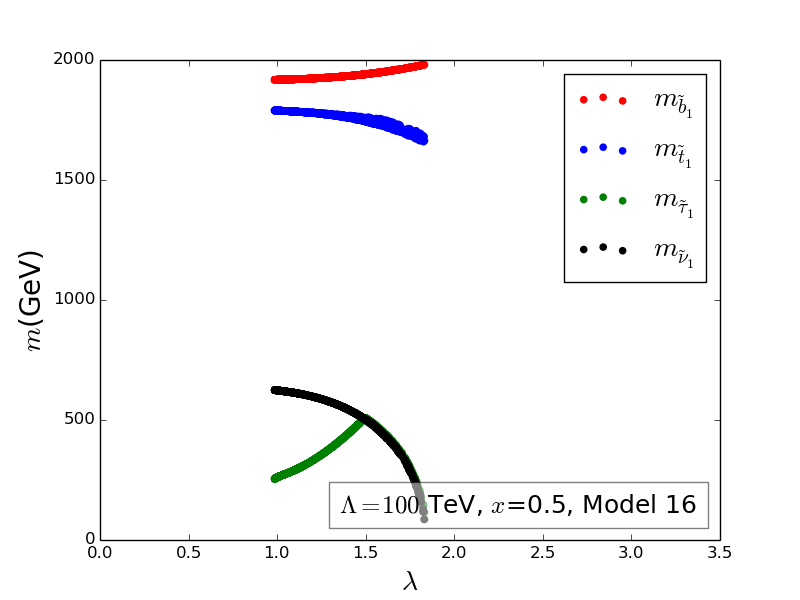}}
\subfigure[]{
\includegraphics[height=5cm, width=7cm]{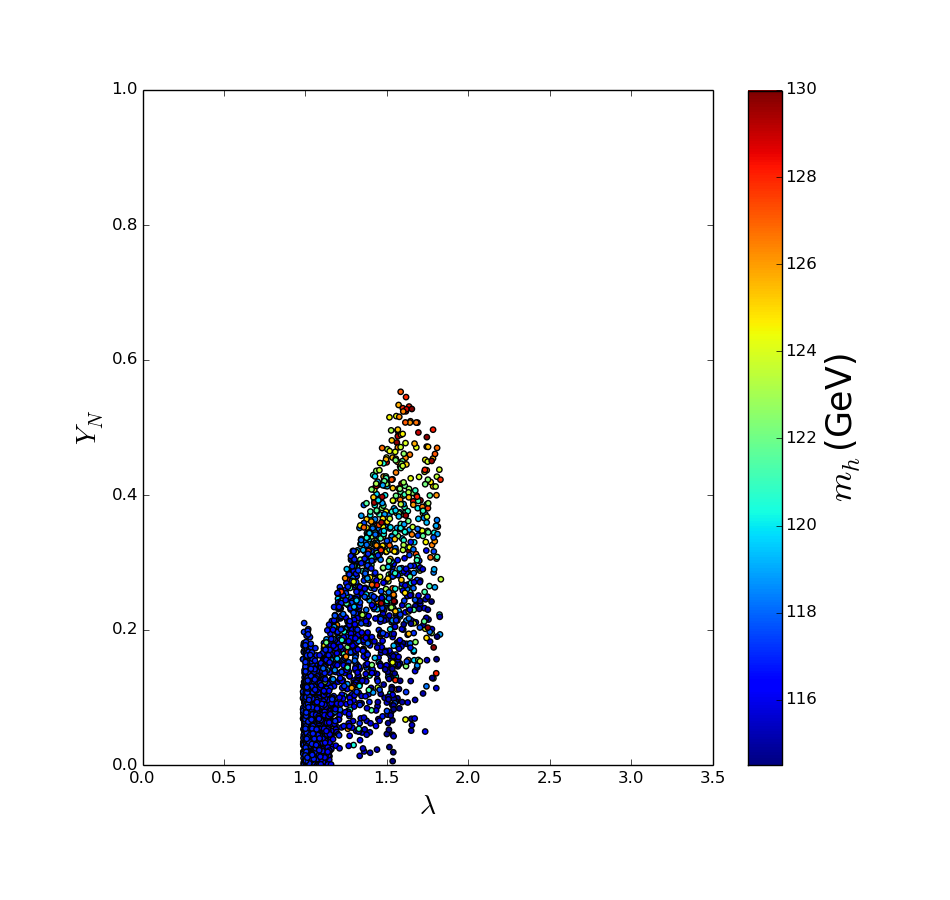}}
\subfigure[]{
\includegraphics[height=5cm, width=7cm]{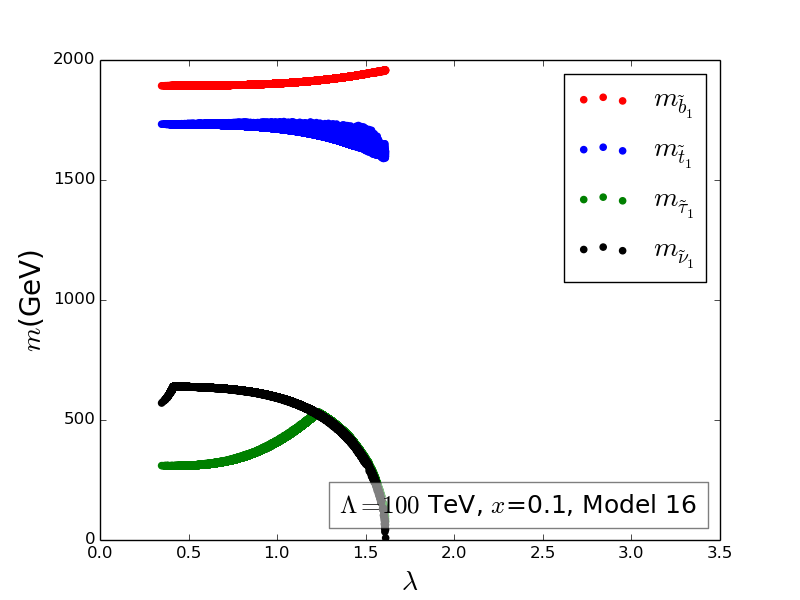}}
\subfigure[]{
\includegraphics[height=5cm, width=7cm]{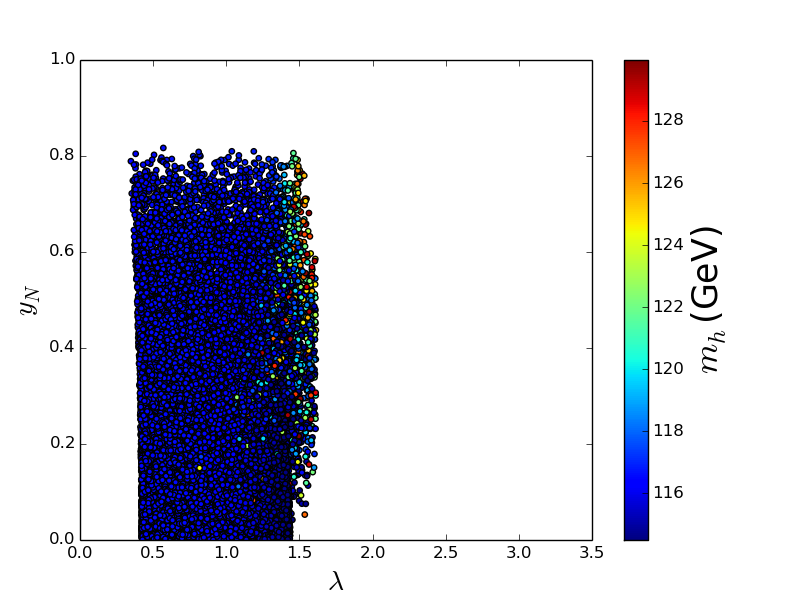}}
\caption{\sf Model 16: Spectrum variation with $\lambda$ and Higgs 
mass values in $y_N$ and $\lambda$ plane. Please see caption of 
\fig{f:mod1} 
for  details of notation.}\label{f:mod16}
\end{figure}

\subsection{Model 16}
This is a Type II model with  $ W_{\rm mix}= \lambda L S H_u^m$. Corrections 
due to this superpotential to the soft mass are shown below. None of the 
squarks gets a correction.  Both $\delta  M^2_{\tilde L }$ and $\delta  
M^2_{\tilde S }$ get negative one-loop and positive two-loop corrections.
\beqa
 \delta  M^2_{\tilde L } &=& \Big[\OOL +  \frac{3 \alpha _{\lambda } \left(-3 
\alpha _1-15 \alpha _2+50 \alpha _{\lambda }\right)}{80 \pi ^2} \Big]\Lambda^2 
\nonumber\\*
 \delta  M^2_{\tilde {e^c} } &=& \Big[ -\frac{3 \alpha _{\lambda } \alpha 
_{\tau 
}}{8 \pi ^2} \Big]\Lambda^2 \nonumber \\
 \delta  M^2_{{H_u} } &=& \Big[ -\frac{3 \alpha _N \alpha _{\lambda }}{16 
\pi ^2} \Big]\Lambda^2 \nonumber\\*
 \delta  M^2_{ {H_d} } &=& \Big[ -\frac{3 \alpha _{\lambda } \alpha _{\tau 
}}{16 \pi ^2} \Big]\Lambda^2 \nonumber\\*
 \delta  M^2_{\tilde {N^c} } &=& \Big[ -\frac{3 \alpha _N \alpha _{\lambda }}{8 
\pi ^2} \Big]\Lambda^2 \nonumber\\*
 \delta  M^2_{\tilde S } &=& \Big[\TwOL + \frac{3 \alpha _{\lambda } \left(5 
\left(-3 \alpha _2+\alpha _N+10 \alpha _{\lambda }+\alpha _{\tau }\right)-3 
\alpha _1\right)}{40 \pi ^2} \Big]\Lambda^2 \nonumber\\*
 \delta  A_{\tau }  &=& \Big[ -\frac{3 \alpha _{\lambda }}{4 \pi } \Big]\Lambda 
\nonumber\\*
 \delta  A_N  &=& \Big[ -\frac{3 \alpha _{\lambda }}{4 \pi } \Big]\Lambda.
\eeqa
Like the model 9, lower values of $\lambda$ of this model are also ruled out by 
requirement of non-zero soft mass of $\tilde S$:
\begin{equation}
\lambda\geq \sqrt{\left(\frac{10}{3} x^2 h(x) \pi + 3\alpha_1+15\alpha_2 
-5\alpha_N 
-5\alpha_\tau\right)\frac{2\pi}{25}},
\end{equation}
where  minimum value of  $\lambda$ has to be 
$\sim 0.34$ and 0.98 for $x=0.1$ and $x=0.5$ respectively. The sparticle 
eigenvalues are shown 
in \fig{f:mod16}(a) and \fig{f:mod16}(c).  

Like other models, in \tabl{t:mod16},  benchmark points, both for 
$x=0.5$ and $x=0.1$ case, are shown. We can see from this table and 
\fig{f:mod16}(b) and \fig{f:mod16}(d) that required Higgs mass can be 
achieved in this model. There is a 8 GeV increase in the Higgs mass from 
its pure MSSM values for all the benchmark points. We can see that all 
the points have heavy sleptons ($\tilde\tau_2$).

 \begin{table}
 $$
 \begin{array}{|c|c|c|c|c||c|c|c|c|c|}
 \hline{\mbox{Parameter}}&{\rm{ x=0.5}}
  &{\rm  x=0.5} & {\rm x=0.1} & {\rm x=0.1} &
 {\mbox{Parameter}}&{\rm{ x=0.5}}&{\rm  x=0.5}
  & {\rm x=0.1} & {\rm x=0.1}\\  \hline
   \lambda &              1.47 &           
   1.25 &              1.48 &              0.96   &
y_N &               0.34 &              0.24 &              0.25 &              
0.23  \\
  \hline
 A_N &           -4031.4 &           -2944.9 &           -4153.3 &           
-1767.2   &
A_t &            -567.3 &           -577.7 &           -690.6 &           
-695.7  \\
 m_R &            5683.3 &            3311.6 &            7557.4 &            
3021.5   &
\mu &               907. &              727. &              838. &              
776.  \\
                                 m_{\tilde\nu_1} &             5544. &          
   3368. &             7325. &             2955.   &
               m_{\tilde\nu_{2,3}}/m_{\nu_{2,3}} &             5687. &          
   3305. &             7558. &             3024.  \\
  \hline \hline
 m_h &            125.86 &            124.33 &            125.30 &            
124.08   &
    m_{H} &              1012. &           
   924. &              880. &              978.  \\
                                  m_h^{0+\Delta} &             114.07 &         
   115.72 &            113.15 &            115.98   &
  m_{A^0}  &             1042. &           
   931. &              902. &              981.  \\
  m_h^{0} &              116.23 &          
  116.23 &             116.33 &            116.33   &
m_{H^\pm} &              1041. &           
   932. &              899. &              984.  \\
  \hline 
    M_{S}  &             1809. &           
  1829. &             1706. &             1782.   &
                                   m_{\tilde g} &              2188. &          
   2188. &             2111. &             2111.  \\
\tilde t_1 &             1757. &           
  1778. &             1636. &             1718.   &
                                     \tilde t_2 &              1862. &          
   1883. &             1778. &             1848.  \\
\tilde b_1 &             1938. &           
  1925. &             1939. &             1900.   &
                                     \tilde b_2 &              1943. &          
   1929. &             1950. &             1907.  \\
                                   \tilde \tau_1 &              488. &          
    354. &              358. &              392.   &
                                  \tilde \tau_2 &               5547. &         
    3362. &            7327. &             2964.  \\
\tilde u_1 &             1887. &           
  1914. &             1806. &             1892.   &
                                     \tilde u_2 &              1979. &          
   1971. &             1963. &             1943.  \\
\tilde d_1 &             1943. &           
  1929. &             1950. &             1907.   &
                                     \tilde d_2 &              1969. &          
   1958. &             1957. &             1928.  \\
\tilde e_1 &              522. &           
   425. &              723. &              437.   &
                                     \tilde e_2 &               580. &          
    593. &              820. &              607.  \\
N_1  &              430. &              429. &              411. &              
411.   &
N_2 &               798. &              704. &              756. &              
727.  \\
 N_3 &              913. &              732. &              843. &              
782.   &
  N_4&              947. &              861. &              887. &              
851.  \\
C_1  &              782. &              695. &              740. &              
716.   &
C_2 &               935. &              843. &              874. &              
837.  \\
 \hline\end{array}
 $$
 \caption{\sf{Benchmark   points for Model 16. See caption of \tabl{t:mod1} for 
details of notation.}}
 \label{t:mod16}\end{table}

\begin{figure}
\centering
\subfigure[]{
\includegraphics[height=5cm, width=7cm]{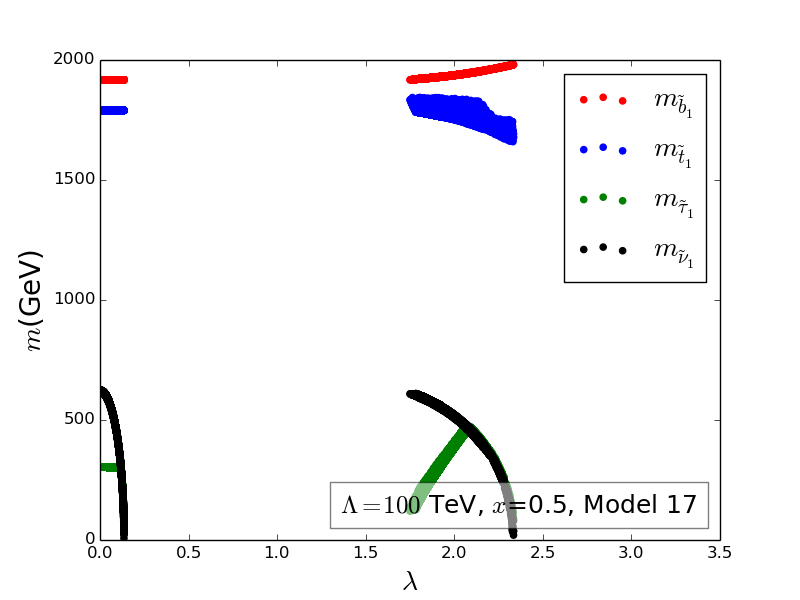}}
\subfigure[]{
\includegraphics[height=5cm, width=7cm]{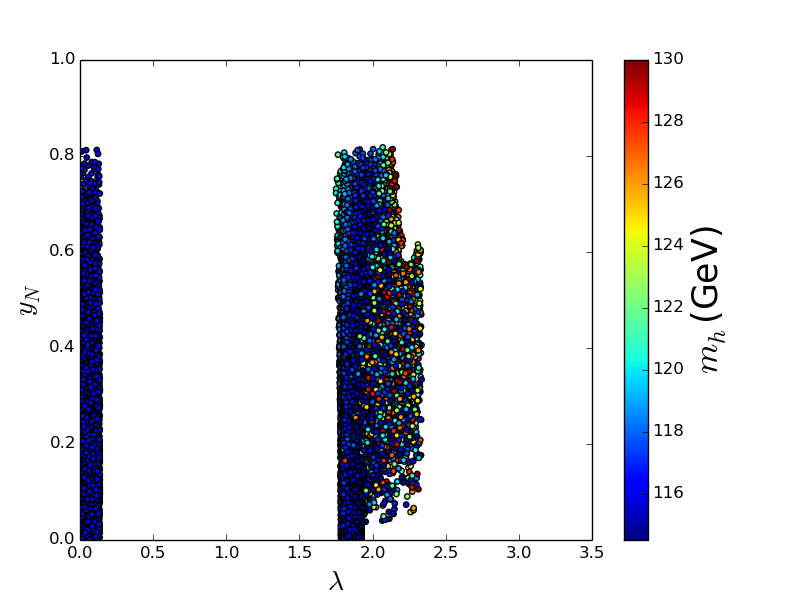}}
\subfigure[]{
\includegraphics[height=5cm, width=7cm]{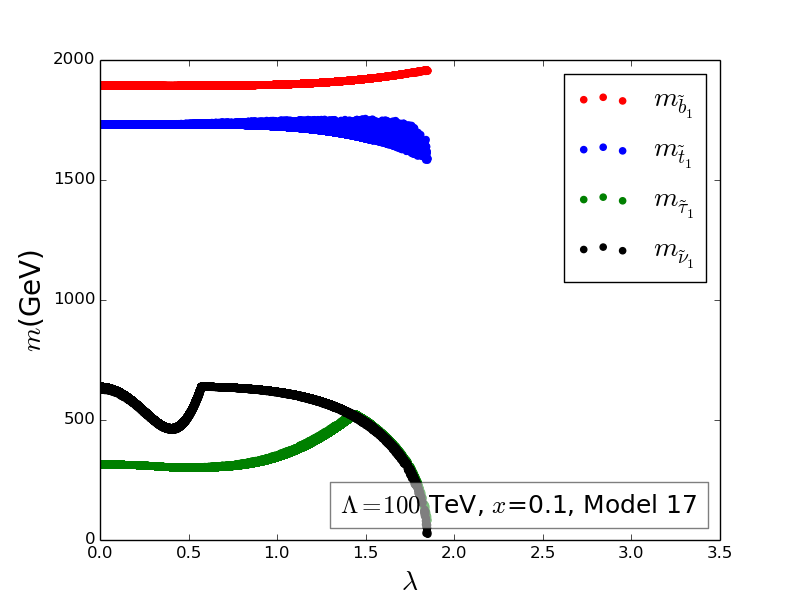}}
\subfigure[]{
\includegraphics[height=5cm, width=7cm]{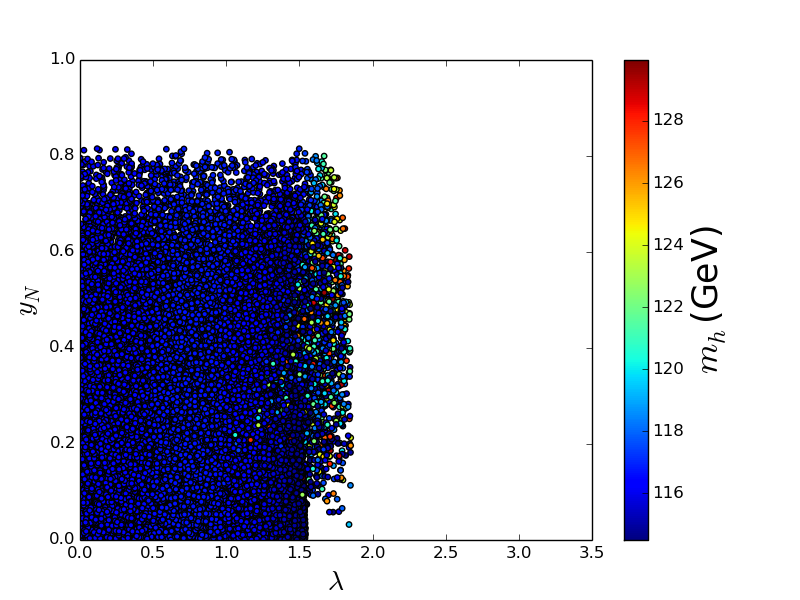}}
\caption{\sf Model 17: Spectrum variation with $\lambda$ and Higgs 
mass data points in $y_N$ and $\lambda$ plane. Please see caption of 
\fig{f:mod1} 
for  details of notation.}\label{f:mod17}
\end{figure}

\subsection{Model 17}
This model belongs to Type I category having interaction  superpotential, 
$\lambda L H^m_u S^m$. The one-loop and  two-loop corrections are given below 
:
\beqa
 \delta  M^2_{\tilde L } &=& \Big[-\frac{\alpha_\lambda}{4\pi}x^2 h(x) + 
\frac{9 
\alpha _{\lambda } \left(-\alpha _1-5 \alpha _2+10 \alpha _{\lambda 
}\right)}{80 
\pi ^2} \Big]\Lambda^2 \nonumber\\*
 \delta  M^2_{\tilde {e^c} } &=& \Big[ -\frac{3 \alpha _{\lambda } \alpha 
_{\tau 
}}{8 \pi ^2} \Big]\Lambda^2 \nonumber\\*
 \delta  M^2_{{H_u} } &=& \Big[ -\frac{3 \alpha _N \alpha _{\lambda }}{16 
\pi ^2} \Big]\Lambda^2 \nonumber\\*
 \delta  M^2_{{H_d} } &=& \Big[ -\frac{3 \alpha _{\lambda } \alpha _{\tau 
}}{16 \pi ^2} \Big]\Lambda^2 \nonumber\\*
 \delta  M^2_{\tilde {N^c} } &=& \Big[ -\frac{3 \alpha _N \alpha _{\lambda }}{8 
\pi ^2} \Big]\Lambda^2 \nonumber\\*
 \delta  A_{\tau } &=& \Big[ -\frac{3 \alpha _{\lambda }}{4 \pi } \Big]\Lambda 
\nonumber\\*
 \delta  A_N  &=& \Big[ -\frac{3 \alpha _{\lambda }}{4 \pi } \Big]\Lambda.
\eeqa
\rplb{Because of one-loop negative contribution to $\delta  M^2_{\tilde L }$, 
lightest stau and sneutrino become tachyonic beyond $\lambda\sim 0.12$ for the 
case of $x=0.5$ as shown in \fig{f:mod17}(a). However for larger values of 
$\lambda$, two loop corrections overtake the one loop corrections and that is 
why we get the upper band of allowed values of $\lambda$. Correct Higgs mass is 
achievable for $\lambda\sim 2.0$ (see \fig{f:mod17}(b)). For $x=0.1$, there is 
only one band of parameter space allowed for $\lambda$ and maximum allowed 
value of $\lambda$ is $\sim1.8$ as shown in \fig{f:mod17}(c). We have a
parameter space near $\lambda\gtrsim 1.5$ and $y_N \gtrsim 0.2$ where we can 
achieve 125 GeV Higgs mass (\fig{f:mod17}(d)).}

Four benchmark points for this model are shown in \tabl{t:mod17}. In the second benchmark 
point, $M_{\tilde L}^2 - m_R^2$ is slightly large, so the contribution to the Higgs 
mass is less as compare to other benchmark points.

\begin{table}
 $$
 \begin{array}{|c|c|c|c|c||c|c|c|c|c|}
 \hline{\mbox{Parameter}}&{\rm{ x=0.5}}
  &{\rm  x=0.5} & {\rm x=0.1} & {\rm x=0.1} &
 {\mbox{Parameter}}&{\rm{ x=0.5}}&{\rm  x=0.5}
  & {\rm x=0.1} & {\rm x=0.1}\\  \hline
\lambda &2.02 &1.81 &1.43 &           
   1.02   &
   y_N & 0.23 &0.28 &0.39 &           
   0.18  \\
  \hline
    A_N &           -7680.1 &           -6180.1 &           -3781.2 &           
-1984.2   &
   A_t &            -569.7 &           -562.9 &           -667.4 &           
-697.2  \\
    m_R &            5380.8 &            3372.8 &            5507.2 &           
 2412.2   &
   \mu & 869. &918. &             1086. &           
   762.  \\
 m_{\tilde\nu_1} &             5663. &           
  2752. &             5054. &             2371.   &
 m_{\tilde\nu_{2,3}}/m_{\nu_{2,3}} &             5378. &          
   3377. &             5509. &             2419.  \\
  \hline \hline
    m_h &            124.50 &            119.16 &            124.46 &           
 124.61   &
 m_{H} & 957. &             1067. &             1164. &           
   976.  \\
  m_h^{0+\Delta} &111.62 &         
  116.60 &             114.22 &            116.17   &
        m_{A^0}  &983. &           
  1080. &             1200. &978.  \\
        m_h^{0} &116.23 &          
  116.23 &            116.33 &            116.33   &
      m_{H^\pm} & 974. &           
  1079. &             1200. &981.  \\
  \hline 
 M_{S}  &             1806. &             1837. &             1762. &           
  1786.   &
   m_{\tilde g} &2188. &           
  2188. &             2112. &             2111.  \\
      \tilde t_1 &             1754. &           
  1788. &             1700. &             1723.   &
     \tilde t_2 &1860. &           
  1888. &             1826. &             1852.  \\
      \tilde b_1 &             1939. &           
  1922. &             1916. &             1898.   &
     \tilde b_2 &1944. &           
  1927. &             1923. &             1904.  \\
   \tilde \tau_1 &399. &           
   180. &523. &349.   &
  \tilde \tau_2 & 5659. &           
   2758. &5058. &2388.  \\
      \tilde u_1 &             1885. &           
  1918. &             1858. &             1897.   &
     \tilde u_2 &1979. &           
  1972. &             1954. &             1942.  \\
      \tilde d_1 &             1944. &           
  1927. &             1923. &             1904.   &
     \tilde d_2 &1969. &           
  1960. &             1944. &             1926.  \\
      \tilde e_1 &517. &           
   393. &524. &397.   &
     \tilde e_2 & 589. &           
   603. &614. &621.  \\
   N_1  &430. &431. &412. &           
   411.   &
   N_2 & 788. &799. &783. &           
   719.  \\
    N_3 &875. &924. &             1093. &           
   768.   &
     N_4&919. &955. &             1105. &           
   845.  \\
   C_1  &773. &785. &765. &           
   709.   &
   C_2 & 906. &944. &             1095. &           
   830.  \\
 \hline\end{array}
 $$
 \caption{\sf{Benchmark   points for Model 17. See caption of \tabl{t:mod1} for 
details of notation.}}
 \label{t:mod17}
 \end{table}

\section{Discussion and conclusions}\label{s:con}

Gauge mediated supersymmetry breaking models are interesting due to their
flavour blindness and unique phenomenological features like gravitino LSP etc. 
The discovery of the Higgs particle and the non-discovery of the coloured 
supersymmetric partners at the LHC has however put strong constraints 
on Gauge Mediated Supersymmetry Breaking models with vanishing 
A-terms, \textit{i.e,} both minimal and general forms of them. 

Allowing for direct matter-messenger interactions in addition to the standard 
gauge interactions is being pursued  as one of the interesting solutions which
allows for a light spectrum $\sim 1 $ TeV and a light CP even Higgs boson
at 125 GeV. Several studies have been presented using these ideas, which
culminated in survey Refs. \cite{Byakti:2013ti,Evans:2013kxa}.  In the present
work, we extended the survey to a particular extension of MSSM including
right handed neutrinos.  We chose the inverse seesaw model because of the
possible large impact on the corrections to the Higgs mass. However with
the usual boundary conditions prescribed for minimal gauge mediation it 
is hard to generate a large enhancement to the Higgs mass through the
neutrino/sneutrino sector. In general gauge mediation framework, a large
separate parameter for the sleptons should be prescribed such that these
corrections become significant. Another way to increase the neutrino/sneutrino 
contributions would be to generate a large sneutrino mixing parameter through
matter-messenger mixing. In the present work, we explored this possibility  in 
great detail and surveyed all possible matter messenger mixing terms with 
leptonic doublet fields (L) and right handed neutrino fields ($N^c$). 
It turns out that in almost all the successful models, in addition to $X_N$, 
a significant correction to $M_{\tilde L}^2$ is also generated.

We classified all possible models by imposing lepton number on the messenger
sector fields also. We found 17 such models. Both of the Type I 
(matter-messenger-messenger) 
and Type II (matter-matter-messenger). We analysed each model in detail by 
doing 
a full numerical analysis. The results are summarized in the 
\tabl{t:conclu}.

\begin{table}
 $$
 \begin{array}{|c|c|c|c|c|c|}
\hline
\mbox{Model} & \mbox{Interaction} & \multicolumn{2}{ |c| }{\mbox{125 GeV 
Higgs}} 
& \multicolumn{2}{ |c| } {\mbox{Range of $\lambda$}}\\
\hline
\hline
\multicolumn{6}{|c|}{\mbox{Models with } N^c}\\
\hline
& &  \mbox{x=0.5} & \mbox{x=0.1} & \mbox{x=0.5} & \mbox{x=0.1} \\ \hline
1. \vphantom{\frac{\frac12}{1}}  & N^c Q \bar Q_m & \times & \times & 1.48 & 
1.24 \\
\hline
2. \vphantom{\frac{\frac12}{1}} & N^c U^c \bar U^c_m & \times & \times & 1.86 & 
1.60 \\
\hline
3. \vphantom{\frac{\frac12}{1}}  & N^c D^c \bar D^c_m & \times & \times & 1.15 & 
0.94 \\
\hline
4. \vphantom{\frac{\frac12}{1}} & N^c L H_u^m & \checkmark & \checkmark & 1.84 & 
1.61  \\
\hline
5. \vphantom{\frac{\frac12}{1}}  & N^c E^c \bar E^c_m & \times & \times & 2.10 & 
1.60 \\
\hline
6. \vphantom{\frac{\frac12}{1}} & N^c H_u H_d^m & \times & \times & 1.0 & 0.57 
\\
\hline
7. \vphantom{\frac{\frac12}{1}} & N^c H_d H_u^m & \checkmark & \checkmark & 1.68 & 
1.52  \\
\hline
8. \vphantom{\frac{\frac12}{1}}  & \frac12 (N^c)^2 S_m  & \times & 
\times & \sqrt{4 \pi} & 
\sqrt{4 \pi} \\
\hline
9. \vphantom{\frac{\frac12}{1}} & N^c S S_m  & \times & \times & 
1.05-\sqrt{4 \pi} & 
0.02-\sqrt{4 \pi} \\ \hline
10. \vphantom{\frac{\frac12}{1}} & N^c H_u^m H_d^m  & \times & \times & 
\sqrt{4 \pi} & \sqrt{4 \pi} \\
\hline
\multicolumn{6}{|c|}{\mbox{Models with L}}\\
\hline
11. \vphantom{\frac{\frac12}{1}} & L Q \bar D^c_m & \checkmark & \checkmark & 
1.21 & 1.09 \\
\hline
12. \vphantom{\frac{\frac12}{1}}  & L D^c \bar Q_m & \checkmark & \checkmark & 
1.21 & 2.23 \\
\hline
13. \vphantom{\frac{\frac12}{1}} & L E^c H_d^m & \checkmark & \checkmark & 1.20 
& 0.93 \\
\hline
14. \vphantom{\frac{\frac12}{1}} & L H_d E^c_m & \checkmark  & \checkmark & 2.63 
& 2.26 \\
\hline
15. \vphantom{\frac{\frac12}{1}}  & L H_u S_m & \checkmark & \times & 1.13 & 
0.62 \\
\hline
16. \vphantom{\frac{\frac12}{1}}  & L S H_u^m & \checkmark  & \checkmark & 0.98-1.83 & 
0.34-1.61 \\
\hline
17. \vphantom{\frac{\frac12}{1}} & L H_u^m S_{m} & \checkmark & \checkmark & 
2.34 & 1.85 \\
\hline
\end{array}
 $$
\caption{\sf Summary of all the models for both $x$=0.5 and $x$=0.1 case. The 
symbol $\checkmark$ shows that a 125 GeV Higgs can be achieved otherwise the 
symbol $\times$ is used. In the last two columns the range of the $\lambda$ is 
shown. We mention the upper allowed value of $\lambda$ and for the models
 where there is a lower bound we mention the allowed range (see model 9 and 
16).}
\label{t:conclu}
 \end{table}

We perform thorough numerical analysis by using SuSeFLAV considering 
$\tan\beta=10$, \rplb{$\mu_s=10^{-4} \times m_e$} and  $\Lambda=100$ TeV (see 
\Eqn{e:lambda}) for $x=0.1$ and $0.5$. \rplb{ The parameter $x$ played an 
important role and the competition between one-loop and two-loop 
messenger-matter corrections is clear from the plots for these two values of 
$x$.  In model 2, 3, 4, 5, 7, 11, 12, 13, 
14, 15 and 17, there are two bands in $\lambda$, for $x=0.5$. Except for the 
model 11 and 12, these bands get merged in to one for $x=0.1$. }
\rpla{{{\sf\bf To be deleted} $\Rightarrow$}In model 8 and 10, only $M_{\tilde 
N^c}$ gets both the positive and negative corrections at the boundary from 
messenger-matter corrections and these corrections are small compared to right 
handed neutrino mass. Thus these models are insensitive to the value of $x$. 
Model 9 is also insensitive to $x$. One exception for this model (with the other 
two) is  that in this case $\lambda$ has a lower bound set by non-tachyonic mass 
for $\tilde S$.}

\rplb{In the models 1, 2, 3, 5, 6, 8, 9, and 10, we do not get correct Higgs 
mass (see \tabl{t:conclu}) because correction to $M_{\tilde L}^2$ is either 
zero or negative in the boundary. In the rest of the models, $L$ directly 
couples to the messenger fields and thus in these model, $M_{\tilde L}^2$ gets 
both the positive two loop and negative one loop corrections at the boundary. 
For $x=0.1$, two loop effect dominates over the one loop effect and  $M_{\tilde 
L}^2$ becomes comparable to $m_R$. As a consequence, Higgs mass gets 
significant correction.}

\rplb{ EWSB also played an important role in several models where $H_u$ 
and $H_d$ field are present in the messenger-matter Interaction. For 
example, in models 14, for $x=0.5$, $M_{H_u}$  gets both the negative and 
positive corrections whereas for $x=0.1$ positive contributions dominate over 
the negative one-loop contribution. Thus with $\lambda$, this mass increases and 
EWSB  can not be achieved beyond some value. }

Models with messenger-matter corrections to $M_{\tilde S}$ are also interesting. 
 The requirement of non-tachyonic $M_{\tilde S}$ puts lower bound on value of 
$\lambda$. In model 16, for $x=0.5$, complete parameter space is ruled out by 
this condition.

In the models where $Q, U^c$ and $H_u$ are interacting with the 
messenger field $A_t$ term is generated at the boundary and these models are 1, 
2, 6, 11 and 15. \rplb{However, in models 1, 2 and 6 one cannot achieve a 125 
GeV Higgs even for $x=0.1$.  Among all the 17 models, model 11   has 
benchmark points with the   lightest $m_{\tilde t_1}$ mass.  Thus this is the 
most promising model in the context of fine-tuning. }

The set of models presented here has several unique features.  The mass spectra
is like the minimal gauge mediation models in the colored sector;  the stops 
are light  with small or negligible mixing.  In the leptonic sector,  the 
charged  sleptons are light (typically except one eigenvalue), whereas the sneutrinos are heavy, close to 
the right handed neutrino mass scale.  The typical 
collider signatures would mimic that of a minimal GMSB model, with large stau mixing, spectrum in LHC accessible range,  and still have the  lightest CP even higgs mass  
at 125 GeV.\\

\paragraph{Acknowledgement:} We thank Biplob Bhattacherjee and Eung Jin Chun 
for discussions. SKV thanks CPHT, Ecole Polytechnique for hospitality during 
the 
final stages of this work. We thank Carlos E. M. Wagner for raising questions 
regarding factor of $1/2$ in the Higgs mass calculation. \\

\appendix
\section{One loop neutrino-sneutrino corrections to the Higgs mass}
Tree level scalar potential ($V_0$) gets modified by one-loop Coleman-Weinberg 
potential ($\triangle V$)\cite{Coleman:1973jx}. In principle one has to 
calculate vacuum by minimizing the complete potential i.e. $ V_0 + \triangle 
V$; 
however in practice VEV of the Higgs fields are calculated by minimizing the 
tree level potential\footnote{There is a conflict of factor of $1/2$ in Eq. (\ref{higgsmass}) between \cite{Elsayed:2011de} 
and \cite{Guo:2013sna}. We resolve this issue and agree with 
\cite{Guo:2013sna}}. Thus the Higgs mass, which should be double derivative, 
contains some extra single derivative terms:
\be \Delta M_{ij}^2= \frac{1}{2} \left( \frac{\partial^2 \Delta V }{\partial 
H_i 
\partial H_j} 
- \frac{\delta_{ij}}{ H_i} \frac{\partial \Delta V}{\partial 
H_i} \right), \label{higgsmass} \ee
where $i,j=u,d$. Corrections from the top-stop sector are well-known. Here we 
are reviewing the correction to the Higgs mass from neutrino-sneutrino sector. 
Mass eigenvalues of these fields are already mentioned in \S\ \ref{s:mes-mix}. 
Plugging these values into Coleman-Weinberg potential, $\Delta M_{\nu ij}^2$ 
has 
to be calculated. We drop suffix $\nu$ from these terms for brevity.
\begin{eqnarray}\label{Higgscor} 
\Delta  M_{\rm uu}^2 &=& \,\frac{1}{32\pi^2}\sum_{\alpha= 1, 2, 3} \left(\tilde 
L_\alpha\, 
\tilde B_{\alpha u}^2 + m_{\tilde \nu_\alpha}^2 \left(\tilde A_{\alpha 
uu}-\frac{\tilde B_{\alpha u}}{ H_u}\right)(\tilde L_\alpha-1) 
\right)\Bigg|_{H_u= 
\langle H_u \rangle }\nonumber\\
&-& \,\frac{1}{32\pi^2}\sum_{\alpha= 1, 2, 3} \left( L_\alpha\, 
 B_{\alpha u}^2 + m_{ \nu_\alpha}^2 \left( A_{\alpha 
uu}-\frac{ B_{\alpha u}}{ H_u}\right)(L_\alpha-1) \right)\Bigg|_{H_u= 
\langle H_u \rangle },\\*
\Delta  M_{\rm dd}^2 &=& \,\frac{1}{32\pi^2}\sum_{\alpha= 1, 2, 3} \left(\tilde 
L_\alpha\, 
\tilde B_{\alpha d}^2 + m_{\tilde \nu_\alpha}^2 \left(\tilde A_{\alpha 
dd}-\frac{\tilde B_{\alpha d}}{ H_d}\right)(\tilde 
L_\alpha-1)\right)\Bigg|_{H_d= 
\langle H_d \rangle 
},\\*
\Delta  M_{\rm ud}^2 &=& \,\frac{1}{32\pi^2}\sum_{\alpha= 1, 2, 3} \left(\tilde 
L_\alpha\, 
\tilde B_{\alpha u} \tilde B_{\alpha d} + m_{\tilde \nu_\alpha}^2 \tilde 
A_{\alpha 
ud}(\tilde L_\alpha-1)\right)\Bigg|_{H_u= \langle H_u \rangle , H_d= \langle 
H_d 
\rangle },
\end{eqnarray}    
where 
\begin{eqnarray}
L_\alpha &=& \log\left(\frac{m_{\nu_\alpha}^2}{M_{\rm SUSY}^2}\right), \quad
\tilde L_\alpha = \log\left( \frac{m_{\tilde\nu_\alpha}^2}{M_{\rm 
SUSY}^2}\right),\\*
B_{\alpha j} &=& \frac{\partial m_{\nu_\alpha}^2 }{\partial H_j}, \quad 
\tilde B_{\alpha j}  =  \frac{\partial m_{\tilde \nu_\alpha}^2 }{\partial 
H_j}, \\*
A_{\alpha jk} &=& \frac{\partial B_{\alpha j}}{\partial H_k},  \quad 
\tilde A_{\alpha jk} = \frac{\partial B_{\alpha j}}{\partial H_k}.
\end{eqnarray}
Explicit expressions for $\tilde B_{\alpha i}$ are as follows:
\begin{eqnarray}
 \tilde B_{1u} &=&\frac{2 v_d \mu X_N y_N^2}{d_1} + 2 v_u 
\left(1+\frac{m_R^2}{d_2} + \frac{X_N^2}{d_1}\right) y_N^2, \\*
\tilde B_{1d} &=& \frac{-2 v_u \mu X_N y_N^2}{d_1}, \\*
\tilde B_{2u} &=& \frac{-2 v_d \mu X_N y_N^2 }{d_1} +2 v_u \left(1 - 
\frac{X_N^2}{d_1}\right) y_N^2, \\*
\tilde B_{2d} &=&\frac{2 v_u \mu X_N y_N^2}{d_1},\\*
\tilde B_{3u} &=&\frac{-2 v_u m_R^2 y_N^2}{d_2},\\*
\tilde B_{3d}&=& 0.
\end{eqnarray}
Similarly one can calculate the $\tilde A_{\alpha ij}$ terms:
\begin{eqnarray}
\tilde A_{1uu} &=& \frac{2 v_d^2 \mu^2 y_N^2}{v_u^2 d_1} +\frac{4 v_d 
\mu X_N y_N^2}{v_u d_1} + 2 \left(1 + \frac{m_R^2}{d_2} + 
\frac{X_N^2}{d_1}\right) y_N^2, \\*
\tilde 
A_{1ud} &=& \frac{-2 v_d \mu^2 y_N^2}{v_u d_1} - \frac{2 \mu X_N 
y_N^2}{d_1}, \\*
\tilde A_{1dd}  &=& \frac{2 \mu^2 y_N^2}{d_1},   \\*
\tilde A_{2uu} &=& \frac{-2 v_d^2 \mu^2 y_N^2}{v_u^2 d_1}
     -\frac{4 v_d \mu X_N y_N^2}{v_u d_1} + 
2 \left(1 - \frac{X_N^2}{d_1}\right) y_N^2 , \\* 
  \tilde A_{2ud} &=& \frac{2 v_d \mu^2 y_N^2}{v_u d_1} +\frac{2 \mu X_N 
y_N^2}{d_1}, \\*
\tilde A_{2dd} &=& \frac{-2 \mu^2 
y_N^2}{d_1},\\*
 \tilde A_{3uu} &=& \frac{-2 m_R^2 y_N^2}{d_2},\\*
  \tilde A_{3ud} &=& 0, \\*
\tilde A_{3dd} &=& 0. 
\end{eqnarray}
Similarly $B_{\alpha i}$ terms are:
\begin{eqnarray}
 B_{1 u }&=& \frac{4 v_u^3 y_N^4 \mu _S^2}{m_R^4}, \\
 B_{2 u} &=& \frac{2 v_u y_N^2 \left(\frac{v_u^2 y_N^2}{2 
m_R}+m_R\right)}{m_R}, \\
 B_{3 u} &=& \frac{2 v_u y_N^2 \left(\frac{v_u^2 y_N^2}{2 
m_R}+m_R\right)}{m_R}, 
\\
 B_{1 d }&=& 0, \\
 B_{2 d} &=& 0, \\
 B_{3 d} &=& 0.
\end{eqnarray}
Finally the $A_{\alpha ij}$ terms are as follows:
\begin{eqnarray}
 A_{1 uu} &=& \frac{12 v_u^2 y_N^4 \mu _S^2}{m_R^4}, \\
 A_{2 uu} &=& \frac{2 v_u^2 y_N^4}{m_R^2} +\frac{2 y_N^2 \left(\frac{v_u^2 
y_N^2}{2 m_R}+m_R\right)}{m_R}, \\
 A_{3 uu} &=& \frac{2 v_u^2 y_N^4}{m_R^2}+\frac{2 \left(\frac{v_u^2 y_N^2}{2 
m_R}+m_R\right) y_N^2}{m_R},\\
 A_{1 {ud}} &=& 0, \\
 A_{2 {ud}} &=& 0, \\
 A_{3 {ud}} &=& 0, \\
 A_{1 {dd}} &=& 0, \\
 A_{2 {dd}} &=& 0, \\
 A_{3 {dd}} &=& 0. 
\end{eqnarray}

Now being equipped with the above formulas one can calculate Higgs mass. We are 
here giving a simple formula for alignment limit:
\begin{equation}
m_h^2 = m_Z^2 \cos^2 2\beta + \mbox{ top-stop correction } + \sin^2 \beta 
\triangle M_{uu}^2 + \sin2\beta \triangle M_{ud}^2 + \cos^2\beta \triangle 
M_{dd}^2.
\end{equation}


\bibliographystyle{JHEP}
\bibliography{NuMessMix}

\end{document}